\documentclass[sigconf, nonacm]{acmart}

\usepackage{algorithm}
\usepackage[noend]{algpseudocode}
\usepackage{subcaption}
\usepackage{optidef}
\usepackage{mdframed}
\usepackage{multirow}
\newmdtheoremenv{theo}{Theorem}

\theoremstyle{definition}
\newtheorem{exmp}{Example}[section]
\newcommand{\secref}[1]{\S\ref{#1}}
\AtBeginDocument{%
  \providecommand\BibTeX{{%
    \normalfont B\kern-0.5em{\scshape i\kern-0.25em b}\kern-0.8em\TeX}}}

\begin{document}
% \input{revisionletter}
\title{ProBE: Proportioning Privacy Budget for Complex Exploratory Decision Support}

\author{Nada Lahjouji}
\email{nlahjouj@uci.edu}
\orcid{0009-0006-8261-1949}
\affiliation{%
  \institution{University of California, Irvine}
  % \streetaddress{P.O. Box 1212}
  \city{Irvine}
  \country{USA}
}

\author{Sameera Ghayyur}
\email{sghayyur@snap.com}
\orcid{0009-0007-8701-2053}
\affiliation{%
  \institution{Snap Inc.}
  % \streetaddress{P.O. Box 1212}
  \city{Los Angeles}
  \country{USA}
}

\author{Xi He}
\email{xi.he@uwaterloo.ca}
\orcid{0000-0002-4999-4937}
\affiliation{%
  \institution{University of Waterloo}
  % \streetaddress{P.O. Box 1212}
  \city{Waterloo}
  \country{Canada}
}

\author{Sharad Mehrotra}
\email{sharad@ics.uci.edu}
\orcid{0000-0003-1667-5435}
\affiliation{%
  \institution{University of California, Irvine}
  % \streetaddress{P.O. Box 1212}
  \city{Irvine}
  \country{USA}
}
\begin{abstract}

This paper studies privacy in the context of complex decision support queries  composed of multiple conditions on different aggregate statistics combined using disjunction and conjunction operators. Utility requirements for 
%decision support 
such queries
necessitate the need for private mechanisms that guarantee a bound on the false negative and false positive errors. 
%while minimizing privacy loss.
%Previous work in decision support does not encompass the full scope of complex decision support nor does it emphasize binding both errors. 
%This
This
paper formally defines complex decision support queries and their accuracy requirements, and provides algorithms that proportion the existing budget to optimally minimize privacy loss while supporting a bounded guarantee on the accuracy. Our experimental results on multiple real-life datasets show that our algorithms successfully maintain such utility guarantees, while also minimizing privacy loss.
\end{abstract}

\maketitle

\newcommand{\xh}[1]{\noindent{ \textcolor{violet}{[{\bf Xi}: #1]}}}

\newcommand{\nl}[1]{\noindent{ \textcolor{red}{[{\bf Nada}: #1]}}}

\newcommand{\sm}[1]{\noindent{ \textcolor{violet}{[{\bf Sharad}: #1]}}}

\newcommand{\stitle}[1]{\smallskip \noindent{\bf #1}}

\newcommand{\eat}[1]{}

\section{Introduction}
% \nl{make motivation stronger
% 1. DS 2. privacy in DS 3. FN vs FP (exploratory data analysis) 4. previous work MIDE + limitations 5. complex DS queries}

% 1. Decision Support
We consider the privacy-preserving execution of complex aggregate queries over $d$-dimensional data. Consider, for instance, a dataset containing medical records of patients and their respective diseases with the following schema: \textit{PATIENT\_DATA(patient\_name, age,}
\textit{gender, disease, disease\_type)}. An analytical query of interest over such data, listed below in SQL,  identifies  prevalent viral diseases  that afflict vulnerable populations such as the elderly (age over 65) or children (age below 5). 
\begin{verbatim}
    SELECT disease, count(*) FROM PATIENT_DATA
    WHERE disease_type = 'viral'
    GROUP BY disease
    HAVING (count(*) > c1 AND avg(age) > 65)
    OR (count(*) > c2 AND avg(age) < 5)
\end{verbatim}
Such queries often arise in decision support (DS) applications \cite{healthDS,carDS,tpcds,med1} as part of online analytical processing (OLAP)  \cite{olapDS}. OLAP plays   a crucial role in exploring data to produce valuable insight that facilitates informed decision-making. For instance, businesses and organizations utilize decision support to evaluate KPIs \cite{kpi} (Key Performance Indicators), metrics which gauge progress towards an intended goal by computing metrics based on aggregate statistics. Multiple KPIs are often used in tandem to evaluate the performance of businesses, e.g., sales volume and retention rate to infer growth. Such KPIs are instances of complex aggregate queries composed of multiple conditions comparing different aggregate statistics to their respective thresholds, combined using AND/OR operators. Other use cases for such complex queries include clinical decision support applications \cite{med1,med2,med3} which use complex queries to diagnose and classify diseases (e.g. the previously defined query), building management systems \cite{bmsds1,bmsds2} that ensure building code compliance by comparing aggregate statistics to policy thresholds, and supply management systems \cite{supply1,supply2} which optimize operations by analyzing statistics as they relate to existing benchmarks or user-defined criteria.

It follows that data sources used by decision support applications often contain sensitive information about individuals, and releasing aggregated statistics from such sources can lead to severe privacy leaks \cite{recons_attack:dinur2003revealing,recons_attack:dwork2008new}. Differential privacy \cite{Dwork:2014:AFD:2693052.2693053, dwork2} is a popular and effective notion that provides a formal guarantee on privacy by hiding individual records while releasing aggregate statistics, but this is done at the expense of accuracy by adding noise to the data. 
Traditionally, privacy-preserving query answering uses a
 "privacy-first" architecture  that provides   formally defined privacy guarantees while maximizing the utility of data
 %privacy-loss is bounded utility is typically maximized
 given \cite{google,census}. Recent work \cite{mide,Apex:2019:RQP:1007568.1007642,dpella,gupt} has argued the advantages of a dual "utility-first" approach instead, wherein a desired level of utility is specified and privacy maximized given this requirement. This approach is far more suitable in decision support setting, as it not only provides a guarantee on the query answer and therefore confidence in the decision made based on it, but it also offers the opportunity for higher privacy provided that the utility requirements are met at a lower level of invasiveness.

In the context of decision support, utility requirements consist of more than one metric. In particular, DS queries answered by a differentially private mechanism, result in two types of errors: false positives (FP) and false negatives (FN). These errors are the basis of statistical hypothesis testing \cite{hypo}, a widely established method used in multiple fields \cite{hypo1,hypo2,hypo3} which determines the validity of a hypothesis based on sample data. The testing results in either a Type I error (FP) or Type II error (FN) which are subsequently compared to pre-set bounds to make a decision about said hypothesis. DS can be considered a direct application of hypothesis testing, as it uses queries based on statistical methods to make an informed decision.
It is thus critical for a differentially-private DS framework to enforce utility bounds on both false positive and negative errors in order to guarantee the validity of its query results.

Prior work \cite{Apex:2019:RQP:1007568.1007642,mide} has studied the utility-first approach 
%in the context of DS queries, 
but their scope is limited, especially in the context of DS queries. Firstly, they only focus on simple queries which do not have multiple conditions comparing different aggregate statistics to their respective thresholds, i.e. can only answer a query with a single condition in the HAVING clause (e.g. $count(*) > c1$). Secondly, they do not properly tackle the dual utility requirements of DS (i.e., FP and FN). APEx \cite{Apex:2019:RQP:1007568.1007642}, which does not differentiate between the two error types (viz., FP and FNs) considering them both as  errors, offers error bounds only for data point  that are far from the threshold specified in queries. Errors (i.e., misclassification of points as FP and FNs) in an uncertainty region  close to the threshold,  remain unbounded. 
%Thus, error bounds,dN/FP and offers bounds and only offers a guarantee when the data point is far from the threshold.
\textbf{MIDE} \cite{mide}, overcomes this limitation of APEx and offers formal utility bounds irrespective of where data points lie, but it only provides bounds on 
false negatives. Given a set bound on FN, it uses a heuristic approach to explore the trade-off between privacy loss and the FP error. It does so by \textit{weakening} the classifying threshold in such a way that a desired FN bound can be reached at a significantly lower privacy loss while incurring a small penalty on the FP.
Moreover, both APEx and MIDE consider simple queries. 
%utility constraint 
%considered the FN/FP utility requirements, but only proposed mechanisms that guarantee a false negative bound for simple queries. 
%It follows that neither APEX, nor MIDE  meet the requirements of complex DS queries, and does not provide appropriate bounds on both the FN and FP errors.

In this paper, we study a  comprehensive approach to solving the problem of answering complex DS queries in a differentially private manner so as to offer dual utility bounds on both FP and FN while minimizing privacy loss. To address the complexity of the query, the intuition behind our approach is to decompose the original query into multiple simple queries with single aggregate functions (i.e. with known query sensitivities) which can be answered by differentially private mechanisms. Such mechanisms can then be composed by using the respective AND/OR operators used in the original query to link the aggregate functions. The main challenge, then, lies in finding a methodology to similarly decompose or proportion the overall utility bounds across the simple queries (i.e. how to assign allowed error bounds per sub-query to meet the overall accuracy requirement of the final query), as well as the privacy budget, to subsequently guarantee these bounds in a way that optimally minimizes privacy loss. One possible solution could be to divide the utility budgets equally across the multiple conditions, but such an approach, as we will see, is sub-optimal, seeing as different aggregate threshold queries may require a higher error tolerance depending on the selected data distribution. 
Alternatively, we propose \textbf{ProBE}
(\textbf{Pro}portioning Privacy \textbf{B}udget in Complex \textbf{E}xploratory Decision Support Queries), a framework that optimally partitions the privacy/utility budget across the individual simple queries such that the required utility bounds are guaranteed at the lowest overall privacy loss possible. We postulate this as a multi-criteria optimization problem which aims to minimize privacy loss given the constraints enforced by the desired FN/FP bounds. This approach faces two main challenges: it firstly requires quantifying the trade-off between FN and FP errors as well as the trade-off between privacy loss itself and the two errors; secondly, it requires the formulation of the utility bounds as well as the privacy loss as differentiable functions in terms of variables derived from the simple queries. We successfully address these challenges in our approach in the context of the disjunction and conjunction of such queries, and solve a multi-variate optimization problem which yields an apportionment technique for such bounds/budgets. We then propose new algorithms which implement this apportionment framework by adapting previously proposed mechanisms in a way that answers complex DS queries with the appropriate utility guarantees. We subsequently discuss the additional complexities and address our approach to solving them. Our main contributions are as follows:
\begin{itemize}
    \item We formally define privacy-preserving complex decision support queries and their accuracy requirements.
    \item We postulate proportioning the privacy budget for complex decision support queries as a multi-criteria optimization problem and solve it using the method of Lagrange Multipliers.
    \item We propose algorithms that build upon and modify previous mechanisms to implement our budget proportioning technique to support complex decision support queries.
    \item We evaluate our approach against real-world datasets in different domains and show the efficacy of our approach.
\end{itemize}
The organization of this paper is as follows: Section 2 provides background on differential privacy. Section 3 defines complex decision support queries and their accuracy requirements. In Section 4, we propose our ProBE technique to optimally apportion the privacy budget for such queries. Section 5 implements ProBE through two algorithms and proposes additional optimizations. Section 6 evaluates our algorithms on multiple real datasets using complex queries. Lastly, we discuss related work in Section 7 and future work directions in Section 8.

\section{Background}
\label{background}

% \subsection{Background in Differential privacy}
We use existing differential privacy concepts as a basis for our work. Given an input dataset $D \in \mathcal{D}$, an algorithm satisfies differential privacy \cite{Dwork:2014:AFD:2693052.2693053} if its output does not significantly change when adding or removing a single tuple in $D$. Formally:
\begin{definition}
[$\epsilon$-Differential Privacy (DP)]\label{def:ar_ep}
A randomized mechanism $M: \mathcal{D} \to \mathcal{O}$ satisfies $\epsilon$-differential privacy if
\begin{equation}
ln \frac{P[M(D)=O]}{P[M(D')=O]} \leq \epsilon(O)
\end{equation}
for any set of outputs $O \subseteq \mathcal{O}$, and any pair of neighboring databases $D$,$D'$ such that $|D \!\setminus\! D' \cup D'\!\setminus\! D| =1$. 
The privacy metric $\epsilon$ represents the privacy budget. A higher $\epsilon$ value implies higher privacy loss, whereas a lower $\epsilon$ implies strong privacy guarantees.
\end{definition}
Differential privacy offers important properties \cite{Dwork:2014:AFD:2693052.2693053,dawa} that allow for composability of multiple DP mechanisms and assessment of their privacy loss.
\begin{theorem}
[Sequential Composition]\label{def:seq_comp}
Let $M_{1},...,M_{k}$ be $k$ algorithms that satisfy $\epsilon_{i}$-differential privacy. The sequence of $M_{1},...,M_{k}$ provides $\sum_{i=1}^{k}\epsilon_i$-differential privacy. 
\end{theorem}
% \xh{Sequential composition for ex-post DP is tricky: \url{https://arxiv.org/pdf/1705.10829.pdf} (Sec 2.1), \url{https://arxiv.org/pdf/1605.08294.pdf} }
 
When a randomized algorithm runs a $\epsilon$-differentially private algorithm repeatedly until a stopping condition is met, it does not satisfy $\epsilon$-differential privacy because the number of iterations is not known prior to its execution. Its overall privacy loss, however, can be determined after the output is returned. A metric used for these algorithms is \textit{ex-post differential privacy} \cite{ex-postdp}. Formally,
\begin{definition}
[Ex-Post Differential Privacy]\label{def:ar_ep}
Let $\mathcal{E} : \mathcal{O} \to (\mathbb{R}_{\geq 0} \cup \{\infty\})$ be a function on the outcome space of mechanism  $M:\mathcal{D} \to O$. Given an outcome $O = M(D)$, $M$ satisfies $\mathcal{E}(O)$-ex-post differential privacy if for all $O \in \mathcal{O}$,

\begin{equation}
    \max_{D,D':D\sim D'}ln \frac{P[M(D)=O]}{P[M(D')=O]} \leq \mathcal{E}(O)
\end{equation}

% \begin{equation}
% P[M(D) \in O] \leq e^{\epsilon}P[M(D') \in O]
% \end{equation}
for any set of outputs $O \subseteq \mathcal{O}$, and any pair of neighboring databases $D$,$D'$ such that $|D \!\setminus\! D' \cup D'\!\setminus\! D| =1$. Ex-post differentially private mechanisms also benefit from composability properties. Specifically, the Sequential Composition theorem (Def. \ref{def:seq_comp}) holds for ex-post DP mechanisms as well \cite{expostcomposition}, i.e. the sequence of ex-post DP mechanisms results in a differentially private mechanism with privacy loss equal to the sum of ex-post privacy losses $\epsilon_i$.
\end{definition}
% \vspace{1}
% \begin{theorem}
% [Parallel Composition]\label{def:par_comp}
% Let $M_{1},...,M_{k}$ be $k$ algorithms that satisfy $\epsilon_{i}$-differential privacy. Let $D_{1},...,D_{k}$ be the $k$ disjoint partitions of the domain $\mathcal{D}$ and each algorithm $M_i$ is executed on the $D_i$th partition.  The parallel execution of $M_{1},...,M_{k}$ provides $\max(\epsilon_i)$-differential privacy. 
% \end{theorem}
% % laplace and randomiz3eed response.... local very post diff. privacy. 
% \xh{If we did not use parallel composition in this paper, we may drop it}

The Laplace mechanism \cite{Dwork:2014:AFD:2693052.2693053} is a widely used differentially private algorithm that achieves $\epsilon$-differential privacy by adding noise drawn from the Laplace distribution. This noise is also calibrated to the \textit{sensitivity} of the query.
\begin{definition}
[Sensitivity]
The sensitivity of a function $g : \mathcal{D} \to \mathbb{R}^d$, denoted $\Delta g$, is defined as the maximum $L_1$ distance between all pairs of neighboring databases $D$ and $D'$ differing in at most one element.
\begin{equation}
    \Delta g = \max_{\forall D,D'}\parallel g(D)-g(D')\parallel_1
\end{equation}
 The sensitivity of a function highly depends on the aggregate statistic queried. For instance, the sensitivity of a counting query is 1.
\end{definition}
\begin{theorem}
[Laplace Mechanism (LM)]\label{def:lmm}
Given a function $f : \mathcal{D} \to \mathbb{R}^d$, the Laplace Mechanism that outputs $f(D)+\eta$ is $\epsilon$-differentially private, where $\eta$ is a d-length vector of independent samples drawn from a Laplace distribution with the probability density function $p(x|\lambda)=\dfrac{1}{2\lambda}e^{-|x|/\lambda}$ where $\lambda=\Delta f/\epsilon$.
\end{theorem}
\section{Problem Definition} 
In this section, we first formalize the decision support queries and their utility requirements, and then present the problem which answers these queries with DP and utility guarantees.
\subsection{Query and Utility Definitions}
A DS query consists of a set of  {\em aggregate threshold queries} described below. These atomic aggregate threshold queries are connected through logical operators (i.e., $\cup$ and $\cap$), which we refer to as the conjunction or disjunction of multiple such queries. Specifically, an aggregate threshold query returns the set of objects in a dataset, deferred to as predicates, whose aggregate values exceed the set thresholds. We present the formal definitions as follows.

\stitle{Aggregate Threshold Query}. An aggregate threshold query, denoted by $Q_{\mathsf{g(.)}>C}^{\Lambda,f}$,  consists of the following: (i) an aggregate function $\mathsf{g(.)}$, which includes any function whose sensitivity can be computed (e.g. AVG or COUNT); (ii) the set of predicates $\Lambda= \{ \lambda_1,\lambda_2,...,\lambda_k\}$ which represent the objects that the query iterates over to check for condition satisfaction (e.g. the set of diseases in the previous query example); (iii) a set of corresponding thresholds $C = \{c_1, c_2, ..., c_k\}$ for each predicate, and (iv) an optional filter $f$ which can be any selection condition on any column of the record. We use $D^{f}$ to denote all the tuples that satisfy the filter. Each predicate $\lambda_i$ takes in a tuple from filtered tuples $D^{f}$  and outputs \emph{True} or \emph{False} based on its value. We let $D^{f}_{\lambda_i}$ be the set of tuples in $D^f$ that evaluate $\lambda_i$ to be \emph{True}. This query returns all the predicates that have an aggregate over their satisfying tuples $g(D^{f}_{\lambda_i})$ greater than their respective threshold $c_i$, i.e., 
  % \nl{change alpha to u in figure} 
  \begin{equation}\label{eq:query}
  Q_{\mathsf{g(.)}>C}^{\Lambda,f}(D) =\{\lambda_i \in \Lambda \mid \mathsf{g}(D^{f}_{\lambda_i})>c_i\}    
  \end{equation}
   This is similar to a group-by-having query in SQL. 
   % \xh{they are not exactly the same, as groupby in SQL does not specify the k values, but all values. maybe we can make a footnote to differentiate them, or change ``equivalent'' to ``similar''} 
   Given a patient dataset with schema $PATIENT\_DATA(\mathit{patient\_name}, \mathit{age},\mathit{gender},$ $
   \mathit{disease},\mathit{disease\_type}$
   , the following is an example of an aggregate threshold query:
 \begin{verbatim}
 SELECT disease FROM PATIENT_DATA
 WHERE disease_type = 'viral'
 GROUP BY disease HAVING count(*) > c
\end{verbatim}
The WHERE clause \textit{disease\_type = 'viral'} is an example of a filter $f$. The set $disease=\{d_i, i \in [1,k]\}$ is an example of a set of $k$ predicates $\Lambda$. $count()$ is the aggregate function $g(.)$ and $c$ is the threshold, which is the same for all the predicates. An aggregate threshold can also include $\mathsf{g}(D^{f}_{\lambda_i})< c_i$ type inequalities by simply modeling such a condition as the negation of Eq. \eqref{eq:query} and thus returning the negation of its results.
% \xh{break the previous line into multiple sentences.} 
In the context of complex DS queries, we refer to a single aggregate threshold query as an \textit{atomic} query $Q_{a_i}$, which is the basic, irreducible form a complex DS query can take.

% \sm{ refeer to abov as an atomic query... We can then define queries supported by us through the following
% context free grammer.     Q <-- atomic query  ;  if q1 and q2 are queries, themn  q1 and q2 is a qeury.
% q1 or q2 is a query  . We ifnoer not q1 since we can flip the presixate.
%  A query in general can coinsuist of multiple atomic queries. We refer to the atomic queries 
%  as atoms(Q)   give an examole. Note tha an atomic query for a given query Q may appear in multiple 
%  places in the  query. 
%  Given atomic querys queries are generated throuvh following rules.
%     Q <-   AQ    Q <- Q1 AND Q2 ; 

%     Now you xan apportionment problem formally.
% }

\stitle{Complex Decision Support Queries}. We consider a complex DS query $Q^{\Lambda,\mathcal{F}}$ to be a set of atomic aggregate threshold queries $Q_{a_1},...,Q_{a_n}$ connected by the AND/OR operators ($\cap,\cup$), where these $n$ atomic queries share the same set of predicates $\Lambda$, but have different filters $\mathcal{F}=\{f_1,...,f_n\}$,  aggregate functions $\mathcal{G} =\{ g_1,...,g_n\}$ and thresholds $\mathcal{C} = \{c_1,...,c_n\}$.

\begin{figure}[t]
    \centering
    \includegraphics[width=.49\textwidth]{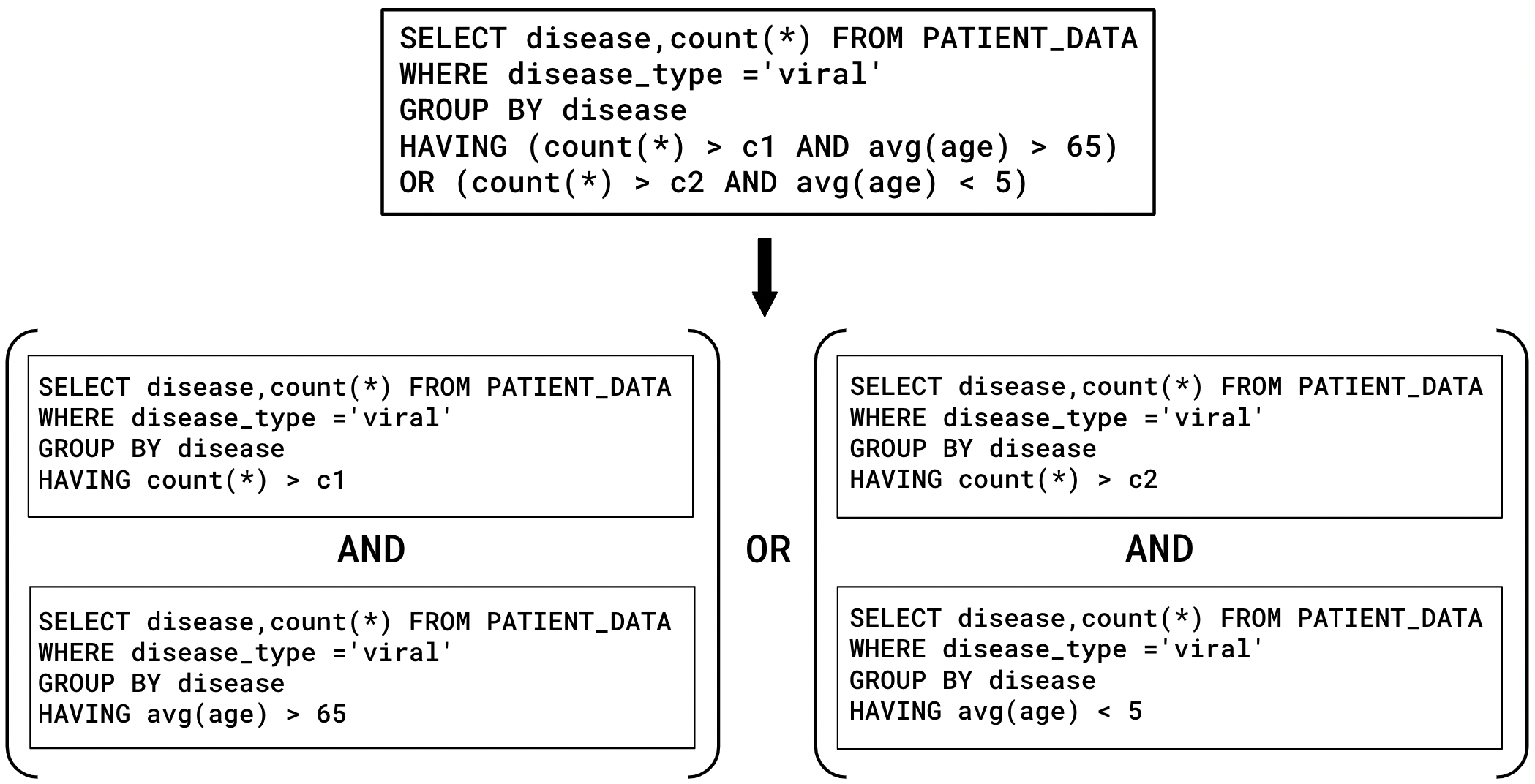}
    \caption{Complex decision support query decomposed into four atomic queries with single HAVING conditions but similar predicates connected by AND/OR operators.}
    \label{fig:exampleq}
\end{figure}

Essentially, we deconstruct a query composed of multiple aggregate functions compared to their respective thresholds into separate, atomic aggregate threshold queries with a single function-threshold pair. Figure \ref{fig:exampleq} shows an example of the decomposition process for the complex DS query previously introduced. The atomic queries contain single conditions comparing an aggregate function $g_i$ to its respective threshold $c_i$, and their results are subsequently composed back together using the AND/OR operators defined in the original query.
To formally define this concept, we make use of a Context-Free Grammar (CFG) as shown below.

\begin{definition}
    [Complex DS Query CFG]\label{def:cfg} Consider the grammar $G = (N,\Sigma,P,S)$ where 
    the complex DS query is a non-terminal symbol $N = \{Q^{\Lambda,\mathcal{F}}\}$, the atomic aggregate threshold query is a terminal symbol $\Sigma = \{Q_a\}$, and $S = Q^{\Lambda,\mathcal{F}}$ is the starting symbol. The production rules $P$ are:
    \begin{subequations}
    \begin{align}
        Q^{\Lambda,\mathcal{F}} &\rightarrow Q_a \label{eq:prod1}\\
        Q^{\Lambda,\mathcal{F}} &\rightarrow Q^{\Lambda,\mathcal{F}_{i_1}} \cap Q^{\Lambda,\mathcal{F}_{i_2}} \label{eq:prod2} \\ 
        Q^{\Lambda,\mathcal{F}} &\rightarrow Q^{\Lambda,\mathcal{F}_{i_1}} \cup Q^{\Lambda,\mathcal{F}_{i_2}}\label{eq:prod3}  
    \end{align}
    \end{subequations}
    % \xh{Does this grammar enforce all atomic queries have the same set of predicates and filters?}
    % \nl{I think so.. I added subscripts for filters to show that they are different, but the predicates have the same $\Lambda$}
\end{definition}
Using this grammar, we can recursively compose complex decision queries with any combination of AND/OR operators. Specifically, by using production rule \ref{eq:prod1} and \ref{eq:prod2} we produce a complex query composed of atomic queries connected only by the AND operator like $Q^{\Lambda,\mathcal{F}}=Q_1\cap Q_2 \cap,\ldots,\cap Q_n$, which we refer to as a \textbf{conjunction query}. Similarly, by using production rules \ref{eq:prod1} and \ref{eq:prod3} we derive a query composed of OR operators only $Q^{\Lambda,\mathcal{F}}=Q_1\cup Q_2 \cup,\ldots,\cup Q_n$, which we refer to as a \textbf{disjunction query}.

\stitle{Utility Measures}.
Decision support applications, as mentioned previously, require setting a bound on the false negative and false positive errors while minimizing privacy loss. We define these bounds formally below.
% Defn  Bounds on FNR & Prob of false Neg.

% Let M be mechansim, Q be a complex query, and D a database. We say M satisfies a beta-bound on false negative rates if for all D, and  Q, and for all 
% pred_i in Q^ the following holds:

%   confitional prob < beta

% Likewise we say M satifies a gamma-bound on prob of false negatives if for all D, and  Q, and for all 
% pred_i in Q^ the following holds:

%    real prob < gamma

%    Note that a mechanism that satisiies b-bound on FNR also satisifes the b-bound on FNR  (due t bayes theorem) but hte other
%    way round not necc.  We define both...
% \noindent
\begin{definition}
[Bound on the False Negative Rate (FNR)/ False Positive Rate(FPR)]\label{def:ar_fn} 
Let $M:\mathcal{D} \to O$ be a randomized mechanism that answers a complex decision support query $Q^{\Lambda,\mathcal{F}}$ composed of $n$ atomic aggregate threshold queries $Q_1,Q_2,\ldots,Q_n$ with the same predicates $\Lambda$ and different filters $F$. We say $M$ satisfies 
(i) a $\beta$-bound on the FNR
if for any database $D\in \mathcal{D}$, for all predicates $\lambda_i \in \Lambda$, the following holds:
\begin{equation}
P[\lambda_i \not \in M(D) \wedge \lambda_i \in Q^{\Lambda,\mathcal{F}}(\mathcal{D} = D)] \leq \beta
\label{eq:fnrdef}
\end{equation}
(ii)  a $\alpha$-bound on the FPR
if for any database $D\in \mathcal{D}$, for all predicates $\lambda_i \in \Lambda$, the following holds:
\begin{equation}
P[\lambda_i \in M(D) \wedge \lambda_i \not \in Q^{\Lambda,\mathcal{F}}(\mathcal{D} = D)] \leq \alpha
\label{eq:fprdef}
\end{equation}
\end{definition}
In other words, Eq. \eqref{eq:fnrdef} represents a bound $\beta$ on the FNR, i.e. the probability that a predicate $\lambda_i$ is not in the result of mechanism $M$ given it is in the result of query $Q^{\Lambda,\mathcal{F}}$ for all predicates. Similarly, Eq. \eqref{eq:fprdef} represents a bound $\alpha$ on the FPR, i.e. the probability that a predicate $\lambda_i$ given it is in the result of $M$ but not in the result of $Q^{\Lambda,\mathcal{F}}$ for all predicates. We specify $\mathcal{D}=D$ as the probability of a predicate being in the result of the query considers the data distribution, but for the rest of the paper we use $D$ for simplicity. 

Note that in our framework, the $\beta$ and $\alpha$ accuracy bounds are user-set parameters and depend largely on the nature of the overarching decision support application; for instance, a medical diagnosis application may choose to emphasize the FNR over the FPR if the absence of disease detection is more crucial than an erroneous detection. This model is motivated by similar models used in approximate query processing \cite{aqp} where users specify confidence intervals, or statistical inference which is used to control false negative and/or positive errors \cite{statinf1}. Additionally, we choose to focus on FNR/FPR over other accuracy measures due to the nature of the queries we tackle, which return a set of objects rather than aggregate values. For instance, a different metric such as variance error may yield a high error value for a specific query whereas the returned set remains unchanged, thus not precisely reflecting the accuracy of the results.
%Other measures exist for quantifying FN/FP, which we discuss in Appendix B, but we focus on the probability in this paper. 
% \xh{We may comment what this probability means, and why we specify $(\mathcal{D}=D$, as the second part of the probability term considers the data distribution, the probability for the predicate returns True for a database $D$? And for simplicity, we simply wrote it as $D$ for the rest of the paper. Can add comments on other types of bound like FNR/FPR, but we focus on the prob for this paper.}

\eat{
\begin{definition}
[Bound on Probability of False Positives (FPR)]\label{def:ar_fp} 
Let $M:\mathcal{D} \to O$ be a randomized mechanism that answers a complex decision support query $Q^{\Lambda,\mathcal{F}}$ composed of $n$ atomic aggregate threshold queries $Q_1,Q_2,\ldots,Q_n$ with the same predicates $\Lambda$ and different filters $F$. We say $M$ satisfies a $\alpha$-bound on the false positive probability
if for any database $D\in \mathcal{D}$, for all predicates $\lambda_i \in \Lambda$, the following holds:
\begin{equation}
P[\lambda_i \in M(D) \wedge \lambda_i \not \in Q^{\Lambda,\mathcal{F}}(\mathcal{D} = D)] \leq \alpha
\end{equation}
\end{definition}

\begin{definition}
[Bound on Probability of False Negatives (FNR)]\label{def:ar_fn} 
Let $M:\mathcal{D} \to O$ be a randomized mechanism that answers a complex decision support query $Q^{\Lambda,\mathcal{F}}$ composed of $n$ atomic aggregate threshold queries $Q_1,Q_2,\ldots,Q_n$ with the same predicates $\Lambda$ and different filters $F$. We say $M$ satisfies a $\beta$-bound on the false negative probability
if for any database $D\in \mathcal{D}$, for all predicates $\lambda_i \in \Lambda$, the following holds:
\begin{equation}
P[\lambda_i \not \in M(D) \wedge \lambda_i \in Q^{\Lambda,\mathcal{F}}(\mathcal{D} = D)] \leq \beta
\end{equation}
\end{definition}
\begin{definition}
[Bound on Probability of False Positives (FPR)]\label{def:ar_fp} 
Let $M:\mathcal{D} \to O$ be a randomized mechanism that answers a complex decision support query $Q^{\Lambda,\mathcal{F}}$ composed of $n$ atomic aggregate threshold queries $Q_1,Q_2,\ldots,Q_n$ with the same predicates $\Lambda$ and different filters $F$. We say $M$ satisfies a $\alpha$-bound on the false positive probability
if for any database $D\in \mathcal{D}$, for all predicates $\lambda_i \in \Lambda$, the following holds:
\begin{equation}
P[\lambda_i \in M(D) \wedge \lambda_i \not \in Q^{\Lambda,\mathcal{F}}(\mathcal{D} = D)] \leq \alpha
\end{equation}
\end{definition}}

\subsection{ProBE Problem Definition}
\label{ssec:probdef}

 Given a complex decision support query $Q^{\Lambda,\mathcal{F}}$ on a dataset $D$ composed of atomic queries $Q_{a_1},...,Q_{a_n}$ in the structure of query tree $T$, we want to develop a differentially private mechanism $M_{\textsc{probe}}(T)$ that answers the overall query such that privacy loss $\epsilon$ is minimized subject to a $\beta$-bound on FNR and $\alpha$-bound on FPR. 

% \xh{I moved the mechanisms as they part of the problem definition rather than the accuracy specification. The transition to the optimization problem is also smoother.}

\stitle{Primitives for Atomic Queries and Limitations.} We show two mechanisms from prior work to illustrate their utility guarantees for a simple atomic aggregate threshold query $Q^{\Lambda,f}_{g(\cdot)>C}$ and then present the problem formulation that builds on top of these mechanisms for complex decision support queries.

\begin{figure}[t]
    \centering
    \includegraphics[width=.38\textwidth]{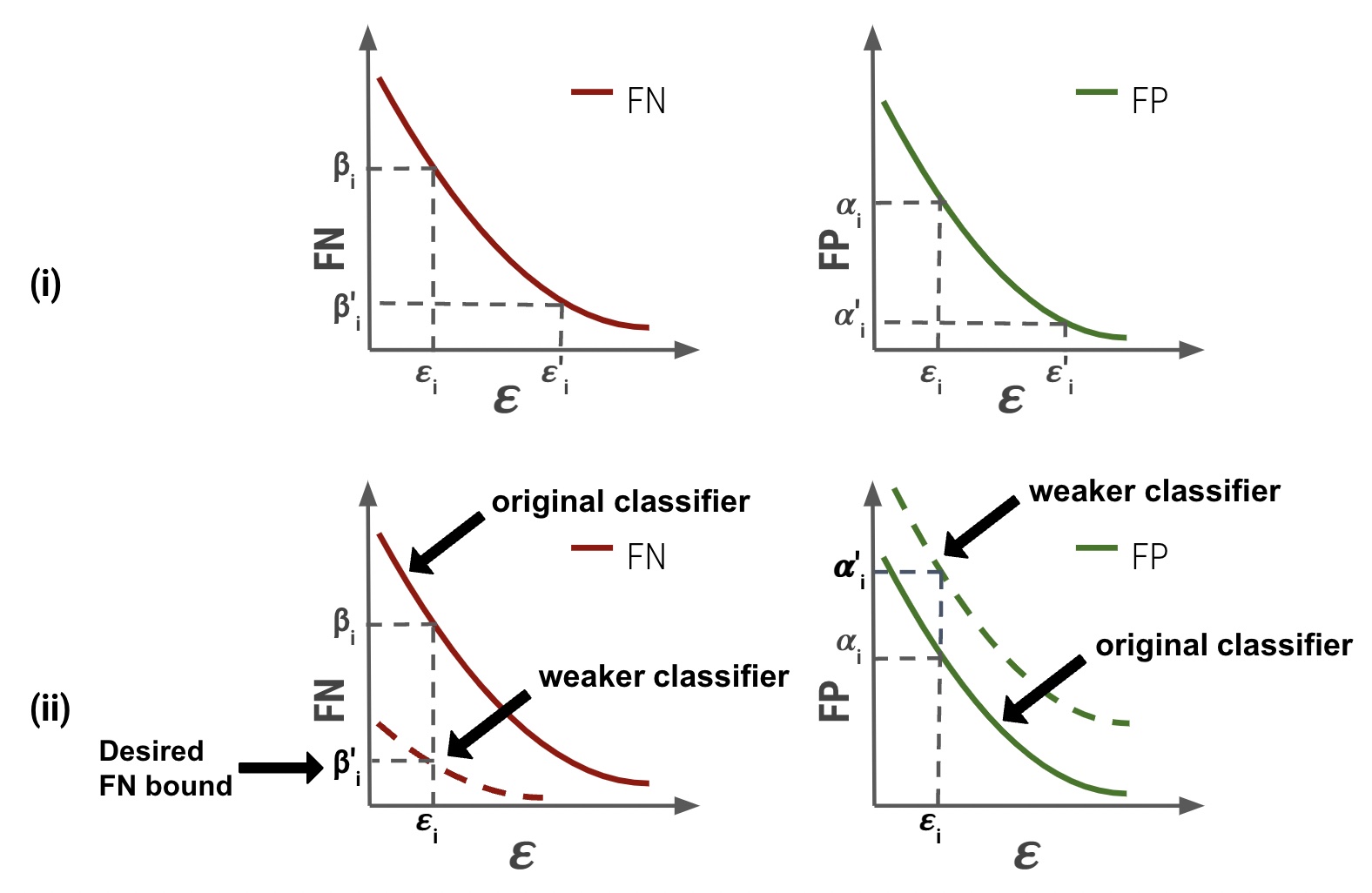}
    \caption{Trade-off between false negatives FN, false positives FP and the privacy budget $\epsilon$ in (i) with APEx and (ii) with MIDE.}
    \label{fig:tradeoff}
\end{figure}

% \nl{I think below might not be completely correct? This mechanism is actually apex, not laplace. standard laplace does not offer the guarantee at all and there is no concept of an uncertain region. Here we can explain the parameters more since there's more space.}
% \nl{Added more text on uncertain region and parameters.}
The first mechanism is the Laplace mechanism used by APEx~\cite{Apex:2019:RQP:1007568.1007642} to answer atomic queries, but \emph{it fails to offer either bounded FPR or bounded FNR}. 
It adds noise to the aggregate value per predicate and compares the result to its corresponding threshold $c_i$ given a $\beta$ bound on the overall error rate and a region $u$ around the threshold in which the error is unbounded as inputs (i.e. error tolerance region). 
However, this mechanism does not offer the $\beta$-bound on FNR guarantee for predicates which have aggregates too close to their thresholds, i.e. their aggregates are in the region 
$[c_i-u,c_i+u]$. We refer to $u$ as the \textit{uncertain region}, which factors into determining the privacy budget needed to achieve the $\beta$ guarantee outside of said region by using 
$\epsilon=\frac{\Delta g\ln(1/(2\beta)}{u}$. It follows that a larger uncertain region $u$ implies smaller privacy loss, at the expense of increased false negatives and positives alike. Conversely, a smaller $u$ would lead to less unbounded errors, but would increase privacy loss as a result. Figure \ref{fig:tradeoff}(i) shows this relationship between FP/FN and $\epsilon$, where increasing $\epsilon$ leads to an equal decrease in FP/FN and vice versa.
\setlength{\textfloatsep}{-0.5pt}

% \xh{change to "The second mechanism, known as Threshold Shift Laplace Mechanism (TSLM) designed by mIDE~\cite{mide} offers \emph{one-sided guarantee to either bounded FNR or bounded FPR} to atomic queries. TSLM generaizes ... ''}
The second mechanism, known as Threshold Shift Laplace Mechanism (TSLM) designed by MIDE~\cite{mide} offers a \emph{one-sided guarantee on either bounded FNR or bounded FPR}. TSLM  generalizes the threshold $c_i$ by shifting $c_i$ to $c_i-u$ so that the entire uncertain region (i.e. the unbounded error) resides on the left side of the old threshold, and compares this new threshold to the noisy aggregate. Figure \ref{fig:tradeoff} (ii) shows the effect of this approach on the trade-off between FN/FP and $\epsilon$, where shifting the threshold by $u$ results in a weaker classifier which achieves a lower bound on FN at the same $\epsilon$ but at a higher FP. By setting the privacy budget $\epsilon=\frac{\Delta g\ln(1/(2\beta)}{u}$ as defined in \cite{Apex:2019:RQP:1007568.1007642}, this solution 
guarantees a $\beta$-bound on FNR for all predicates. 

\vspace{1.5mm}
\setlength{\fboxrule}{1.5pt}
\noindent
\fbox{%
\parbox{.96\linewidth}{\textbf{Threshold Shift Laplace Mechanism (TSLM):} Given an atomic aggregate threshold query $Q^{\Lambda,f}_{g(\cdot)>C}$, by setting the privacy budget to $\epsilon=\frac{\Delta g\ln(1/(2\beta)}{u}$ where $u$ is the generalized parameter used to shift threshold $C$ to $C-u$, the Threshold Shift mechanism achieves a $\beta$-bound on FNR.
}
}
\vspace{1.5mm}

Through the threshold shift mechanism, the uncertain region is shifted to $[c_i-2u,c_i]$, thus providing a formal bound $\beta$ on the FNR without incurring additional privacy cost, but doing so at the expense of the FPR due to the unbounded error being entirely made of false positives. Due to the nature of this algorithm, it does not provide any sort of bound on the FPR but rather increases them due to the trade-off between the FNR and FPR resulting from the shift. This is due to the fact that the increase in FPR resulting from the shift depends entirely on the data distribution, hence the FPR cannot be pre-determined before running the mechanism. 
% \xh{may comment that ``The changes in the false positives due to the shift depend on the data distribution. Hence, we cannot pre-determine the FPR/FPR before running TSLM.''}
% MIDE \cite{mide}  further proposes multi-step algorithms that make use of predicate-wise DP to minimize privacy loss. We extend these mechanisms to answer complex decision support queries, which will be explained in Section 5.

\stitle{Optimization Problem for Complex Queries.}
Minimizing the privacy loss while bounding both FPR and FNR before running a DP mechanism is difficult to achieve unless data distribution is known ahead. Hence, we formulate a hybrid approach instead, wherein one of the two constraints is fixed in our optimization problem and chosen to generate possible solutions, whereas the second constraint is used post-optimization to algorithmically relax the solution in such a way that ensures its bound is upheld. For the rest of our paper, we consider the constraint on FNR as our optimization problem constraint, and FPR to be our post-optimization constraint, but a mirrored problem (i.e. flipping the constraints) is supported by our approach as well. The choice of constraints relies on the nature of the DS application and its intended use cases.

As each atomic query $Q_{a_i}$ can be individually answered using a randomized mechanism 
% \xh{the equation should take in $\beta_i$ here right?} 
$M_a(\beta_i):\mathcal{D} \to \mathcal{O}_i \times \mathbb{R}^+$
% \xh{change $O_i$ to $\mathcal{O}_i \times \mathbb{R}^+$} 
that takes an FNR bound $\beta_i$ and outputs a query answer and
% \xh{add ``a query answer and ''} 
 ex-post privacy loss $\epsilon_i$  (e.g. mechanisms presented in \cite{mide}), we want to use the outputs of such mechanisms $O_I$, and privacy budgets $\epsilon_i$ as inputs for our mechanism $M_{\textsc{probe}}(T)$.
% \xh{may remove subscript $q$ as it is not used anywhere?}
 Thus, the aim of ProBE is to generate functions that apportion the overall FNR bound in terms of each atomic query's FNR bound to formulate a minimization problem for privacy loss $\epsilon_{\textsc{probe}}=f_\epsilon(\epsilon_1,..,\epsilon_n)$. 
 Specifically, we want to generate a function $f_\beta(\beta_1,...,\beta_n)$ which apportions the $\beta$ bound into each $\beta_i$ such that $FNR \leq \beta$, and  $\epsilon(\epsilon_1,..,\epsilon_n)$ is minimized.
We therefore obtain a constrained optimization problem defined as:
\begin{mini}|l|[0]
{\epsilon_1..\epsilon_n}{f_\epsilon(\epsilon_1,..,\epsilon_n)}
{}{}
\addConstraint{f_\beta(\beta_1,...,\beta_n) \leq \beta}
\label{eq:prelimoptprob}
\end{mini}
We first solve this optimization problem given the single constraint on FNR in Section 4, where we first develop ProBE instantiated with the $\beta$-bound on FNR. We then relax our solution by enforcing the post-optimization constraint on FPR in Section 5.
% \xh{I re-organize the text. Please double-check the transitions and ensure important points are not missed. Also, check the section refs in the later part.}

\eat{
\subsection{ProBE Problem Definition}
\label{ssec:probdef}
% \xh{I feel the 2nd paragraph belongs to the solution section, instead of the problem definition. How about we directly start this section with the third paragraph and equation (8) will include the constraint on both the FNR and FNR. After fully state the problem to the problem with just FNR with the post-optimization. 
% }
% \xh{Can we add pointer to where the post-optimization in the paper?}
% \nl{does this make more sense?}

Given a complex decision support query $Q^{\Lambda,\mathcal{F}}$ on a dataset $D$ composed of atomic queries $Q_{a_1},...,Q_{a_n}$ in the structure of query tree $T$, we want to develop a novel mechanism $M_{\textsc{probe}}(T)$
% \xh{may remove subscript $q$ as it is not used anywhere?}
that answers the overall query such that privacy loss $\epsilon$ is minimized given a bound on FN and FP. 
% \xh{may say ``We focus on FNR and FPR for our problem analysis from now onward. FNR and FPR can be similarly derived.'' or even somewhere earlier in the end of sec 3.1} 
As each atomic query $Q_{a_i}$ can be individually answered using a randomized mechanism 
% \xh{the equation should take in $\beta_i$ here right?} 
$M_a(\beta_i):\mathcal{D} \to \mathcal{O}_i \times \mathbb{R}^+$
% \xh{change $O_i$ to $\mathcal{O}_i \times \mathbb{R}^+$} 
that takes an FN bound $\beta_i$ and outputs a query answer and
% \xh{add ``a query answer and ''} 
 ex-post privacy loss $\epsilon_i$  (e.g. mechanisms presented in \cite{mide}), we want to use the outputs of such mechanisms $O_I$, and privacy budgets $\epsilon_i$ as inputs for our mechanism $M_{\textsc{probe}}(T)$.
% \xh{may remove subscript $q$ as it is not used anywhere?}
 Thus, the aim of ProBE is to generate functions that apportion the overall FN bound in terms of each atomic query's FN bound to formulate a minimization problem for privacy loss $\epsilon_{\textsc{probe}}=f_\epsilon(\epsilon_1,..,\epsilon_n)$, while also guaranteeing an additional bound on FP. 
 Specifically, we want to generate a function $f_\beta(\beta_1,...,\beta_n)$ which apportions the $\beta$ bound into each $\beta_i$ such that $FNR \leq \beta$, $FPR \leq \alpha$, and  $\epsilon(\epsilon_1,..,\epsilon_n)$ is minimized.
We therefore obtain a constrained optimization problem defined as:
\begin{mini}|l|[0]
{\epsilon_1..\epsilon_n}{f_\epsilon(\epsilon_1,..,\epsilon_n)}
{}{}
\addConstraint{f_\beta(\beta_1,...,\beta_n) \leq \beta}
\addConstraint{FPR \leq \alpha}
\label{eq:prelimoptprob}
\end{mini}
% Likewise, we can express our optimization problem in terms the FN/FP rates by apportioning each atomic query $Q_a$'s individual false negative probability bound $\beta_i$ guaranteed by a mechanism $M_a$ which is used as the constraint function.
 % Previous work did not tackle the problem of answering complex decision support queries in a way that encompasses the full scope of their utility requirements, i.e. providing a bound on both false negatives and false positives, but can nonetheless be used as a basis for providing a theoretical bound on error in a differentially-private framework. Our ProBE solution thus aims to develop differentially-private mechanisms which answer complex decision support queries by minimizing privacy loss $\epsilon$ subject to constraints $ FNR \leq \beta $ and $FPR \leq \alpha$.
 
 Such an approach, however, is difficult to achieve. The challenge lies in that there is no way to formulate the FP bound due to the individual mechanisms $M_i$ not guaranteeing such a bound on the respective sub-queries. This is attributed to the unbounded error which occurs in the uncertain region $u$ (explained in \secref{ssec:accreq}), which causes the binding of a specific error type on one side of the threshold (e.g. FN on the right side) to incur a cost on the opposite error (e.g. FP on the left side) and vice versa, meaning that generating solutions for our optimization problem given the two constraints on FN and FP is not possible unless data distribution is known.

We formulate a hybrid approach instead, wherein one of the two constraints is fixed in our optimization problem and chosen to generate possible solutions, whereas the second constraint is used  post-optimization to algorithmically relax the solution in such a way that ensures its bound is upheld. For the rest of our paper, we consider the constraint on FN as our optimization problem constraint, and FP to be our post-optimization constraint, but a mirrored problem (i.e. flipping the constraints) is supported by our approach as well. The choice of constraints relies on the nature of the DS application and its intended use cases.
Thus, we formalize and solve our optimization problem given the single constraint on FN in Section 4, where we first develop ProBE instantiated with the $\beta$-bound on FNR. We then relax our solution by enforcing the post-optimization constraint on FP in Section 5.}
\section{ProBE Framework}
In this section, we first present optimization techniques for complex decision support queries that consist of a single connection operator (conjunction or disjunction) with the purpose of guaranteeing a $\beta$-bound on FNR. We then generalize our approach to queries combining both conjunction/disjunction operators for $n$ atomic queries.
 % \xh{we can remove this line. instead, we can provide detailed pointers to the appendix since we can include appendix to this pdf.} \textbf{Proofs of lemmas and equations in this section can be found in Appendix A.}

% \xh{After I read this section, I think the intro of this section may be updated simply as this ``In this section, we first present optimization techniques for complex decision support queries that consist of a single connection operator (conjunction or disjunction) that links two atomic aggregate threshold queries. Then we generalize the techniques to queries combining both conjunction and disjunctions for $n$ atomic queries. \textbf{Formal ... Appendix}. ''}

\subsection{Query Conjunction Mechanism}
Consider a conjunction query composed of a conjunction operator that links two atomic aggregate threshold queries $Q=Q_1\cap Q_2$. Our optimization problem has two differentially private mechanisms, $M_1$ and $M_2$ of the same type, that answer $Q_1$ and $Q_2$, respectively. One way to combine the independent outputs of $M_1$ and $M_2$ for the final answer of $Q$ is to find their intersection $M(D)=M_1(D)\cap M_2(D)$. We define this step as \emph{query conjunction mechanism}.

\begin{definition}[Query Conjunction Mechanism]
\label{def:M_conj}

Let randomized mechanism $M_i:\mathcal{D} \to O_i$ with a differential privacy guarantee satisfy a $\beta_i$-bound on FNR for aggregate threshold query $Q_i$. We can answer query $Q$ which is a conjunction of 2 aggregate threshold queries $Q_1 \cap Q_2$ using mechanism $M$ where $M(D)=M_1(D)\cap M_2(D)$. 
\end{definition} 
% \xh{may update this to a mechanism for a single conjunction operator, like a primitive mechanism that can be used in more complex queries.}

\begin{figure}
    \centering
    \includegraphics[width=.48\textwidth]{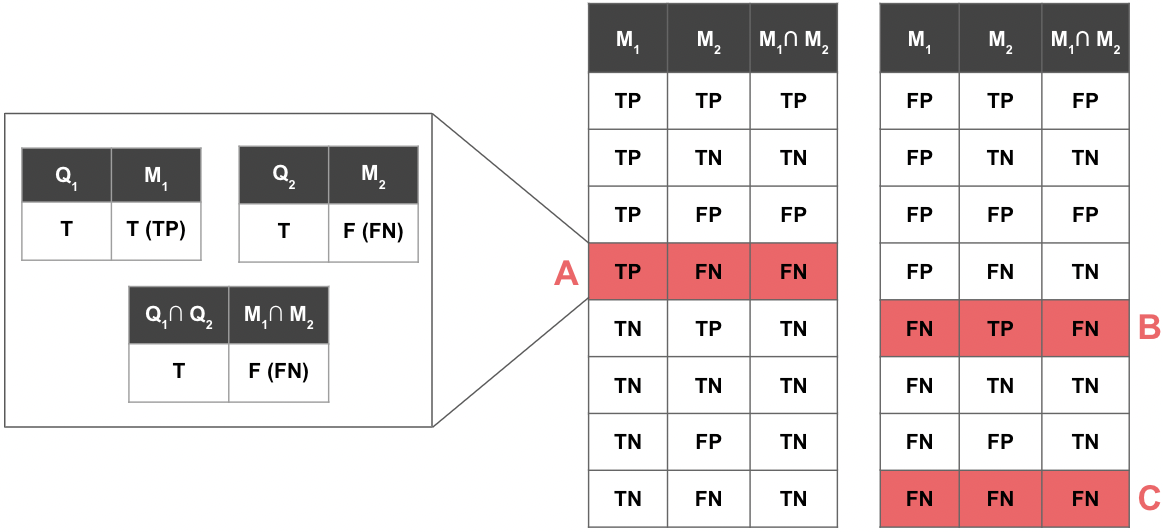}
    \caption{Classification of conjunction of outputs $M_1$ and $M_2$ resulting from running the mechanisms on $Q_1$, $Q_2$. False negative (FN) outcomes are highlighted in red. }
    \label{fig:truthtable}
\end{figure}

\stitle{$\boldsymbol\beta$-Bound on False Negative Rate}.
As per Sec. 3.2, we know that a mechanism $M_i$ that answers an individual aggregate threshold query has an associated $\beta_i$-bound on FNR. To derive the apportioning function $f_\beta$ for the FNR for $Q$ in terms of $\beta_1$ and $\beta_2$, we generate a confusion matrix (Figure \ref{fig:truthtable}) by running $M_1$ and $M_2$ on $Q_1$ and $Q_2$ respectively and classifying their outcomes, as well as classifying their conjunction and disjunction. As seen in Figure \ref{fig:truthtable}, the conjunction mechanism $M= M_1 \cap M_2$ results in a false negative in any of the three cases (A,B,C). As $M_1$ and $M_2$ are mechanisms of independent randomness, their outcomes are independent from one another given the true query answer, though the atomic queries themselves $Q_1$ and $Q_2$ may not be independent. Note that because the three cases are mutually exclusive, we can simply add their probabilities to obtain the overall probability of M resulting in a false negative. Thus, we can deduce the overall FNR as the three lines below.  
% We denote each sub-query's selectivity, i.e. $P[\lambda_j \in Q_i]$ as $s_i$. 
For any predicate $\lambda_j\in \Lambda$, 
\begin{eqnarray}
 && P[\lambda_j\!\notin\! M(D) | \lambda_i\!\in\! Q^{\Lambda,F}(D)] \nonumber\\
 &=&  P[\lambda_j \! \in M_1(D)  |  \lambda_j \! \in \! Q_1(D)] \! \cdot\! P[\lambda_j \! \notin \! M_2(D)  |  \lambda_j \! \in \! Q_2(D)]    \nonumber\\
 &+& P[\lambda_j \! \notin M_1(D)  |  \lambda_j \! \in \! Q_1(D)] \! \cdot\! P[\lambda_j \! \in \! M_2(D)  |  \lambda_j \! \in \! Q_2(D)]   \nonumber\\
  &+& P[\lambda_j \! \notin M_1(D)  |  \lambda_j \! \in \! Q_1(D)] \! \cdot\! P[\lambda_j \! \notin \! M_2(D)  |  \lambda_j \! \in \! Q_2(D)]  \nonumber\\
  &=&(1-FNR_1) \! \cdot \! FNR_2 \!+\! (1-FNR_2) \! \cdot \! FNR_1 \! + \! {FNR}_1 \! \cdot \! FNR_2\label{eqwithin}\label{eqwithin}\\
    &=& FNR_1+FNR_2- FNR_1 \cdot FNR_2\\
    &\leq& FNR_1+FNR_2 \leq \beta_1+\beta_2 \nonumber
  \end{eqnarray}

The last inequality holds due to the $\beta_i$-bound guarantee on $FNR_i$ provided by $M_i$. Hence, we obtain the function: 
\begin{eqnarray}
    f_{\beta}(\beta_1,\beta_2)=\beta_1+\beta_2
    \label{eq:beta12conj}
\end{eqnarray} 

The detailed analysis can be found in Appendix~\ref{app:fnrbound}.
% As per Def \ref{def:ar_fn}, we want to bound the resulting false negative rate FNR by $\beta$. We use Eq. (\ref{eq:fnr_conj}) to derive an upper bound on the false negative rate FNR:
% \begin{equation*}
% FNR \leq \beta_1 + \beta_2 - \beta_1\beta_2 \leq \beta
% \end{equation*}
% This can be further simplified by providing an upper bound on FNR by removing the clause $-\beta_1\beta_2$ since it is always negative. We thus obtain the apportioning function:
% \begin{equation}   
% FNR = f_\beta(\beta_1,\beta_2) = \beta_1 + \beta_2 \leq \beta 
% \label{eq:final_fnr_conj}
% \end{equation}

\stitle{Privacy Loss}. 
We apply the sequential composition theorem of differential privacy (Def. \ref{def:seq_comp}) to compose the privacy loss of the two sub-mechanisms of $M$. Hence, the budget apportioning function of $M$ is
% Because the mechanism we are using is composed of a set of sub-mechanisms i.e. $M(D)=M_1(D)\cap M_2(D), ... , M_n(D)$ ran sequentially, we can use the Sequential Composition theorem (Def. \ref{def:seq_comp}) to compute the overall $\epsilon$. Thus, the overall privacy loss can be written as a sum of the individual privacy losses $\epsilon_i$ resulting from their respective mechanisms. In the case of a 2-query conjunction, we obtain the function: 
% \xh{shorten to ``We apply sequential composition of differential privacy (Def. \ref{def:seq_comp}) to compose the privacy loss of the two sub-mechanisms of $M$. Hence, the budget apportioning function of $M$ is}
\begin{equation}
    \epsilon=f_\epsilon(\epsilon_1, \epsilon_2) = \epsilon_1 + \epsilon_2
\end{equation}
This means that our optimization problem is now as follows.
\begin{mini}|l|[0]
{}{\epsilon_1 + \epsilon_2}
{}{}
\addConstraint{\beta_1+\beta_2  \leq \beta}
\label{eq:optprob}
\end{mini}
We consider the primitive mechanism for atomic queries, TSLM~\cite{mide}, described in Sec.~\ref{ssec:probdef} to illustrate the optimization.
% For each aggregate threshold query, we consider the aggregate function $g(.)$ to be a counting function, meaning that its sensitivity $\Delta g$ is always 1.
In this case, we can use the privacy budget $\epsilon_i =\frac{\Delta g_i\ln(1/(2\beta)}{u_i}$ which guarantees a $\beta_i$-bound on FNR. Thus, we rewrite the $\epsilon$ apportioning function as follows:
\begin{eqnarray}
f_\epsilon(\epsilon_1, \epsilon_2)&=& \epsilon_1 +\epsilon_2  \nonumber\\ 
 &=& \frac{\Delta g_1\ln(1/(2\beta_1))}{u_1} +\frac{\Delta g_2\ln(1/(2\beta_2))}{u_2}\nonumber
 \end{eqnarray} 
 \begin{eqnarray}
 &=&  \ln(1/2\beta_1)^{\frac{\Delta g_1}{u_1}} +\ln(1/2\beta_2)^{\frac{\Delta g_2}{u_2}}   \nonumber\\ 
 & =& - \ln((2\beta_1)^{\frac{\Delta g_1}{u_1}}(2\beta_2)^{\frac{\Delta g_2}{u_2}}) 
 \end{eqnarray}  
% \textnormal{To minimize $\epsilon$ we thus need to maximize: } \nonumber \\

To minimize $\epsilon$, we thus need to maximize 
\begin{eqnarray}    
(\beta_1)^{\frac{\Delta g_1}{u_1}} (\beta_2)^{\frac{\Delta g_2}{u_2}}
\label{eq:maxconj}
\end{eqnarray}

Our optimization problem is now as follows. We want to minimize $\epsilon$ by maximizing the expression defined in Eq. \eqref{eq:maxconj}, subject to the $\beta$-bound constraint from problem (\ref{eq:optprob}). In other words, we aim to find the local maxima of such a multivariate function given the inequality constraint on its variables. This type of constrained optimization problem can thus be solved with the Lagrange Multipliers method \cite{lagrange}, as this method aims to determine the extrema of a function composed of multiple variables given an equality or inequality constraint. The Lagrange method achieves this by reformulating the optimization problem into a set function called the Lagrangian function. Solving this function yields the proportioning technique below. A complete proof can be found in Appendix~\ref{app:fnrbound}.
\begin{theorem}
\label{def:fnr_conj_theorem}  
Given a conjunction query $Q=Q_1 \cap Q_2$ answered by a conjunction mechanism
% \xh{I think $M_i$ should be the particular ``Treshold shift Laplace mechanism'' rather than any general DP mechanism. You may want to say ``answered by a conjunction mechanism $M(D)=M_1(D)\cap M_2(D)$ where $M_i$ is a threshold shift Laplace mechanism, we achieve minimum privacy loss by ...''}
$M(D) = M_1(D) \cap M_2(D)$ where $M_i$ implements TSLM and $\epsilon = \epsilon_1 + \epsilon_2$, we achieve minimum privacy loss $\epsilon$ by budgeting the $\beta$-bound on FNR as:
\begin{eqnarray}
    \beta_1=\frac{u_2\Delta g_1\beta}{u_1\Delta g_2+u_2\Delta g_1}, 
    \beta_2=\frac{u_1\Delta g_2\beta}{u_1\Delta g_2+u_2\Delta g_1}
    \label{eq:2conj}
\end{eqnarray}
\end{theorem}

\stitle{Generalized $n$-Query Conjunction}. We extend the previous approach to generalize it over the conjunction of $n$-aggregate threshold queries. Consider a query $Q = Q_1 \cap ... \cap Q_n$, where each atomic query $Q_i$ is answered with TSLM $M_i$ that has an associated FNR bound $\beta_i$. By decomposing the conjunctions into $(n-1)$ 2-way conjunctions, we can generalize our previous $f_\epsilon$ and $f_\beta$ functions into: 
\begin{mini}|l|[0]
{}{\Sigma_{i=1}^n\epsilon_i}
{}{}
\addConstraint{\Sigma_{i=1}^n \beta_i \leq \beta}
\label{eq:genbetaconj}
\end{mini}
Using the Lagrange Multipliers method again, we obtain the apportionment technique below:
\begin{theorem}
    Given a complex DS query $Q^{\Lambda,F}$ composed of $n$ aggregate threshold queries connected by $n-1$ conjunctions, we achieve minimum privacy loss $\epsilon$ by budgeting the $\beta$-bound on FNR as:
    \begin{eqnarray}
    \beta_i =\frac{\Delta g_i \beta  \prod_{x=1}^{n,x\neq i } (u_x )}{\sum_{y=1}^{n}\prod_{x=1}^{n,x\neq y } (u_x \Delta g_y )},  \forall i=\{1,2,...,n\}
    \label{eq:gen_beta}
    \end{eqnarray}
    % \xh{change the subscript for $\beta$ from $j$ to $i$ }
    \label{def:generalbeta}
\end{theorem}
% \sm{ Above considered binary... the theorem above generalzies
% to conjunctions of n atomic queries.  Let Q^ =  ^_i(Q_i). , the theorem below shows the correxponding optimal values for B_i
% theeom 4.5 
% Proofs of the theorem above in appendiux. By setting indiigfual false negatibves... for all Qi
% }

Thus, by setting the individual $\beta_i$ bounds for sub-queries $Q_i$ according to Eq. \eqref{eq:gen_beta}, we provide a mathematical guarantee that privacy loss $\epsilon$ is optimally minimized while also maintaining the $\beta$-bound guarantee on FNR.

% \xh{Can we state the above as theorem, and fold the derivation as part of the proof?}
Note that other DP composition theorems could be used to express $\epsilon$ as a function of the individual $\epsilon_i$, including the Advanced Composition Theorem \cite{Dwork:2014:AFD:2693052.2693053}, but this would change the formulation of our optimization problem. Similarly, other mechanisms may be used to provide DP guarantees presuming that they also offer a $\beta$-bound on FNR, but will require different derivations to solve the optimization problem. Using relaxations of DP such as Rényi differential privacy \cite{renyi} or $\mathit{f}$-differential privacy \cite{fdp} is, however, non-trivial, as the expression of privacy loss $\epsilon$ in terms of the $\beta$ or $\alpha$ error rates used (i.e. as defined in \cite{mide}) may not necessarily hold. Thus, they require extensive analysis to formulate a baseline expression of privacy loss before solving the optimization problem.

\subsection{Query Disjunction Mechanism}
% \xh{Similary, we update this section for query of a single disjunction operator}
Similarly to the previous section, we model a privacy-preserving mechanism on a query composed of two atomic aggregate threshold queries linked by the disjunction operator $Q^{\Lambda,F} = Q_1 \cup Q_2$, where $Q_1$ and $Q_2$ are answered by two differentially private mechanisms $M_1$ and $M_2$ respectively. The overall mechanism is the union of the two sub-mechanisms $M(D)=M_1(D)\cup M_2(D)$.

\begin{definition}[Query Disjunction Mechanism]
\label{def:M_disj}

Let mechanism $M_i:\mathcal{D} \to O_i$  with a differential privacy guarantee satisfy a $\beta_i$-bound on FNR for aggregate threshold query $Q_i$. We can answer query $Q$ which is a disjunction of 2 aggregate threshold queries $Q_1 \cup Q_2$ using mechanism $M$ where $M(D)=M_1(D)\cup M_2(D)$. 
\end{definition}

\stitle{$\boldsymbol\beta$-Bound on False Negative Rate}. Similarly to the conjunction mechanism, each randomized mechanism $M_i$ ran on an individual aggregate threshold query in a disjunction has an associated $\beta_i$-bound on FNR. We thus derive the apportioning function $f_\beta$ for the overall false negative rate FNR for $Q$ in terms of $\beta_i$ by deriving and using a confusion matrix similar to Figure \ref{fig:truthtable} for disjunction.  Thus, by similar analysis (shown in Appendix~\ref{app:fnrbound}), the overall FNR can be upper bounded by the three lines below. For any predicate $\lambda_i\in \Lambda$,
% \begin{eqnarray*}
% && P[\lambda_i\notin M(D)|\lambda_i\in Q^{\Lambda,F}(D)]  \\
% &=&  P[\lambda\notin (M_1(D)\cup M_2(D))|\lambda\in(Q_1(D)\cup (Q_2(D))]    \\
% &=& (P[\lambda\notin M_1(D)|\lambda\notin Q_1(D)]\cdot P[\lambda\notin M_2(D)|\lambda\in Q_2(D)] \\
% && \cdot \ P[\lambda\notin Q_1(D)] \cdot P[\lambda\in Q_2(D)] + \\
% && P[\lambda\notin M_1(D)|\lambda\in Q_1(D)]\cdot P[\lambda\notin M_2(D)|\lambda\notin Q_2(D)] \\
% && \cdot \ P[\lambda\in Q_1(D)] \cdot P[\lambda\notin Q_2(D)] + \\
% && P[\lambda\notin M_1(D)|\lambda\in Q_1(D)]\cdot P[\lambda\notin M_2(D)|\lambda\in Q_2(D)] \\
% && \cdot \ P[\lambda\in Q_1(D)] \cdot P[\lambda\in Q_2(D)]) \\
% && \cdot \ (P[\lambda\in Q_1(D)]+ P[\lambda\in Q_2(D)]-P[\lambda\in Q_1(D)] \\
% &&\cdot \ P[\lambda\in Q_2(D)])^{-1}
% \end{eqnarray*}
\begin{eqnarray}
 && P[\lambda_j\!\notin\! M(D) | \lambda_i\!\in\! Q^{\Lambda,F}(D)] \nonumber\\
 &\leq&  P[\lambda_j \! \notin M_1(D) | \lambda_j \! \notin \! Q_1(D)] \! \cdot\! P[\lambda_j \! \notin \! M_2(D) | \lambda_j \! \in \! Q_2(D)]  \nonumber \\
 &+& P[\lambda_j \! \notin M_1(D) | \lambda_j \! \in \! Q_1(D)] \! \cdot\! P[\lambda_j \! \notin \! M_2(D) |\lambda_j \! \notin \! Q_2(D)]  \nonumber\\
  &+& P[\lambda_j \! \notin M_1(D) | \lambda_j \! \in \! Q_1(D)] \! \cdot\! P[\lambda_j \! \notin \! M_2(D) | \lambda_j \! \in \! Q_2(D)]\nonumber\\
  &=&  TNR_1 \cdot FNR_2 + FNR_1 \cdot TNR_2 +FNR_1FNR_2  \nonumber\\
    &=& TNR_1 \! \cdot \! FNR_2+FNR_1 (TNR_2 + FNR_2) \nonumber\\
    &\leq& FNR_2 + FNR_1 \leq \beta_1 + \beta_2 \nonumber
  \end{eqnarray}
  % Since $Q_1$ and $Q_2$ are independent, $P[\lambda_j \!\in\! Q^{\Lambda,F}(D)] = s_1\!+\!s_2\!-\!s_1s_2$.
%   \begin{eqnarray}
% &=&  (TNR_1 \cdot FNR_2 \cdot (1-s_1)s_2 + FNR_1 \cdot TNR_2 \cdot s_1(1-s_2) \nonumber\\
% &+& FNR_1FNR_2s_1s_2)/(s_1+s_2-s_1s_2)\\
% &=& FNR_1 + FNR_2 - [  FNR_2FPR_1s_2(1-s_1)\nonumber\\
% &+& FNR_1FPR_2s_1(1-s_2) + FNR_1s_2(1-FNR_2s_1)\nonumber\\
% &+& FNR_2s_1  ] / (s_1+s_2-s_1s_2) \\
% &\leq& FNR_1 + FNR_2 \leq \beta_1 + \beta_2 \nonumber
% \end{eqnarray}

 Again, we simplify this expression by providing an upper bound on the FNR by removing negative clauses. We thus obtain the apportioning function:
\begin{equation}   
FNR \leq f_\beta(\beta,\beta) = \beta_1 + \beta_2 \leq \beta 
\label{eq:final_fnr_disj}
\end{equation}

\stitle{Privacy Loss}. Similar to conjunction, we use the sequential composition theorem of differential privacy (Def. \ref{def:seq_comp}) to compose the privacy loss of the two sub-mechanisms of $M$. Assuming that each sub-mechanism uses TSLM, we can again use the budget $\epsilon_i = \frac{ \Delta g_i\ln(1/(2\beta_i)}{u_i}$, which guarantees a $\beta_i$-bound on FNR.  Thus, overall privacy loss can be minimized by maximizing
\begin{equation}
(\beta_1)^{\frac{\Delta g_1}{u_1}} (\beta_2)^{\frac{\Delta g_2}{u_2}}
\label{eq:ep_min_function}
\end{equation}
Since our $f_\epsilon$ and $f_\beta$ functions in the case of disjunctions are identical to that of conjunctions, our optimization problem for both are the same. Using the Lagrange method therefore yields the same apportionment technique. Thus, Theorems \ref{def:fnr_conj_theorem} and \ref{def:generalbeta} hold for disjunction as well, which are formally shown in Appendix \ref{app:fnrbound} and \ref{app:genbound} respectively.

\subsection{Combined Conjunctions/Disjunctions} 
\label{treestructure}
Consider a complex decision support query $Q^{\Lambda,F}$ comprised of a set of $n$ atomic aggregate threshold queries $Q_{a_1},...,Q_{a_n}$ connected by disjunctions or conjunctions. Such a query corresponds to a binary operator tree $T$ where each operator is either a conjunction ($\cap$) or a disjunction ($\cup$). To execute $Q^{\Lambda,F}$, we first evaluate each of the atomic queries, then recursively combine 
their outcomes based on the connecting operators in
the operator tree $T$  to determine the results. 
% \xh{Question: are we running a TLSM on each of the leaf nodes of the operator tree or a TSLM on the distinct set of atomic queries? For example, in Fig4(b), do we run the DP mechanism twice on Q1? Based on the privacy analysis, we only run a single DP on Q1. But then in our beta-analysis, we assume the randomness of the DP mechanisms are independent. but in Fig 4(b), node n3 and node n5 share correlated randomness, then our analysis earlier on may not work. }
% \xh{Maybe a fix is to consider running a TLSM on each of the leaf nodes of the operator tree, and this will add $o_j$ to $\epsilon_j$ as well. How will this affect the solution and our experiment results?}
% Note that the non-leaf nodes may not be independent despite the leaf nodes (i.e. atomic queries) being so. For example, in Figure \ref{fig:q-tree}(b), sub-query $(Q_1 \cup Q_2)$ is not independent of $(Q_1 \cup Q_3)$. Thus, we choose to provide a $\beta$-bound  guarantee on FNR for such queries, through which a $\beta$-bound on FNR is achieved as well.

Our challenge, therefore, lies in generalizing our apportionment technique such that,  given a  query tree $T$, we  determine the $\beta$ budget distribution
across all atomic queries  such that the $\beta$-bound is guaranteed while minimizing privacy loss $\epsilon$.
We show through the example below how to express the $\beta$ as a function of the $\beta_i$ given
a tree structure, then formalize the problem of optimal apportionment
given any tree.

\begin{figure}[t]
\centering
\includegraphics[width=.35\textwidth]{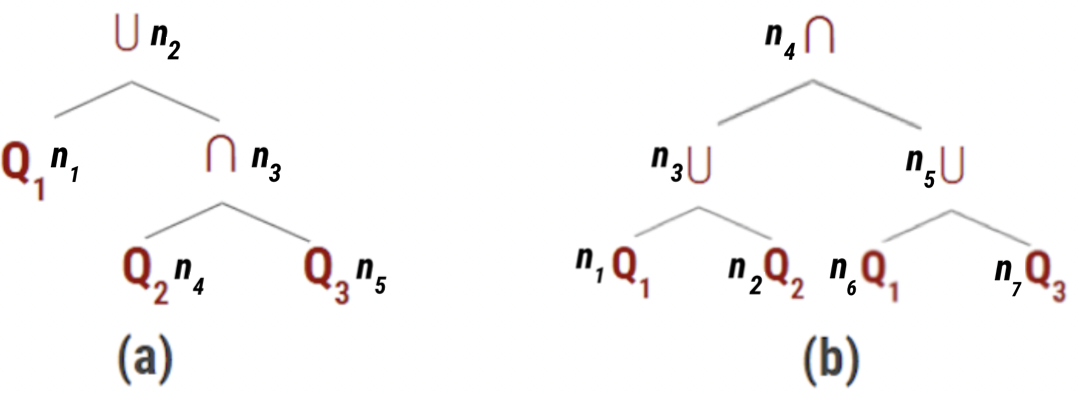}
\caption{The figure shows the query trees for (a) $Q_{T1} = Q_1\cup( Q_2\cap Q_3)$, and (b) $Q_{T2} = ( Q_1\cup Q_2)\cap( Q_1\cup Q_3)$
}
\label{fig:q-tree}
\end{figure}

\begin{exmp} 
Consider an example query $Q_{T1} = Q_1\cup( Q_2\cap Q_3)$ shown in Figure \ref{fig:q-tree}(a). We refer to the sub-query associated with a node
by its node-id (e.g., in Figure \ref{fig:q-tree}(a) node $n_1$ corresponds to sub-query $Q_1$.).
For $Q_{T2}$, we can derive the FNR by first deriving individual FNRs for the sub-tree $Q_{n_3}=Q_2\cup Q_3$ using the 2-conjunction mechanism, where the two sub-trees $Q_{n_3}$ and $Q_{n_1}$ can in turn be executed using the 2-disjunction mechanism. We thus obtain 
$\beta_{n_3}=\beta_2 +\beta_3$, $\beta_{n_1}=\beta_1$ and $\beta_{n_1}+\beta_{n_3}\leq \beta
$.
By substitution, we obtain the false negative rate constraint
\begin{equation}
    \beta_1 + \beta_2 + \beta_3 \leq \beta
    \label{eq:distributed}
\end{equation}
Let us further consider query $Q_{T2}= (Q_1\cup Q_2)\cap(Q_1 \cup Q_3)$ shown in Figure \ref{fig:q-tree}(b).
Executing the same recursive steps for $Q_{T2}$ yields
\begin{equation}
    2\beta_1 + \beta_2 + \beta_3 \leq \beta
    \label{eq:notdistributed}
\end{equation}
\vspace{-5mm}
\label{def:example41}
\end{exmp}
In the example above,  note that for $Q_{T1}$, leaf nodes
in $T1$ correspond to  unique atomic queries while in 
$Q_{T2}$, the atomic query   $Q_1$ appears twice in $T2$. 
Let us denote the number of occurrences of a atomic query $Q_i$
in the leaf nodes of a tree by $o_i$. 
Thus, in $T1$, the values of $o_1,o_2$, and $o_3$ are all 1, whereas
in $T2$ $o_2$, and $o_3$ are 1, while the value of $o_1$ is 2, leading
to the difference in the FNR constraint for $T1$ and $T2$.
More specifically, given a tree $T$
with $Q_{a_1},...,Q_{a_n}$ sub-queries and corresponding $o_1,...,o_n$ occurrences, the overall FNR for query $Q^{\Lambda,F}$ can be
expressed as
\begin{equation}
    f_\beta( \beta_1, \beta_2, ...,\beta_n) = o_1\beta_1 + o_2\beta_2 + ...+o_n\beta_n \leq \beta
\end{equation}
A formal proof of the above is provided in Appendix~\ref{app:genbound}. As for privacy loss, we must run the atomic mechanisms $M_i$ on each leaf node in order to maintain their independent randomness. Therefore, the $\epsilon_i$ budget allocated for each sub-query must be divided across its occurrences as well, i.e. $o_i \epsilon_i$. 
Thus, our apportionment problem for a given tree is now
\begin{mini}|l|[0]
{}{\Sigma_{i=1}^n o_i\epsilon_i}
{}{}
\addConstraint{\Sigma_{i=1}^n o_i \beta_i \leq \beta}
\label{eq:prelimoptoi}
\end{mini}
We use the Lagrange Multipliers method to yield the theorem below.
% i.e. the number of occurrences $o_i$ for each sub-query $Q_i$ is 1, the $\beta_i$ budgets are given equal weight 1. However, for $Q_{T2}$ whose tree $T2$ has $Q_1$ appear twice, i.e. $o_1 = 2$, $\beta_1$ budget is given $o_1$-times the weight. It is thus clear that the number of occurrences $o_i$ for a given atomic query $Q_{a_i}$ in a query tree $T$ is inversely proportional to the $\beta_i$ budget allocated to it.  We can thus generalize our false negative rate apportioning function $f_\beta$ to:
\begin{theorem}
    Given a complex decision support query $Q^{\Lambda,F}$ with query tree $T$ composed of $n$ aggregate threshold queries with an associated $o_i$ number of occurrences within the tree,  we achieve minimum privacy loss $\epsilon$ by budgeting the $\beta$-bound on FNR as:
    \begin{eqnarray}
    \beta_i =\frac{\Delta g_i \beta  \prod_{x=1}^{n,x\neq i } (u_x )}{\sum_{y=1}^{n}\prod_{x=1}^{n,x\neq y } (u_x o_y\Delta g_y )},  \forall i=\{1,2,...,n\}
    \label{eq:gen_beta_occ}
    \end{eqnarray}
        % \xh{change the subscript for $\beta$ from $j$ to $i$}
    \label{def:beta_occ}

\end{theorem}
%     \nl{ the epsilon becomes $\frac{o_i \Delta g_i ln(1/2\beta_i)}{u_i}$ so to minimize $\sum \epsilon_i$ we need to maximize $(\beta_1)^{\frac{o_1\Delta g_1}{u_1}} (\beta_2)^{\frac{o_2\Delta g_2}{u_2}}...$ subject to $\sum o_i\beta_i \leq \beta$. so from lagrange we get:
%     \begin{eqnarray}
%     \beta_j =\frac{o_j\Delta g_j \beta  \prod_{x=1}^{n,x\neq j } (u_x )}{o_j\sum_{y=1}^{n}\prod_{x=1}^{n,x\neq y } (u_x o_y\Delta g_y )},  \forall j=\{1,2,...,n\}
%     \label{eq:gen_beta_occ}
%     \end{eqnarray}
%     we can eliminate the common $o_j$ so we get
%     \begin{eqnarray}
%     \beta_j =\frac{\Delta g_j \beta  \prod_{x=1}^{n,x\neq j } (u_x )}{\sum_{y=1}^{n}\prod_{x=1}^{n,x\neq y } (u_x o_y\Delta g_y )},  \forall j=\{1,2,...,n\}
%     \label{eq:gen_beta_occ}
%     \end{eqnarray}
%     }

% \xh{I think we need to write it as $\min \sum_i o_i\epsilon_i$, but it should give the same derivation you had above.}
% As mentioned previously, we can guarantee a $\beta_i$-bound on FNR for each $M_i$ by setting $\beta_i=\beta_i$ since a $\beta$-bound on FNR offers a $\beta$-bound on FNR as well for the atomic queries. 

Note that a query $Q^{\Lambda,F}$ may be represented by multiple equivalent trees, due to the distributive property of logical operators. In Example \ref{def:example41} for instance, $Q_{T1}$ and $Q_{T2}$ have equivalent query trees $T1$ and $T2$, where $T2$ is obtained from distributing the OR operator over the AND operator. From Eq. \eqref{eq:distributed} and \eqref{eq:notdistributed}, however, we can infer that a distributed query tree, i.e. a query with a higher number of nodes, may cause the overall privacy loss of the query to increase, because it allocates a lower $\beta_i$ to one or more given sub-queries $Q_i$. 

Thus, the ProBE mechanism aims to generate the optimal query tree with the least number of leaf nodes to minimize any additional distribution of the $\beta$ budget. We can use Boolean function minimization algorithms
such as \cite{mccluskey} 
%For this purpose, $M_{\textsc{probe}}(T)$ use the Quine–McCluskey Boolean function minimization algorithm \cite{mccluskey} $A(T)$ 
to return the compact tree representation of the query $T_c$. Our mechanism subsequently extracts the number of occurrences $o_i$ from $T_c$ for each atomic query which will be used to compute the $\beta_i$ budget as in Eq. \eqref{eq:gen_beta_occ}.

\begin{algorithm}[t] 
\caption{ProBE Mechanism Overview. DS query $Q^{\Lambda,\mathcal{F}}=[T, \{Q_1(g_1,c_1),\ldots,Q_n(g_n,c_n)\}]$, maximum privacy budget $\epsilon_{max}$, dataset $D$, accuracy requirements ($\beta$-FNR, $\alpha$-FPR)}
\label{algo:probe}
\begin{algorithmic}[1]
\Procedure{ProBE}{$Q^{\Lambda,\mathcal{F}},\epsilon_{max}, D, \beta,\alpha$}
\State Tree minimization $(Q_c, o) \leftarrow minimize(Q)$ \cite{mccluskey}
\State $(O_f, \epsilon_f, Q_c) \gets \textsc{PhaseOne}(Q_c,o, \epsilon_{max}, D,\beta)$
\State $(O_f,\epsilon_f) \gets \textsc{PhaseTwo}(Q_c,o,\epsilon_{max}, D, \beta, \alpha,\epsilon_f, O_f)$
\State \Return $O_f,\epsilon_f$
\EndProcedure
\end{algorithmic}
\end{algorithm}

\section{Implementing PROBE}
Now that we have determined our $\beta$ budget apportionment technique, we can implement ProBE by individually running mechanisms $M_i$ on the atomic queries $Q_i$ based on
apportionment discussed in the previous section, then combining the results following the conjunction/disjunction operators as depicted in the query tree. However, simply implementing this framework incurs an arbitrary cost on the FPR when bounding the FNR as previously explained in \secref{ssec:probdef}. To address this, we develop an additional optimization which aims to provide an algorithmic bound on false positives. 
% Given results of the individual mechanisms, we can then compute the final outcome of the complex query following the given query tree structure.

In this section, we propose a two-phase algorithm that implements the ProBE apportionment framework to answer complex decision support queries with minimal privacy loss and bounds on utility. The overall algorithm is depicted in Algorithm \ref{algo:probe}. We first use a boolean minimization algorithm (Quine-McCluskey) \cite{mccluskey} to minimize our query tree in order to obtain the occurrences of each sub-query within the query tree $o$ and the tokenized version of this minimal tree $Q_c$ (line 2). We subsequently run the two phases (lines 3-4). The first phase of the algorithm traverses the query tree and implements the apportionment framework from Section 4, which solves the optimization problem to guarantee the $\beta$-bound on FNR. The second phase then relaxes this solution by providing a post-optimization bound on the FPR. We first discuss Phase One, wherein we set the initial uncertain region parameter $u$ to a large value in order to potentially obtain minimal privacy loss. 
Phase Two subsequently checks if the resulting false positives in Phase One exceed the FPR bound. If not, it uses intermediate results from Phase One to determine the next optimal uncertain region  $u_{opt}$ such that the $\alpha$ bound on FPR is met. Finally, we propose an iterative, entropy-based variant of the ProBE algorithm which further optimizes privacy loss in terms of Min-Entropy \cite{mide}.

\eat{
\begin{algorithm}[t] 
\caption{False Positives Estimation\xh{still working} }\label{algo:estFP}
\begin{algorithmic}[1]
\Procedure{EstimateFPs}{$Q_i(g_i,f_i,c_i), u_i, G_i,\beta_i$}
\State $O_{pp} \gets \{\lambda_j \in \Lambda \ | \ G_i[j] > c_i\}$
\State $O_p \gets \{\lambda_j \in \Lambda \ | \ G_i[j] > c_i - u_i\}$
\State $O_n \gets \{\lambda_j \in \Lambda \ | \ G_i[j] < c_i - u_i\}$
\State $F \gets O_p - O_{pp}$
\State $f_{est} = |F| + |O_{pp}|*\beta_i$, $r_{est} = \frac{|O_n| - \beta_i |G_i|}{1-\beta_i}$
%\For{$j= 1,..., |F| $}
%\If{($t_i=0$ and $F[j] \! < \! c_i-2u_i$) or ($t_i=1$ and $F[j] \! > \! c_i $) }
%\State remove $F[j]$ from F
%\EndIf
%\EndFor
\State \Return $f_{est}, r_{est}, F$
\EndProcedure
\end{algorithmic}
\end{algorithm}
}

\eat{

\Procedure{PhaseTwo}{$Q, N, R, F, O_{one}, \beta, \alpha, c_i,t_i, \epsilon_{one}, \epsilon_{max}$}
\State Let $O_f \gets \{ \}, \epsilon_f \gets \epsilon_{one}$
\For{query node $Q_i\in Q_c$}
\State compute allowed false positives $f_{max}\gets \frac{\alpha}{n}R[i]$ 
\State compute new u if estimated false positives are more than $f_{max}$ \xh{clarify more here}
\State $l$ is set to be the sorted $G_i[j*]$ of the predicates in $F$ , where $j*=N[i] - f_{max}-1$
\State $l \gets F[N[i] - f_{max}-1] $
\State $u_{opt} \gets (c_i - l) \ / \ 2 $
\State $flag_i \gets True$ [a flag to indicate re-run this TSLM]
\EndFor
\State $Q_f \gets \textsc{Travers}(Q_c.root)$
\State \Return $O_f, \epsilon_f$
\EndProcedure

\Function{Traverse}{node}
  \If{node is conjunction}
  \State $O_l\gets \textsc{Traverse}$(node.left)
  \State if $O_l$ is empty, then skip node.right and \Return $O_l$ 
  \State else, $O_r \gets \textsc{Traverse}$(node.right)
  \State \Return $O_l\cap O_r$
  \ElsIf{node is disjunction}
  \State $O_l\gets \textsc{Traverse}$(node.left)
  \State $O_r \gets \textsc{Traverse}$(node.right)
  \State \Return $O_l\cup O_r$
  \ElsIf{node is a query $Q_i$}
  \State \xh{only rerun it if fail FPs}
  \State $O_i,\epsilon_i,G_i \gets$ \textsc{Tslm} ($Q_i,u_i,\beta_i, D$) 
  \State Update global budget $\epsilon_f \gets \epsilon_f+\epsilon_i$
  \If{$\epsilon_f>\epsilon_{max}$}
  \State Terminate program and \Return `Query Denied'.
  \State \xh{also re-estimate the FPs again, if fails, then query deny as well. }
  \EndIf
  \Return $O_i$
  \EndIf
\EndFunction

}

\begin{algorithm}[t] 
\caption{First Phase of ProBE Mechanism.}\label{algo:ssprobe}
\begin{algorithmic}[1]
\Procedure{PhaseOne}{$Q_c, o, \epsilon_{max}, D, \beta$}
\State Initialize global budget variable $\epsilon_f \gets 0$
\For{query node $Q_i\in Q_c$}
    \State $u_i \gets 0.3*range(Q_i)$
    \State $\beta_i \gets \frac{\Delta g_i (\beta/2)  \prod_{x=1}^{n,x\neq i } (u_x )}{\sum_{y=1}^{n}\prod_{x=1}^{n,x\neq y } (u_x o_y\Delta g_y )}$ 
    \State $flag_i=True$
\EndFor
\State $O_f \gets \textsc{Traverse}(Q_c.root)$
\State \Return $O_f,\epsilon_f, Q_c$
\EndProcedure

\Function{Traverse}{node}
  \If{node is conjunction}
  \State $O_l\gets \textsc{Traverse}$(node.left)
  \State if $O_l$ is empty, then skip node.right and \Return $O_l$ 
  \State else, $O_r \gets \textsc{Traverse}$(node.right)
  \State \Return $O_l\cap O_r$
  \ElsIf{node is disjunction}
  \State $O_l\gets \textsc{Traverse}$(node.left)
  \State $O_r \gets \textsc{Traverse}$(node.right)
  \State \Return $O_l\cup O_r$
  \ElsIf{node is a query $Q_i$}
  \If{$flag=True$}
  \State $O_i,\epsilon_i,G_i \gets$ 
  \textsc{Tslm} ($Q_i,u_i,\beta_i, D$)
  \State Update global budget $\epsilon_f \gets \epsilon_f+\epsilon_i$
  \State $flag_i=False$
  % \State \xh{TODO: check FPR again to deny query (may write it as a comment) for phase 2.}
  \EndIf
  \If{$\epsilon_f>\epsilon_{max}$}
  \State Terminate program and \Return `Query Denied'.
  
  \EndIf
  \Return $O_i$
  \EndIf
\EndFunction
\end{algorithmic}

\end{algorithm}

\subsection{Phase One of ProBE}
The first phase of the ProBE algorithm is detailed in Algorithm \ref{algo:ssprobe}.  This algorithm takes the minimized query $Q_c$, the set of occurrences of each sub-query $o$ as well as the maximum privacy loss allowed $\epsilon_{max}$ and the FNR bound $\beta$. 
 For each query node in the query tree, we first compute the initial uncertain region, which we set to a large percentage (30\%) of its range of values. We also compute its corresponding FNR bound $\beta_i$ according to Eq. \eqref{eq:gen_beta_occ} (lines 3-5). Note that because we are using a two-phase algorithm, we allocate half of the $\beta$ budget to each phase. However, we explore different $\beta$ allocation strategies across the two phases in Appendix F (Additional Experiments). We also store a general flag parameter $flag_i$, which indicates if the DP mechanism (in this case, TSLM) should be run if the sub-query is encountered in the traversal. These parameters are stored within the node itself $Q_i$. 
 
 The algorithm can now begin traversing the tree in a pre-order traversal ( i.e. starting from the root and executing the leaf nodes left to right). We first check if a leaf node is either a query  or an operator. If it is an operator, we recursively call the traversal function in order to reach the leftmost leaf query node and then the rightmost leaf query node (lines 10-13/15-17). Depending on the nature of the operator (conjunction or disjunction), we either intersect the results of the left and right traversal (line 14) or union the results (line 18).  
 
 If the current node is a query $Q_i$, then we execute the TSLM mechanism provided that the $flag_i$ is True.  Thus, TSLM is run with the appropriate $\beta_i$ and $u_i$, which returns predicates whose noisy values are greater than their respective shifted thresholds $c_i - u_i$ into the result set $O_i$, the noisy values set $G_i$, as well as the resulting privacy loss $\epsilon_i$ computed with the equation $\epsilon_i = \frac{\Delta g_i ln(1/2\beta_i)}{u_i}$ (line 21).  The resulting $\epsilon_i$ is subsequently accumulated into the global privacy budget variable $\epsilon_f$, and the execution flag $flag_i$ is changed to False (lines 22-23).
 % Else if an operator has already been seen (i.e. a previous query has been evaluated), then we either union or intersect the results and add the $\epsilon$s (lines 12-18). 
 If the current privacy loss $\epsilon_f$ is above our maximum tolerated privacy loss $\epsilon_{max}$, the query is denied. Otherwise, the output of the current node $O_i$ is returned (line 25).
% the results are either union-ed or intersected with the previous results into the current result set $O_f$ depending on the type of operator $t_i$ (lines 7-12), 
% If the privacy loss exceeds the limit of $\epsilon_{max}$, the query is denied (line 22). 
% Note that though this algorithm is designed to guarantee a $\beta$-bound on FNR by apportioning $\beta$ into each $\beta_i$ appropriately, the same algorithm can be used to guarantee a $\beta$-bound on FNR assuming the independence of sub-queries by replacing $\beta$ with $\beta$ due to the apportionment techniques being the same (i.e. Eq. \ref{eq:gen_beta_occ} and Eq. \ref{eq:betaeq}) \xh{missing ref, using ``\eqref{}''}. 

\stitle{Skipping for conjunction queries.} For the conjunction of $n$ aggregate threshold queries, we note that the result of running a privacy-preserving mechanism like $M(D)=M_1(D)\cap M_2(D) ...$ $\cap \ M_n(D)$ on such a query can be determined as false if one sub-mechanism $M_i(D)$ is evaluated as false due to the nature of the intersection operator. We exploit this fact by adding an optimization which skips the evaluation of sub-queries in conjunctions if any of the previous sub-queries return an empty set, thus further minimizing overall privacy loss. We implement this in line 12, where, upon returning the traversal results of the left node, if the output $O_l$ is an empty set, we can automatically skip the execution of the right node and return the $O_l$ result itself. Note that due to the recursive nature of this traversal, any subsequent conjunctions will also be skipped until a disjunction operator is met or the query terminates.

\subsection{Phase Two of ProBE}

The second phase of the ProBE Mechanism is depicted in Algorithm \ref{algo:secondstep}. This algorithm takes the $\beta, \alpha$ bounds, as well as the maximum budget $\epsilon_{max}$. It also takes the resulting output $O_{one}$ and privacy loss $\epsilon_{one}$ from Phase One. The query $Q_i$ also includes the previously derived parameters (e.g. the noisy aggregates $ G_i$ and the $\beta_i$ bound) resulting from Phase One. The algorithm starts by estimating the number of false positives as a result of running Phase One for each query node $Q_i$ (line 4). We first describe the approach to derive the FPR estimate below.

\stitle{Estimating the Bound on FPR}.
At the beginning of Phase Two, we first determine an upper bound on the FPR resulting from Phase One. If the latter is within the user-specified $\alpha$ bound, we can skip the second phase of the algorithm, otherwise we rerun some of the query nodes with an additional privacy budget. 
We empirically measure the FPR of each mechanism $M_i$ for sub-query $Q_i$ by the ratio between the number of false positives $|FP|$ and the number of negatives $|N|$. However, we cannot compute the truthful number of false positives and negatives without looking at the data, which consumes an additional privacy budget. Hence, we derive (i) an upper bound for the number of false positives and (ii) a lower bound for the number of negatives, to get an upper bound for FPR.

We first obtain the following observed results based on the noisy aggregates from Phase One:
\[O_{pp} \gets \{\lambda_j \in \Lambda \ | \ G_i[j] > c_i\}\]
\[O_p \gets \{\lambda_j \in \Lambda \ | \ G_i[j] > c_i - u_i\}\]
\[O_n \gets \{\lambda_j \in \Lambda \ | \ G_i[j] < c_i - u_i\}\]
%\[O_{nn} \gets \{\lambda_j \in \Lambda \ | \ G_i[j] < c_i-2u_i\}\]
where $G_i[j]$ is the noisy aggregate for predicate $j$ in the $i$th sub-query $Q_i$. 
In particular, $O_p$ and $O_n$ are the reported positives and negatives for $Q_i$ by Phase One. 
$O_{pp}$ are the predicates with large noisy counts which are ``definitely positive''.

A naive upper bound for the number of false positives is $|O_p|$, which includes all the reported positives. However, among them, the predicates in $O_{pp}$ have noisy aggregates much larger than the testing threshold $c_i-u_i$ in TSLM, which are unlikely to be false positives. If all the definitely positive predicates in $O_{pp}$ had true counts $<c-2u_i$ (at worst case), they would have a noisy aggregate $>c-u_i$ and thus become a false positive with a probability $\leq \beta_i$  by the property of Laplace noise.
Therefore, we can have an upper bound for the number of false positives:
\begin{equation}
  |FP| \leq |O_p-O_{pp}| + |O_{pp}|\cdot \beta_i 
\label{eq:fps}
\end{equation}

The number of negatives $|N|$ is greater than $|O_n|-|FN|$, where $|FN|$ is the truthful number of false negatives. By the $\beta_i$-FNR property of TSLM, we have $|FN|\leq \beta_i (|G_i|-|N|)$, where $|G_i|$ is the total number of predicates in the input to query $Q_i$.
Hence, we have this inequality 
\[ |N| \geq |O_n| - |FN|
\geq |O_n| - \beta_i (|G_i|-|N|).\]
Solving this inequality by moving all the terms involving the unknown $|N|$, we have  a lower bound to $|N|$, 
\begin{equation}
    |N| \geq \frac{|O_n|-\beta_i |G_i|}{1-\beta_i}.
    \label{eq:negatives}
\end{equation}

\begin{algorithm}[t] 
\caption{Second Phase of ProBE Mechanism.}\label{algo:secondstep}

\begin{algorithmic}[1]
\Procedure{PhaseTwo}{$Q_c,o,\epsilon_{max},\beta, \alpha,\epsilon_{one}, O_{one}$}
\State Let $O_f \gets \{ \}, \epsilon_f \gets \epsilon_{one}$
\For{query node $Q_i(G_i,c_i,u_i,\beta_i,flag_i)\in Q_c$}
\State  $f_{est}, r_{est}\gets \textsc{EstimateFPs}(Q_i, O_{one})$

\State Compute allowed false positives $f_{max}\gets \frac{\alpha}{n}r_i$ 
\If{$f_{est}>f_{max}$}
\State Search $u_{opt}\gets $
the largest $u$ such that running $\textsc{EstimateFPs}(Q_i, O_{one})$ returns $f_{est}\leq f_{max}$
%\State $l \gets F[N[i] - f_{max}-1] $
%\State $u_{opt} \gets (c_i - l) \ / \ 2 $
\State Set $flag_i \gets True$ to rerun TSLM
\EndIf
\EndFor
\State $O_f \gets \textsc{Traverse}(Q_c.root)$
\State \textbf{if} updated $f_{est} > f_{max}$ for any $Q_i$ \textbf{then} Terminate program and \Return `Query Denied' 
\State \Return $O_f, \epsilon_f$
\EndProcedure

\end{algorithmic}
\end{algorithm}

\begin{algorithm}

\caption{Estimating FPs.}\label{algo:estimatefps}
    \begin{algorithmic}
  
        \Function{EstimateFPs}{$Q_i(G_i,c_i,u_i,\beta_i,flag_i), O_{one}$}
            \State $O_{pp} \gets \{\lambda_j \in \Lambda \ | \ G_i[j] > c_i\}$
            \State $O_p \gets \{\lambda_j \in \Lambda \ | \ G_i[j] > c_i - u_i\}$ \
            \State $O_n \gets \{\lambda_j \in \Lambda \ | \ G_i[j] < c_i - u_i\}$
            \State$O_{pp}\gets O_{pp}\cap O_{one}, O_{p}\gets O_p\cap O_{one}, O_n \gets O_n-O_{one}$
            %\State $F \gets O_p - O_{pp}$
            \State Upper bound for FPs $f_{est}= |O_p-O_{pp}| + |O_{pp}|*\beta_i$
            \State Lower bound for Negatives $r_{est} = \frac{|O_n| - \beta_i |G_i|}{1-\beta_i}$
            %\For{$j= 1,..., |F| $}
            %\If{($t_i=0$ and $F[j] \! < \! c_i-2u_i$) or ($t_i=1$ and $F[j] \! > \! c_i $) }
            %\State remove $F[j]$ from F
            %\EndIf
            %\EndFor
            \State \Return $f_{est}, r_{est}$
        \EndFunction
    \end{algorithmic}
\end{algorithm}

\begin{figure}[t]
    \centering
    \includegraphics[width=0.4\textwidth]{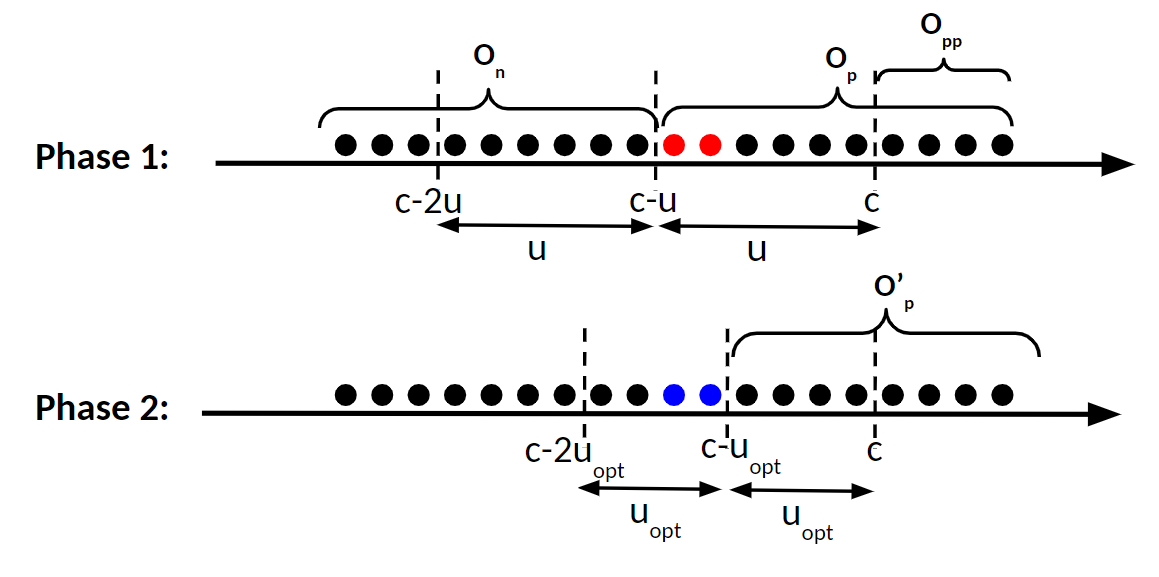}%\includegraphics[width=0.32\textwidth]{figures/sets.png}
    \vspace{-5mm}
    \caption{ Illustration of sets $O_{p}$, $O_{pp}$, ${O_n}$ on noisy values of predicates for Phase One. In Phase Two, we identify a new $u_{opt}$ such that the newly observed positive set $O'_p$ based on $>c-u_{opt}$ is reduced from $O_p$ by a size of $f_{est}-f_{max}$ (indicated by dots changing from red to blue)    
    if the number of estimated false positives $f_{est}$ in Phase One is greater than the allowed number of false positives $f_{max}$.
    }
    \label{fig:sets}
\end{figure}

The process of deriving an upper bound on the FPR is formalized in the function $\textsc{EstimateFPs}$.
We retrieve the $O_{pp}$, $O_p$ $O_n$. We then compute the upper bound on FP and lower bound on N according to Eq. \eqref{eq:fps} and \eqref{eq:negatives} (lines 17-18).
This algorithm also offers an additional optimization based on the nature of the operator (i.e. conjunction/disjunction), through which certain elements in the uncertain region can be eliminated in the first step. We illustrate this through the example below. 
\begin{exmp} Consider an example query $Q = Q_1 \cap Q_2$. Consider a predicate $\lambda_1$ s.t. its noisy value falls within the uncertain region $[c_1-u_1,c_1]$ of $Q_1$ i.e. is undecided, but falls within $[-\infty,c_2-2u_2]$ of $Q_2$ i.e. decidedly negative. We know that due to the nature of the $\cap$ operator, the result of intersecting the values for $\lambda_1$ from $Q_1$ and $Q_2$ will be negative without classifying $\lambda_1$ in $Q_1$. We can therefore eliminate $\lambda_1$ from the set of potential false positives 
within the uncertain region to be used for the second phase.
Conversely, consider the query $Q = Q_1 \cup Q_2$ and a predicate $\lambda_1$ s.t. $\lambda_1$ is reported as negative, i.e. falling into $[-\infty, c_1-u_1]$ (it is in $O_n$), but its noisy value for $Q_2$ falls within $[c_2,\infty]$ of $Q_2$ i.e. decidedly positive. We know that the $\cup$ operator only requires one element to be positive for the result to be positive as well, so $\lambda_1$ can be classified as positive and thus removed from  $O_n$.
\end{exmp}
The above optimization is illustrated in line 16, where we only keep 
% \sm{this is the first mention of positive and decidedly positive terms which we have not explained earlier. I would suggest do not use these terms... instead state 
% predicates in  $O_{p}$ and $O_{pp}$ instead}
predicates in  $O_p$ and $O_{pp}$ which have appeared in the results of Phase One $O_{one}$, and we similarly only keep the negative predicates which are not in $O_{one}$ . 
The estimated upper bound on false positives is stored in $f_{est}$ and the lower bound on negatives in $r_{est}$. With these estimates for the FPR, the rest of the second step can be executed.
\eat{
\begin{figure}[t]
    \centering
\includegraphics[width=0.39\textwidth]{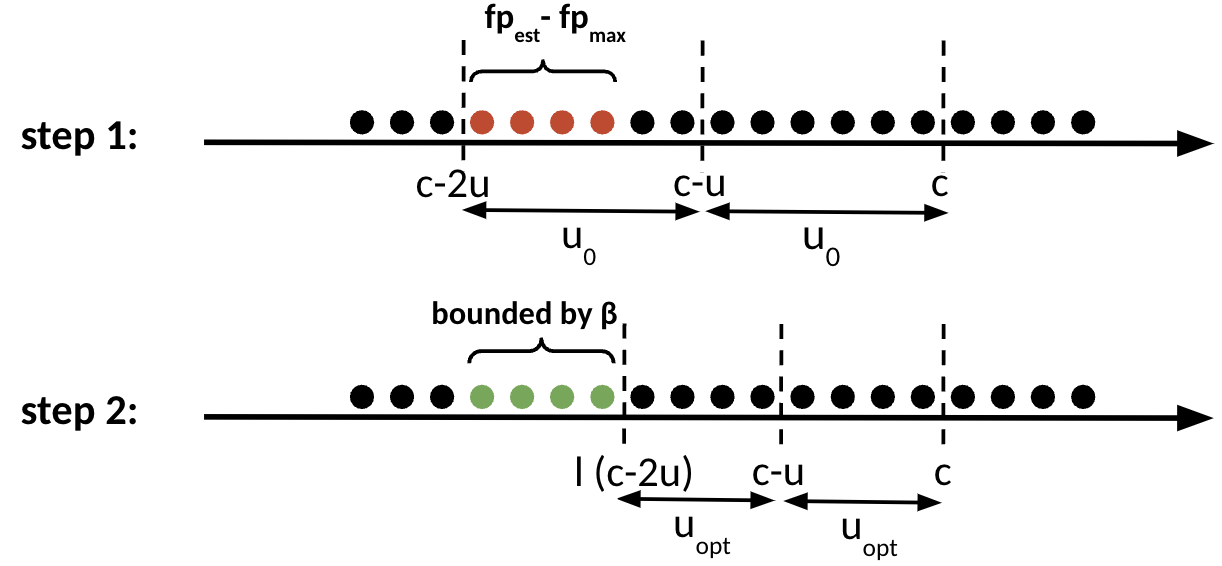}
    \caption{ProBE on a single sub-query $Q_i$. The first step is a threshold shift with an initially large $u_0$. The noisy values are used to estimate the amount of FPs $fp_{est}$ which is compared to the bound $fp_{max}$. The $fp_{est}-fp_{max}$ extra elements (red in step 1) are removed by computing the new uncertain region $u_{opt}$ s.t. they fall outside $[c-2u, c]$ (now green in step 2), thereby bounding their error by $\beta$. 
    % \xh{we need to replace gamma by beta; what are the red and green dots are not clear based on the caption.}
    }
    \label{fig:probets}
\end{figure}
}

\stitle{Resetting the Uncertain Region $u_{opt}$}. After retrieving the upper bound on the current FPR, the second step algorithm first checks if the upper bound on FPs $f_{est}$ is higher than the allowed number of FPs $f_{max}$. The latter value is derived from the bound $\alpha$, which is divided equally amongst sub-queries i.e. $\alpha/n$. We prove that this allocation ensures overall $\alpha$ in Appendix~\ref{app:fprbound}.
% \xh{how can this FPR per sub-query ensure the final FPR? do we need to prove this?}
The bound is then multiplied by the estimated number of negatives stored in $r_{est}$. 

If the bound is exceeded, we compute the next $u_{opt}$ at which the "extra" false positives i.e. $f_{est} - f_{max}$ would reside outside the new threshold $c-u_{opt}$ as shown in Figure~\ref{fig:sets}. In this way, the newly observed positives become $O'_p$, and its size is $f_{est}-f_{max}$ smaller than $O_p$. Hence, the newly estimated false positive negatives based on Eq.~\eqref{eq:fps} are also reduced by that if $|O_p-O_{pp}|\geq f_{est}-f_{max}$. We do so by retrieving the noisy value of the cutoff predicate where the number of extra FPs would reside to the left of.  This noisy value becomes our new lower bound for the uncertain region, i.e. $c_i - u_{opt}$. We thus solve for $u_{opt}$ from this equality.
% \xh{what do the $l$ refer to? We may use fig 5 to illustrate the meaning of $l$.} \xh{I also cannot understand the math expression on line 6. Need clarify}\nl{essentially i am getting the value in F at which we want $c-2u$ to be at, so that the "extra" FPs are outside c, c-2u }
Upon obtaining the new uncertain region, we can now rerun the TSLM algorithm with $u_{opt}$ and $\beta_i$  as parameters and accumulate the privacy loss resulting from this second phase. We do so by setting each execution flag $flag_i$ for the appropriate query nodes, then running the $\textsc{Traverse}$ function again (lines 8-9). 
% \xh{We only rerun TSLM for the leaf node fail $f_max$, right? We may try to reassign $u_i$ for them together, but then run the TLSM when traverse the tree, for the node with smaller enough $N[i]$, we do not need to run it again.}\nl{oh yes, so i think the solution is instead of for i in subqs we go through the i in query nodes.}
We recompute the estimated FPs $f_{est}$ after execution (line 10) by calling the estimation function again, and if the $f_{max}$ FP bound is not met, then the query is denied and the algorithm exits, otherwise the resulting output and privacy loss $O_f,\epsilon_f$ are returned. 

\begin{theorem}
Algorithm 1 satisfies $\epsilon_{\max}$-DP, a $\beta$-bound on the False Negative Rate and an $\alpha$-bound on the False Positive Rate if the query is not denied.
\label{def:mspwlm}
\end{theorem}
Proof of the theorem above can be found in Appendix~\ref{app:algo}.

\subsection{Multi-Step Entropy-based Algorithm}
\label{ssec:probeent}

ProBE assigns different privacy levels to different predicates due to the early elimination of data points at the first step, meaning predicates that go on to the second step have a higher privacy loss $\epsilon$. This concept is captured through the definition of Predicate-wise DP (PWDP)\cite{mide}, a fine-grained extension of differential privacy which quantifies the different levels of privacy loss data points may have in multi-step algorithms. 

Naturally, a measure to quantify this new definition of privacy is needed. Previous work tackles this problem by proposing a new privacy metric for PWDP entitled \textit{Min-Entropy} \cite{mide} $\mathcal{H}_{min}$ which measures a lower bound on the level of uncertainty given the set of predicates and their respective privacy levels $\Theta = \{(\lambda_1,\epsilon_1),...,(\lambda_k,\epsilon_k)\}$. 
the Data Dependent Predicate-Wise Laplace Mechanism (DDPWLM) also introduced in \cite{mide} maximizes min-entropy (i.e. maximizing the lower bound on uncertainty) in an iterative algorithm by also using the noisy aggregate values obtained from previous iterations to compute the best privacy level $\epsilon$ at which Min-Entropy is maximized for the set of elements in the uncertain region. This algorithm similarly guarantees a $\beta$-bound on FNR for a single aggregate threshold query by setting the privacy loss to $\epsilon = \frac{\Delta g\ln(1/(2\beta))}{u}$, but it does so by setting a starting privacy level $\epsilon_s$ as well as a maximum level $\epsilon_m$ prior to execution which is spent across a fixed number of iterations $m$. In each iteration, the privacy budget is further distributed across fine-grained steps $m_f$ and the privacy level with the highest min-entropy is chosen as the next iteration's budget. It follows that the uncertain region parameter $u$ is, again, chosen statically at the beginning, and therefore does not provide a bound on false positives.

We propose ProBE-Ent, a multi-step entropy-based algorithm which integrates DDPWLM with ProBE in order to not only answer complex DS queries with minimized entropy, but also to provide a post-facto bound on FPs. Instead of having a fixed starting $\epsilon_s$, we internally choose a starting $u_0$ and compute the initial privacy level. Within the first step of DDPWLM, we run the FP estimation algorithm (Algorithm \ref{algo:estimatefps}) in order to compute the optimal uncertain region parameter $u_{opt}$. We use this new uncertain region as an upper bound for the algorithm, i.e. we compute the $\epsilon_m$ upper bound that represents the exit condition for the algorithm. Additionally, we provide a $\beta$ budget optimization which exploits the multi-step nature of the algorithm in a way that provides a potentially higher budget if previous sub-queries exit early. This is done by redistributing any remaining $\beta_i$ which was not used in the current sub-query (due to early exit) to subsequent sub-queries, thus fully utilizing the $\beta$ bound given and consequently minimizing the overall privacy loss. We provide a more detailed version of this algorithm with complete definitions in Appendix C.

\section{Experiments}
In this section, we evaluate the ProBE algorithms previously explained on different real-life datasets based on various decision support application types. We assess their performance 
(in terms of resulting 
privacy and utility) over a varying number of complex queries, and 
over %provide 
examples of queries modeled after frequently used KPIs.
%and show their resulting privacy and accuracy levels.
Our experiments prove that all ProBE achieves its utility guarantees, while also successfully minimizing privacy loss for different levels of query complexity.

\subsection{Experimental Setup}
\textbf{Datasets}. To evaluate our approach, we use three real-world datasets from different domains. The first dataset, $\mathsf{NYCTaxi}$, is comprised of New York City yellow taxi trip records in 2020 \cite{nytaxi}, where the data consists of 17 attributes and approximately 3 million records. The second data set, $\mathsf{UCI Dataset}$, is comprised of occupancy data in 24 buildings at the University of California, Irvine campus collected from April to May 2019 \cite{tippers}. This data consists of 18 attributes and 5 million records. The final dataset, $\mathsf{TurkishMarketSales}$, stores records of sale transactions at a chain supermarket across Turkey in 2017 \cite{turkishsales}, where the data consists of 1.2 million records and 26 attributes.
% \vspace{-5pt}
\begin{table*}[!h]
\small
% \resizebox{\textwidth}{!}{
\begin{tabular} { |c|c|c|c| }
\hline
{\bf Dataset} & {\bf Attributes used} & {\bf Predicates} & {\bf \# of predicates $p$} \\
\hline
UCI Dataset & *, age, userType, gender, groupName, office & room, date  & 41(rooms)*14(days)=574
 \\ 
\hline
Turkish Market Sales & *, region, netTotal, customerId, category, gender  & city, date &  43 (cities)*14(days)=602
\\ 
\hline
NYTaxi & *, fareAmnt, totalAmnt, paymentType, location, storeFwdFlag & location, date  & 34(location)*15(days)=476  \\
\hline
\end{tabular}
\caption{Datasets used for experiments and respective attributes/predicates used}
\label{tab:datasets}
\end{table*}
\begin{table*}[tb]
\small
% \resizebox{\textwidth}{!}{
\begin{tabular} { |c|c|c| }
\hline
{\bf Sales KPI} & {\bf Definition} & {\bf Query Equivalent}\\
\hline
Sales Volume & \# of sales & $\mathbf{Q_1}$: SELECT a  GROUP BY a HAVING COUNT(*) > C
 \\ 
\hline
Regional Sales Volume & \# of sales in region & $\mathbf{Q_2}$: SELECT a WHERE region=`A'  GROUP BY a HAVING COUNT(*) >C
\\ 
\hline
User Retention Rate & \# of customers per period & $\mathbf{Q_3}$: SELECT a  GROUP BY a  HAVING COUNT(DISTINCT(customerId)) > C
\\
\hline
Conversion Rate & \# of sales / \# of customers & $\mathbf{Q_4}$: SELECT a  GROUP BY a HAVING COUNT(DISTINCT(customerId)) > C
\\ 
\hline
\end{tabular}
\caption{KPI-based Queries used for Sales Dataset experiments}
% \vspace{-10pt}
\label{tab:kpi}
\end{table*}

\noindent
\textbf{Query Benchmarks}. For each dataset, we model several meaningful aggregate threshold queries as summarized in Table \ref{tab:datasets}. In this paper, we present our experimental results for the COUNT aggregate function only. Queries with other aggregate functions (e.g. AVERAGE), which show similar results, are included in Appendix E.
As per our previous definition of conjunction/disjunction queries, all of our sub-queries have the same predicates but different filters on certain attributes (e.g. sub-queries on the UCI dataset all check the same 41 rooms every day for 14 days for the count of records with filters on different attributes). For the Sales dataset specifically, we model our queries based on frequently used retail and sales KPIs. Table \ref{tab:kpi} summarizes the KPIs used, their respective definitions, and the aggregate functions used to represent them. For example, to illustrate the User Retention Rate KPI, we evaluate the distinct count of customers who visited a specific business in a specific time period. For the UCI and NYTaxi datasets, which do not have clearly defined KPIs, we select meaningful queries that are modeled based on anomaly detection or performance evaluation scenarios typically used in decision support applications, for example, selecting statistics based on specific demographics (e.g. age, gender, etc.) or specific performance criteria  (e.g. fare amount is above the norm for a specific location). For each query, we select the corresponding threshold using the Z-score outlier detection method, to further emulate the concept of anomaly detection that decision support applications implement.

\noindent
\textbf{Algorithms}. We test our ProBE approach and compare it against the Naive approach mentioned in the Introduction. This Naive approach simply proportions the FNR bound $\beta$ into equal parts across the sub-queries $Q_i$ and calls sub-mechanisms with this $\beta_i$. This approach does not optimize the privacy budget given the $\beta$ constraint, nor does it offer utility guarantees on both $\beta$ or $\alpha$. In other words, this baseline is an extension of \cite{mide} where the $\beta$ budget is split evenly across sub-queries. We do not evaluate our approach against the algorithm in \cite{Apex:2019:RQP:1007568.1007642}, as \cite{mide} is directly based on it with the additional focus on bounding the FNR only. We also evaluate two variations of our ProBE framework; the two-step approach of ProBE using the Naive approach (i.e. splitting the $\beta$ budget equally between sub-queries) as its Phase One entitled ProBE-Naive, and the entropy-based iterative variation described in \secref{ssec:probeent} entitled ProBE-Ent.

\noindent
\textbf{Parameters}.
 For all four algorithms, we set a FNR bound of $\beta = 0.05$, a maximum privacy loss of $\epsilon_{max} = 5$. For ProBE-based algorithms, we choose a large starting uncertain region of $u_0=30\%$ of the value range, and set a FPR bound of $\alpha= 0.1$. For the Naive algorithm we choose $u=12\%$ as a default but explore other values in the experiments; we set this smaller default value due to the fact that the FPR is not bound in the Naive approach, meaning that a large $u$ would result in a very high FPR.
% \sm{should we explain why we chose these parameters? e.g., We choose initial u to be very large to be conservative.
% Also, explain why you choose u for naive approach to be larger? }
% \sm{Also, have you defined naive before.. or is this the first mention. 
% Should you not add a small para about approaches tested/compared?}
We run each algorithm over 100 iterations.
% \vspace{-2mm}
\subsection{Experimental Results}
\begin{figure*}
\captionsetup[subfigure]{justification=centering}
    \centering
    \raisebox{-0.7mm}{
    \begin{subfigure}[t]{.217\textwidth}
    \centering
    \includegraphics[width=\textwidth]{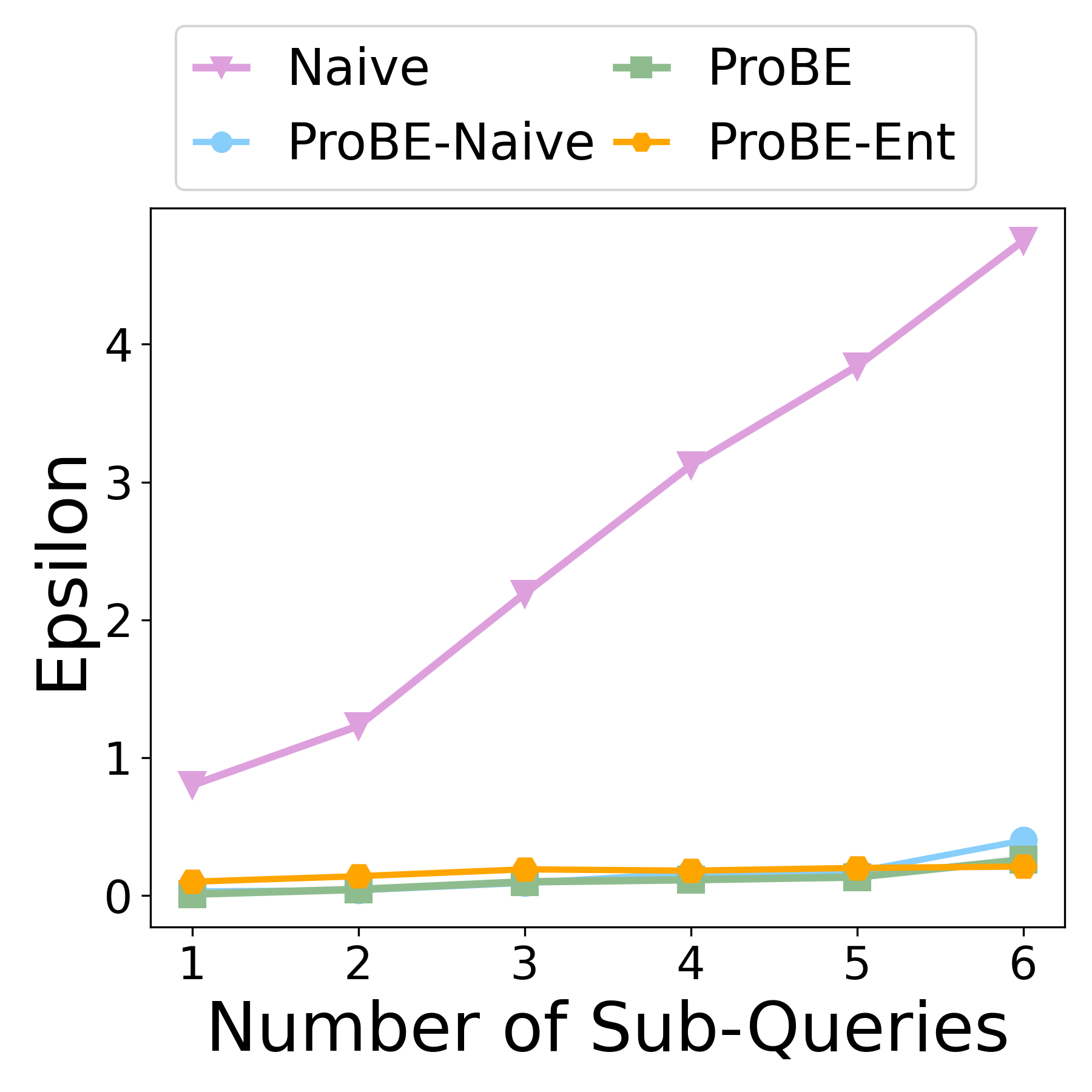}\\[4.7mm]
     
     \includegraphics[width=\textwidth]{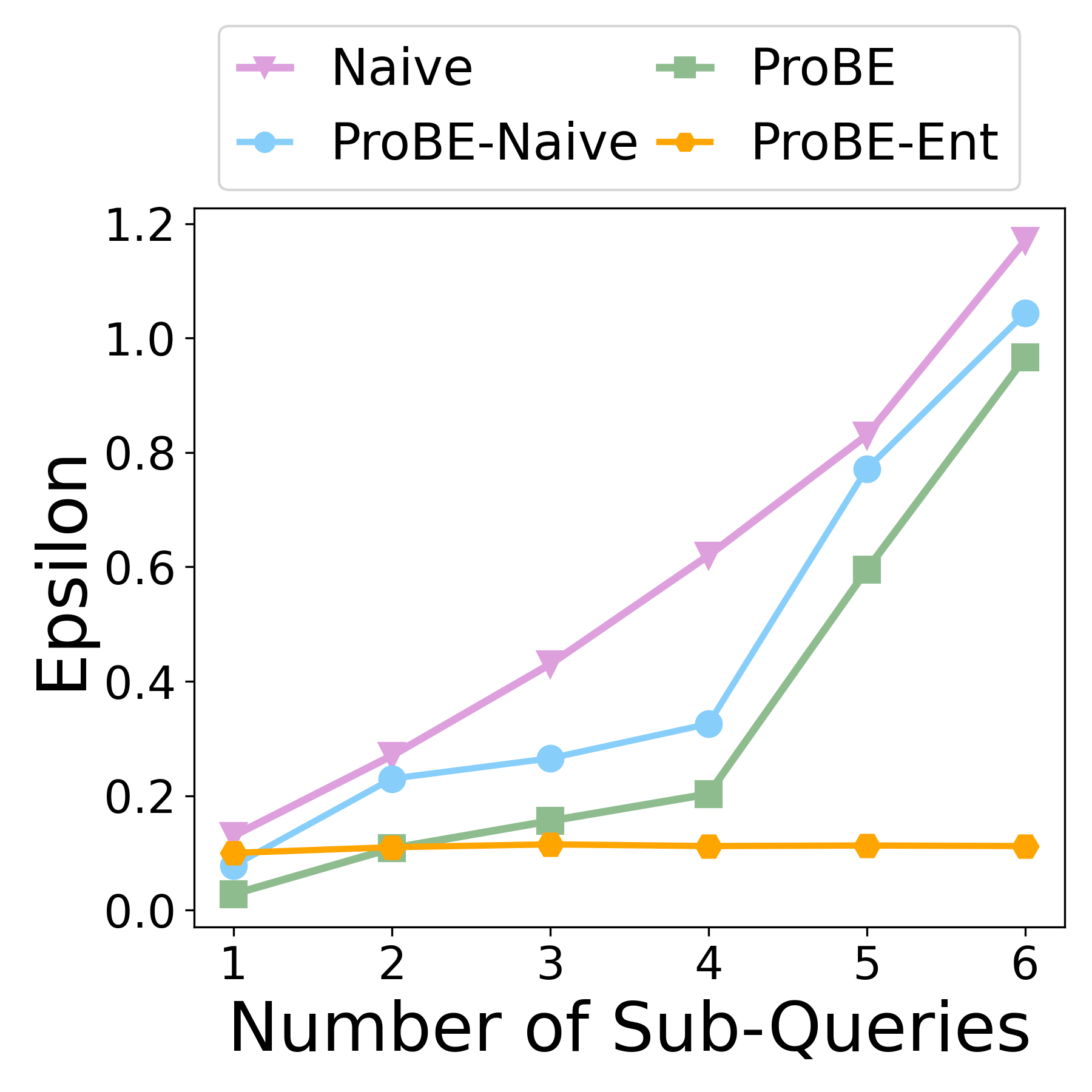}\\[4.5mm]
      
     \includegraphics[width=\textwidth]{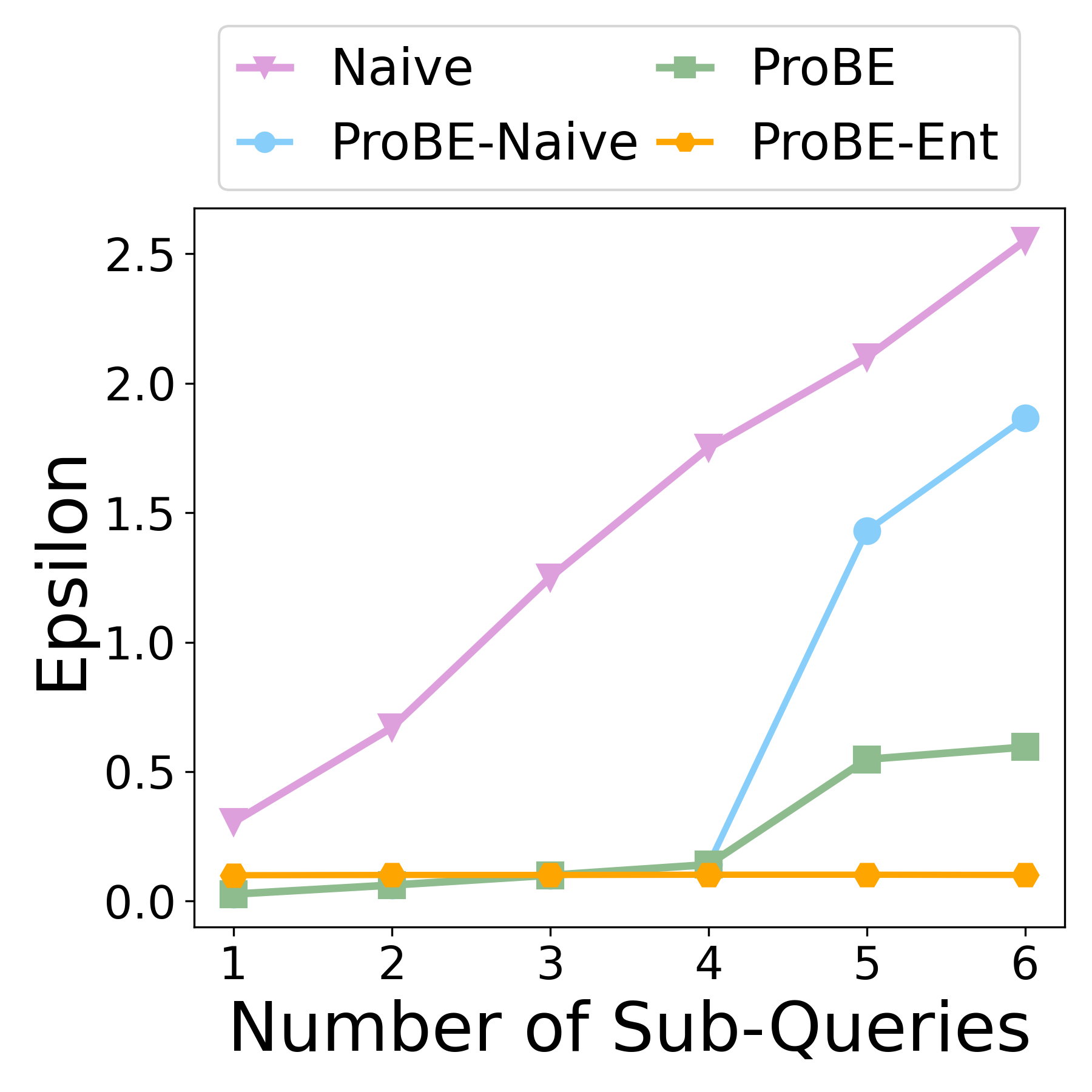}
    % \caption{NYTaxi data Epsilon (top) and \\ Min-Entropy (bottom)}
    
    \end{subfigure}
    }
    \raisebox{-0.7mm}{
    \begin{subfigure}[t]{.217\textwidth}
    \centering
    
    \includegraphics[width=\textwidth]{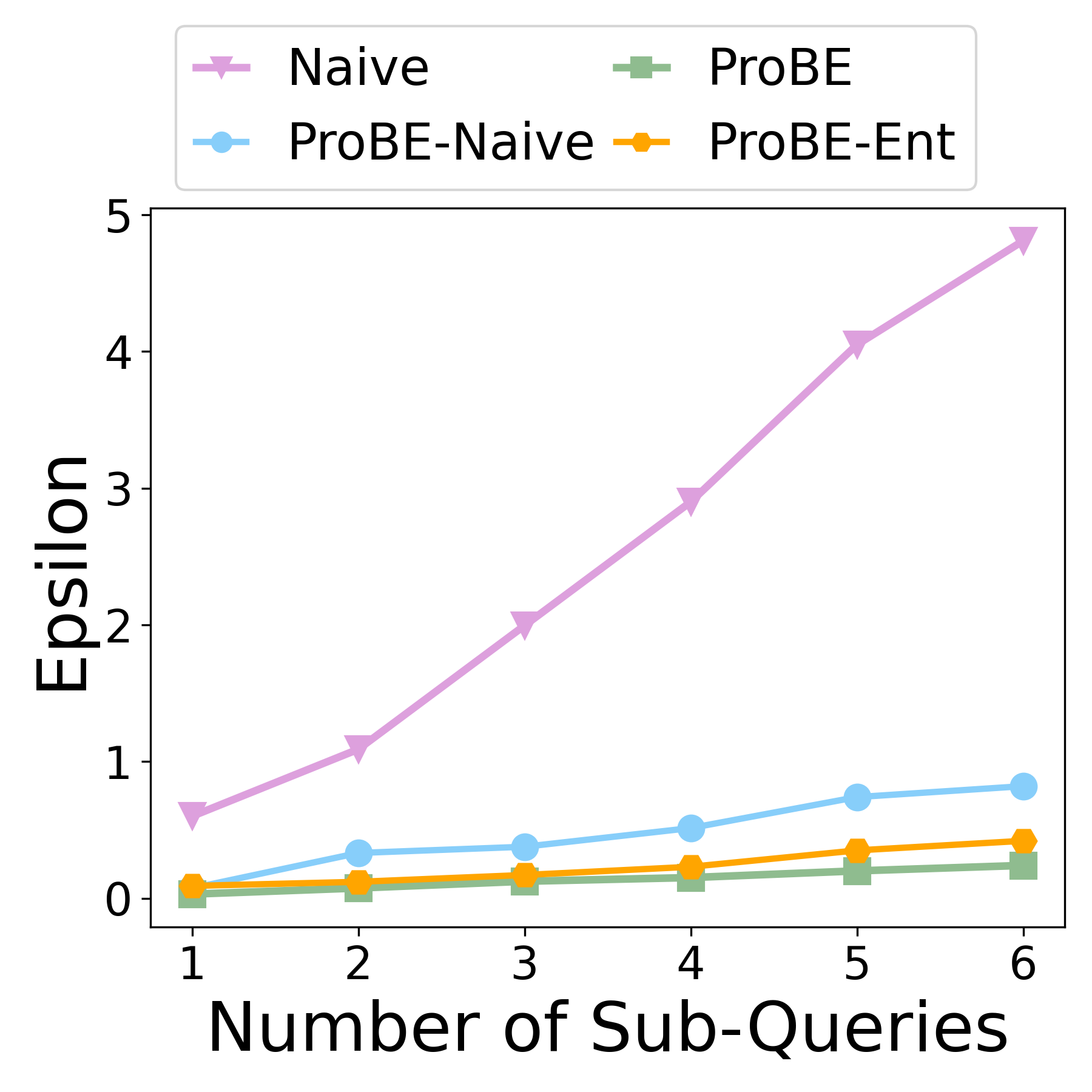}\\[4.7mm]
     \includegraphics[width=\textwidth]{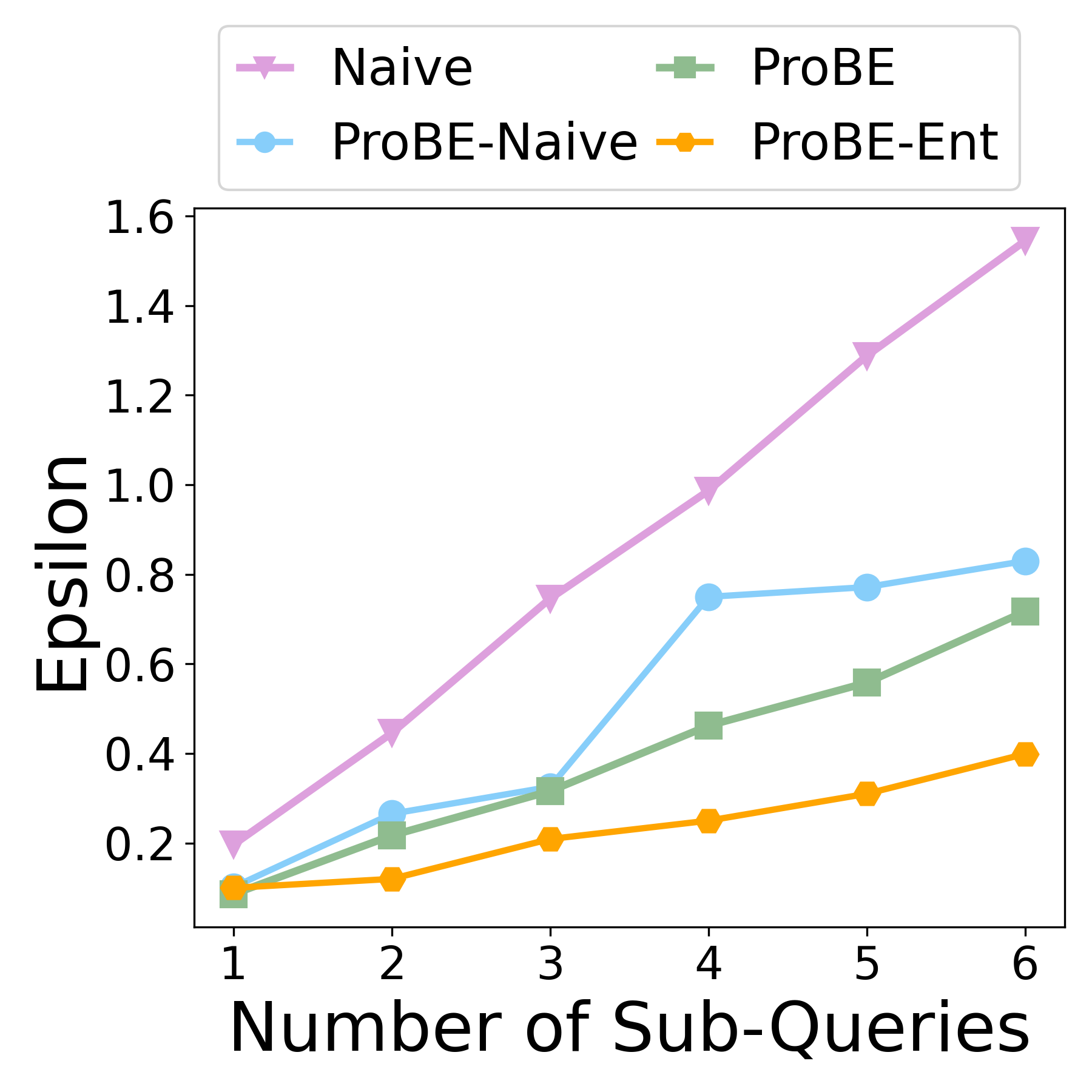}\\[4.5mm]
     \includegraphics[width=\textwidth]{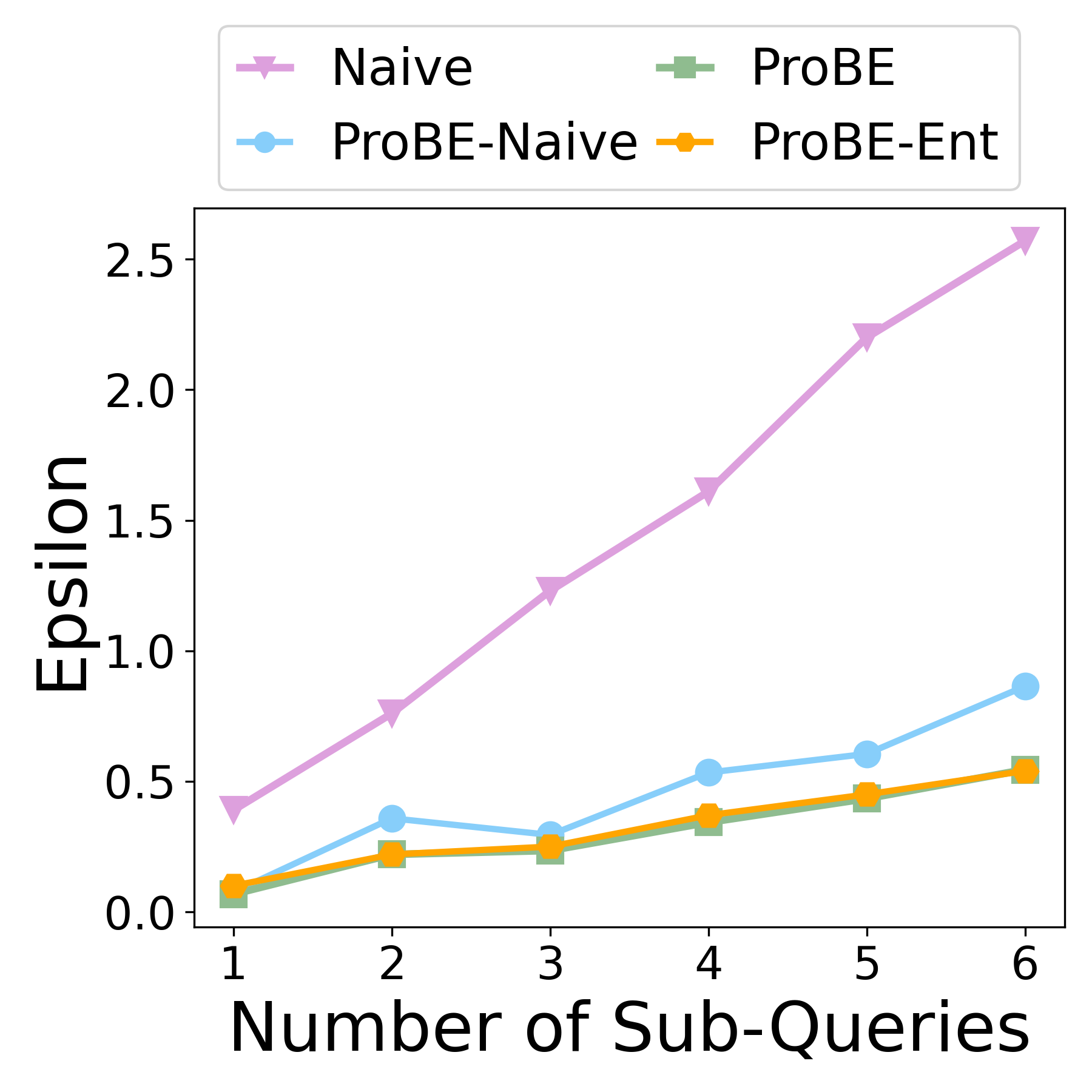}
    % \caption{NYTaxi data Epsilon (top) and \\ Min-Entropy (bottom)}
    \end{subfigure}
    }
    \begin{subfigure}[t]{.244\textwidth}
    \centering
    \includegraphics[width=\textwidth]{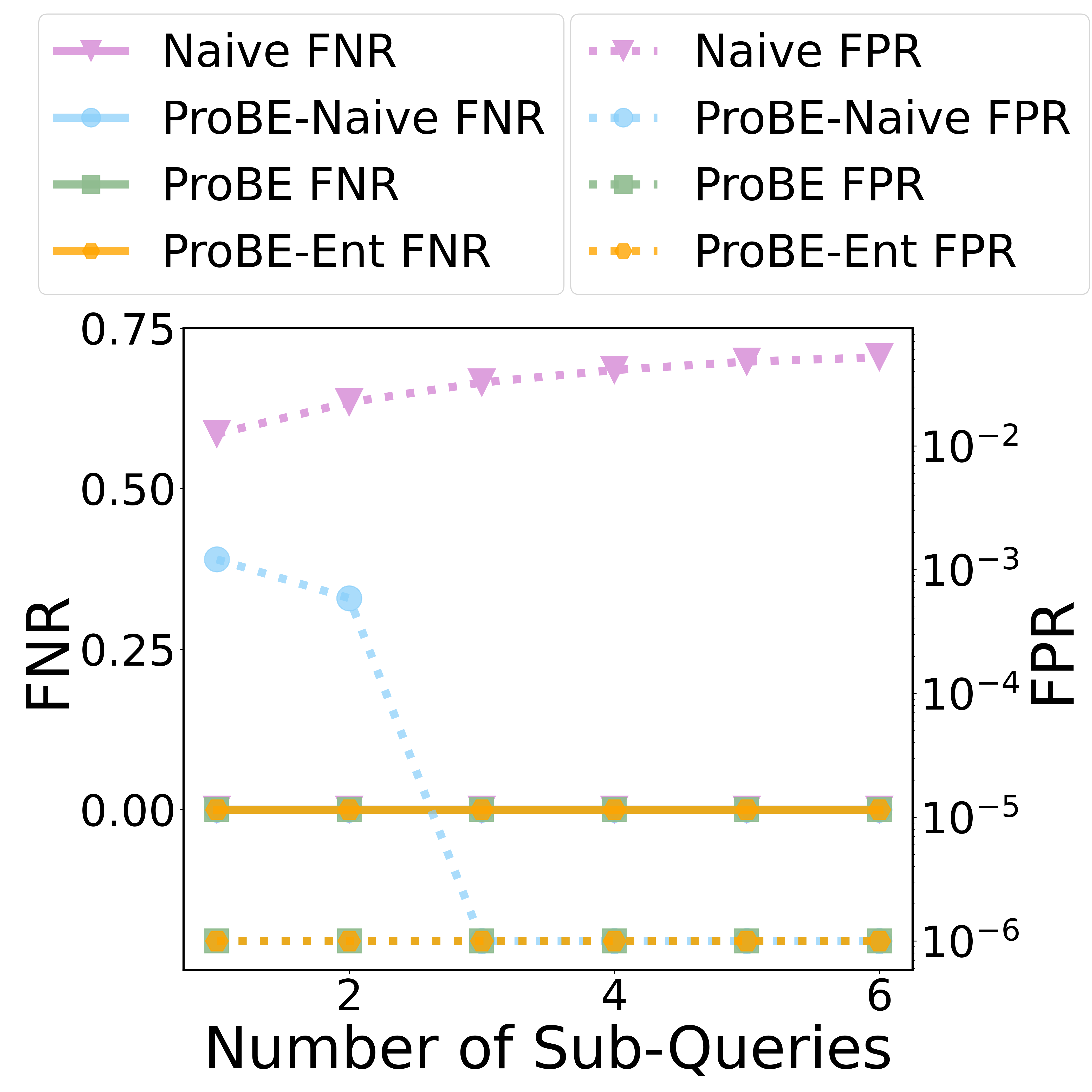}
     \includegraphics[width=\textwidth]{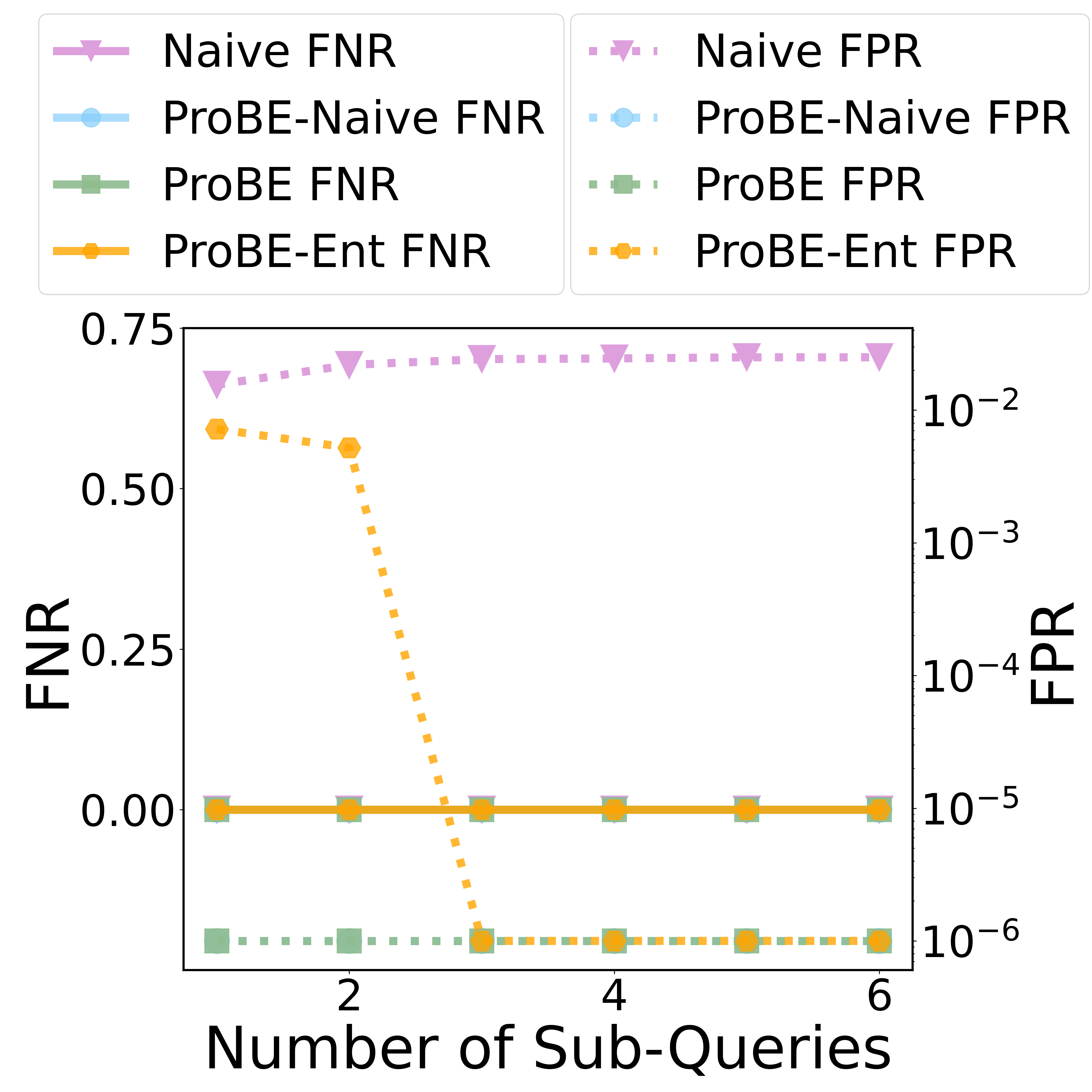}
     \includegraphics[width=\textwidth]{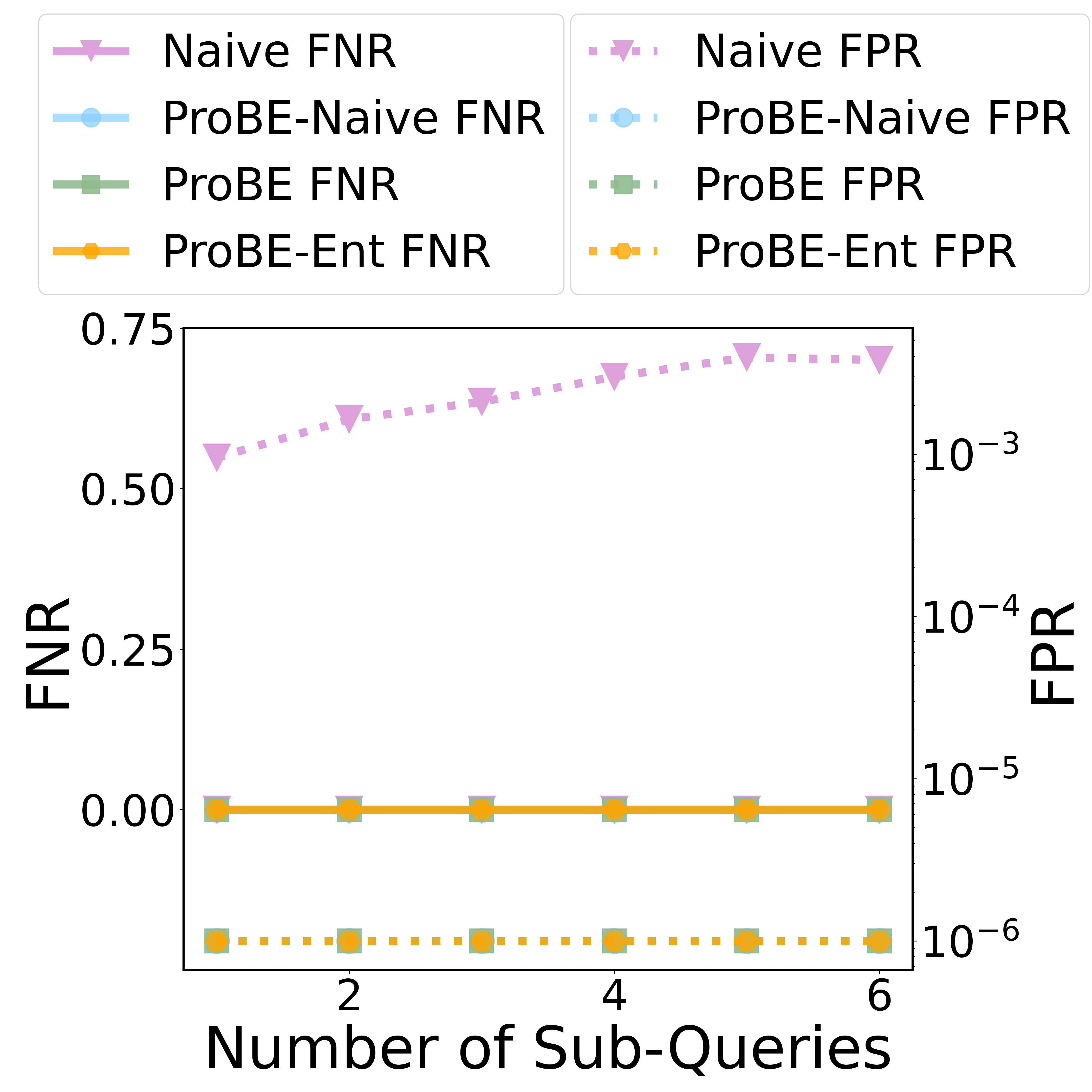}
    % \caption{NYTaxi data Epsilon (top) and \\ Min-Entropy (bottom)}
    \end{subfigure}
        \begin{subfigure}[t]{.244\textwidth}
    \centering
    \includegraphics[width=\textwidth]{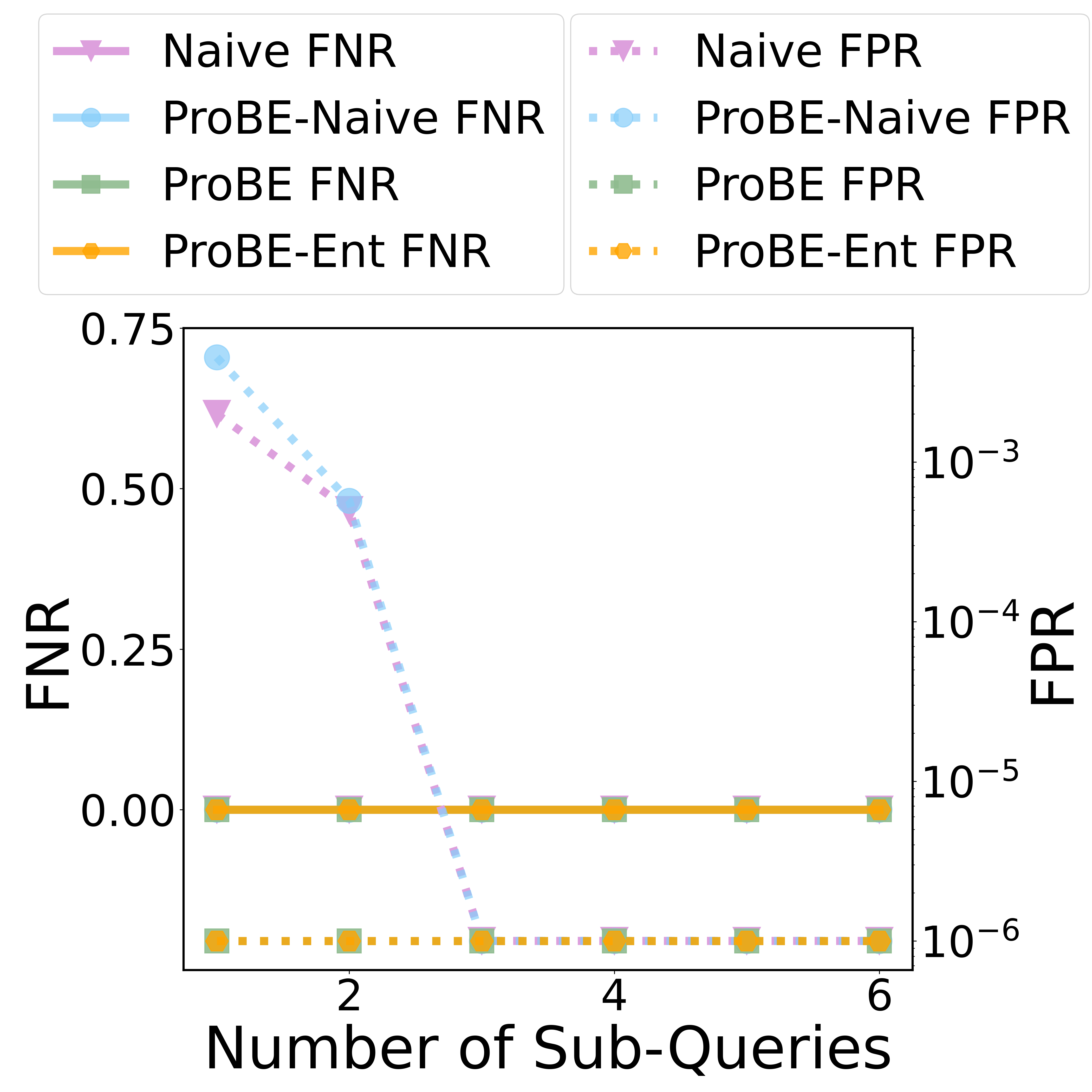}
     \includegraphics[width=\textwidth]{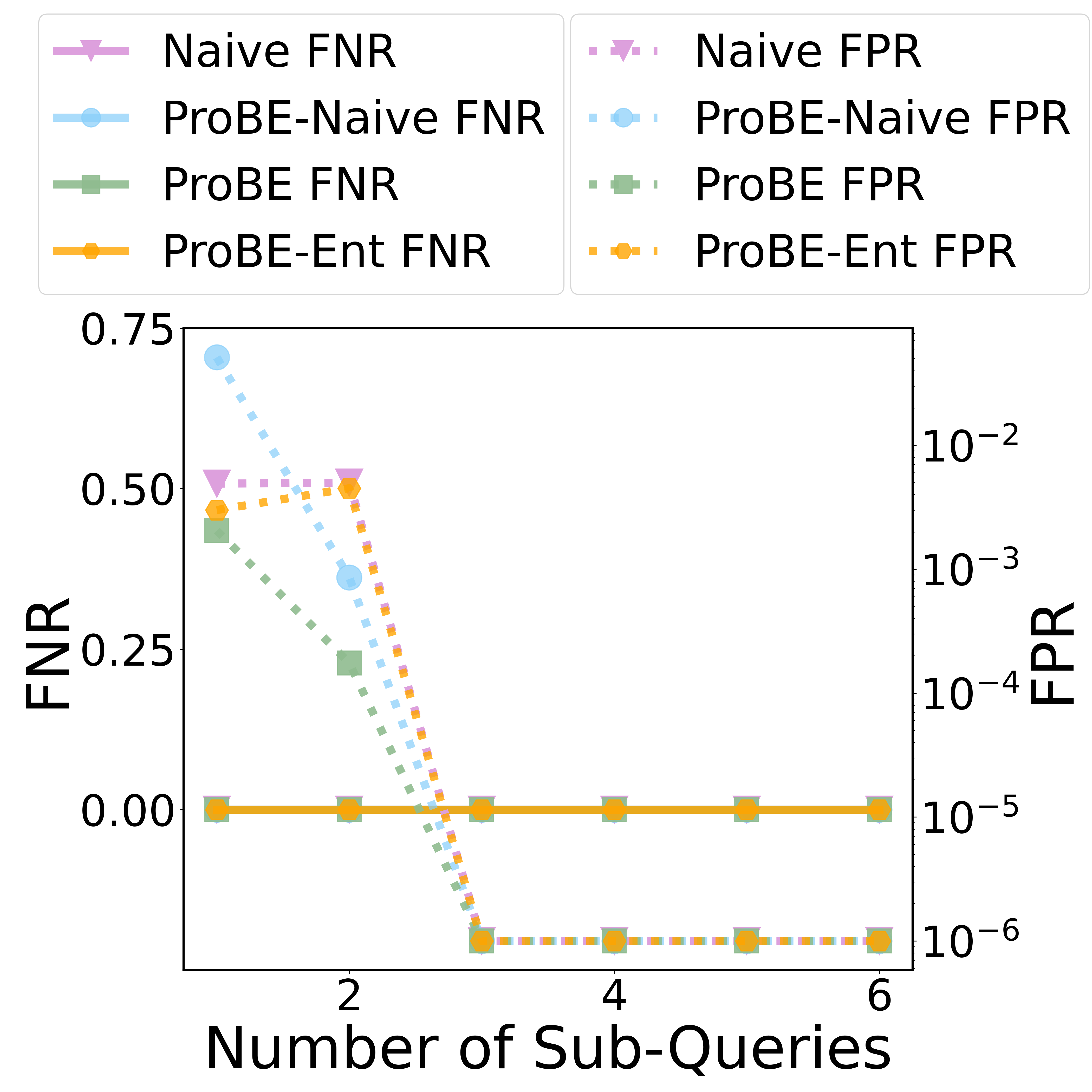}
     \includegraphics[width=\textwidth]{figures/experiments/combo4/comboAccUCI.png}
    % \caption{NYTaxi data Epsilon (top) and \\ Min-Entropy (bottom)}
    \end{subfigure}
    \caption{Privacy Loss in terms of Ex-Post DP $\epsilon$ (cols 1,2) and Accuracy in terms of FNR and FPR (cols 3,4) for 1-6 sub-queries at $\beta=0.05$ for Conjunction (cols 1,3) and Combined Conjunction/Disjunction (cols 2,4) using NYTaxi (row 1), Sales (row 2) and UCI (row 3) data.  The Accuracy plots have two axes, left for FNR (with a range of $[0,0.75]$) and right for FPR (with a range of $[10^{-6},10^{-2}]$ ). }
    \label{fig:privacyres}
\end{figure*}

\textbf{Privacy Results}. We use ex-post differential privacy, denoted by $\epsilon$, as our privacy metric to evaluate the performance of our ProBE optimization as implemented in the previously mentioned algorithms\footnote{We include the additional privacy results in terms of Min-Entropy in the Appendix.}. We assess all four algorithms on exclusive conjunction, exclusive disjunction, and a randomized combination of both for a varying number of sub-queries (1 to 6) in order to evaluate the effect of the operator on privacy loss and utility. We only include conjunction and the combination in Figure~\ref{fig:privacyres}, as disjunction yields the same results. Note that in the case of a single sub-query, the Naive and ProBE algorithms would yield the same privacy results provided that the second step of ProBE is not triggered.
% \nl{we might need to remove the conjunction only results and move them to the appendix}

As expected, the experiments show that ProBE-based algorithms achieve their respective bounds on the FPR and FNR. Furthermore, they achieve these bounds with minimal privacy loss. While the FPR and FNR bounds are set to $0.1$ and $0.05$ respectively, the actual rates are even smaller, while the maximum $\epsilon$ ranges from 1.5 in the Sales dataset to about 5 for the NYTaxi dataset. This difference is due to the underlying data distribution, which might trigger more predicates to be reran in the second step in order to guarantee the $\alpha$ bound. In contrast, the Naive algorithm results in a linear increase on the privacy budget, as well as high values specifically for the FPR. The iterative, entropy-based algorithm ProBE-Ent achieves close or the lowest privacy loss overall. This is due to the fact that ProBE-Ent makes use of the underlying data distribution to progressively compute the results of the query at the lowest privacy cost, while also allowing for early stopping at a much lower privacy loss if all predicates are properly classified outside of the uncertain region.
As expected, the privacy loss increases as the number of sub-queries increases, but the Naive algorithm has a linear increase, whereas ProBE is less affected by the increasing complexity. 
% The data distribution has a significant effect on privacy loss, as each dataset has a different range of privacy loss, with the Sales data set seeing the lowest $\epsilon$ overall. 
The nature of the query (conjunction versus combination) does not cause a significant difference in pattern for privacy loss for the same data set.

\textbf{Comparison of Naive versus Optimized Phase One}. By analyzing the difference in results between the  ProBE-Naive and ProBE algorithms, we are able to highlight the importance of the $\beta$ budget optimization performed by ProBE in Phase One. In cases where query sensitivities and domain sizes dramatically differ (i.e. resulting in vastly different uncertain regions) across sub-queries, the optimal distribution would be to allocate a smaller $\beta$ budget for the sub-query with the larger uncertain region $u$ as they are inversely correlated (a direct reflection of the trade-off between FN and FP), and allocate a larger $\beta$ to the sub-query with a smaller $u$. In cases such as this, the Naive approach would result in a non-optimal equal distribution of the $\beta$ budget and subsequently a higher privacy loss. We see this difference in privacy loss in Figure \ref{fig:privacyres} in specific cases such as the Sales (row 2) and UCI (row 3) datasets where the addition of specific sub-queries results in a much higher privacy loss depending on underlying the data distribution.

\noindent
\textbf{Accuracy Results}. To evaluate our algorithms, we use two accuracy metrics, the FNR, which is defined as the number of false negatives divided by the total number of positives,  and the FPR, defined as the number of false positives divided by the total number of negatives. These metrics are averaged over the number of iterations run per algorithm, which we set to 100. We again use the default value of $\beta=0.05$ for the bound on FNR and  $\alpha =0.1$ for the bound on FPR. We use the same uncertain region parameters as previously mentioned.
\begin{figure*}[t]
\captionsetup[subfigure]{justification=centering}
    \centering
    % \begin{subfigure}[b]{.25\textwidth}
    %  \centering
    % \includegraphics[width=.59\textwidth] 
    % {figures/experiments/varyingSALES/varyep.png}
    % \includegraphics[width=.49\textwidth] 
    % {figures/experiments/varyingSALES/varyacc.png}
    % \end{subfigure}
    \begin{subfigure}[b]{.22\textwidth}
    \centering
    \includegraphics[width=\textwidth]{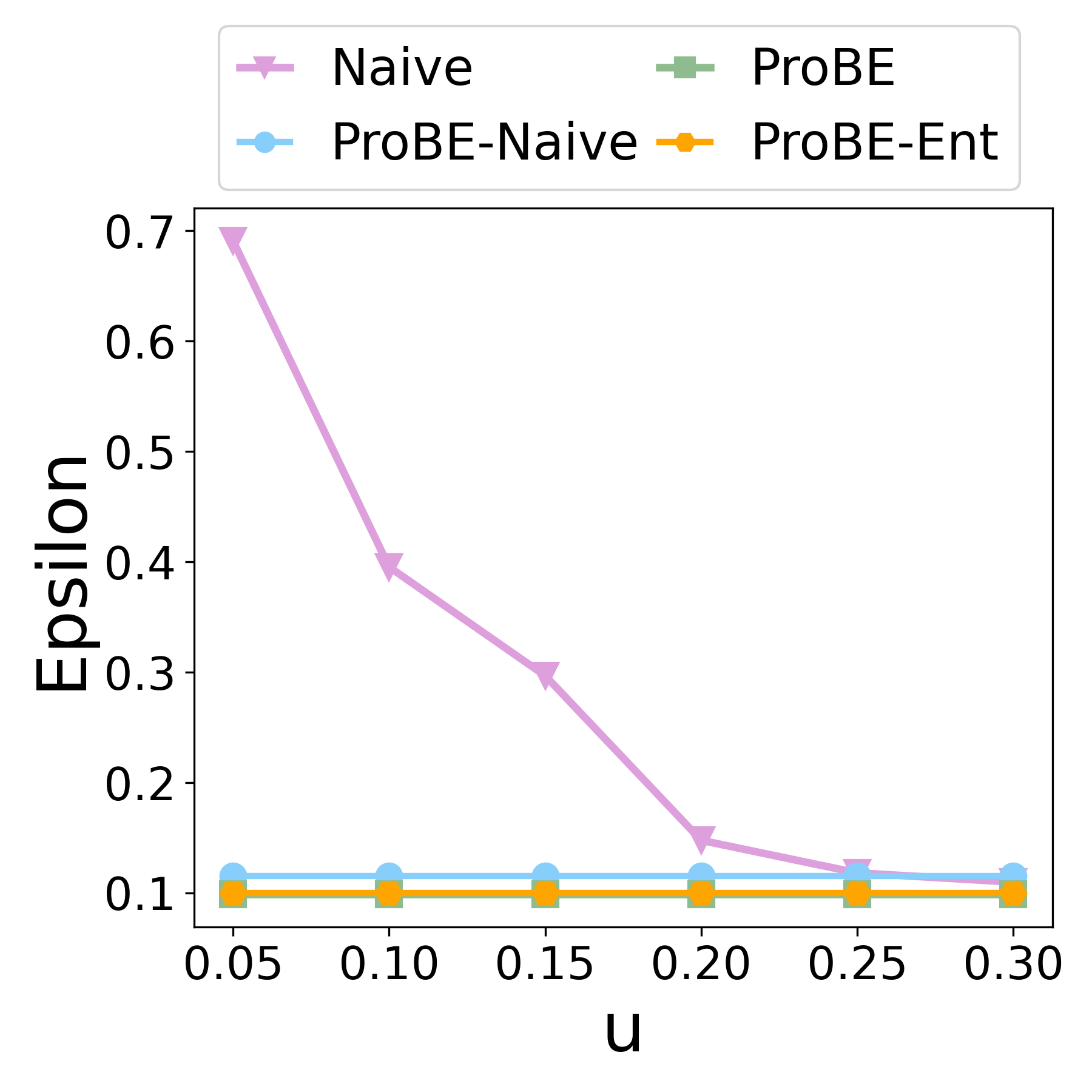}
    \caption{Ex-Post DP $\epsilon$}
    \label{fig:varEp}
    \end{subfigure}
    \begin{subfigure}[b]{.255\textwidth}
    \centering
    \includegraphics[width=\textwidth]{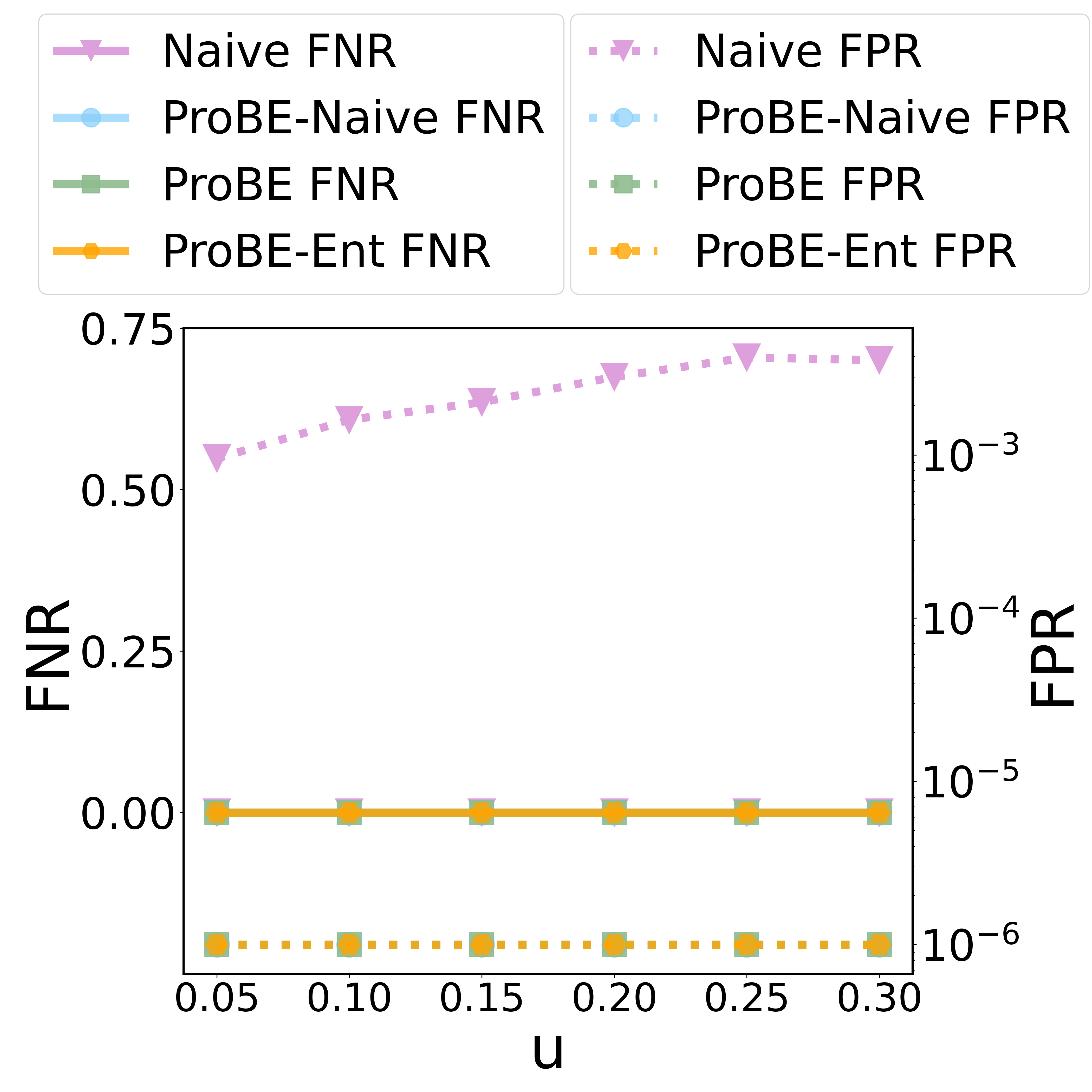}
    \caption{FNR and FPR}
    \label{fig:varEnt}
    \end{subfigure}
    % \raisebox{-2mm}{
    \begin{subfigure}[b]{.22\textwidth}
    \centering
    \includegraphics[width=\textwidth]{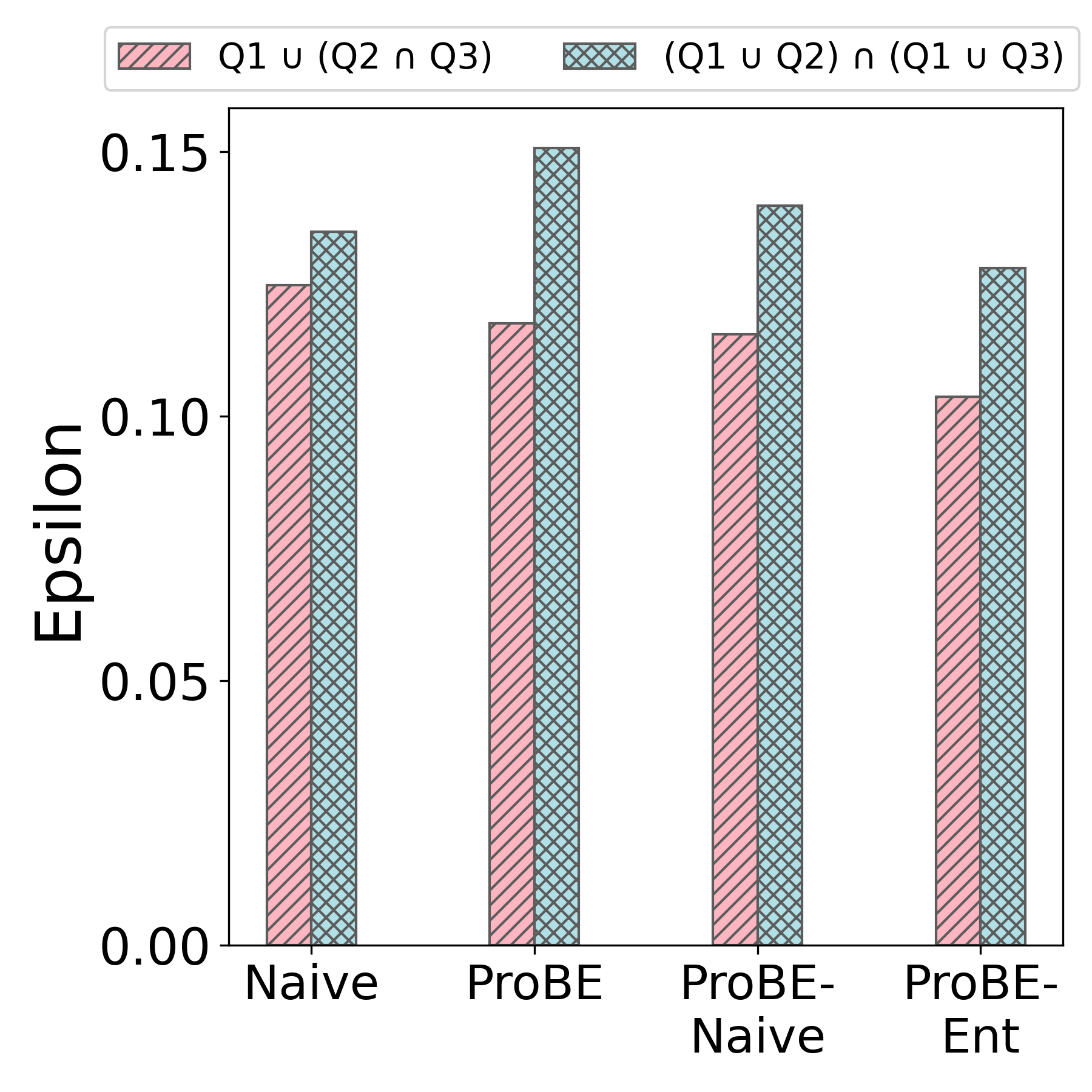}
    % \vspace*{-6mm}
    \caption{Ex-Post DP $\epsilon$}
    
    \label{fig:distEp}
    \end{subfigure}
    % }
    \begin{subfigure}[b]{.27\textwidth}
    \centering
    \includegraphics[width=\textwidth]{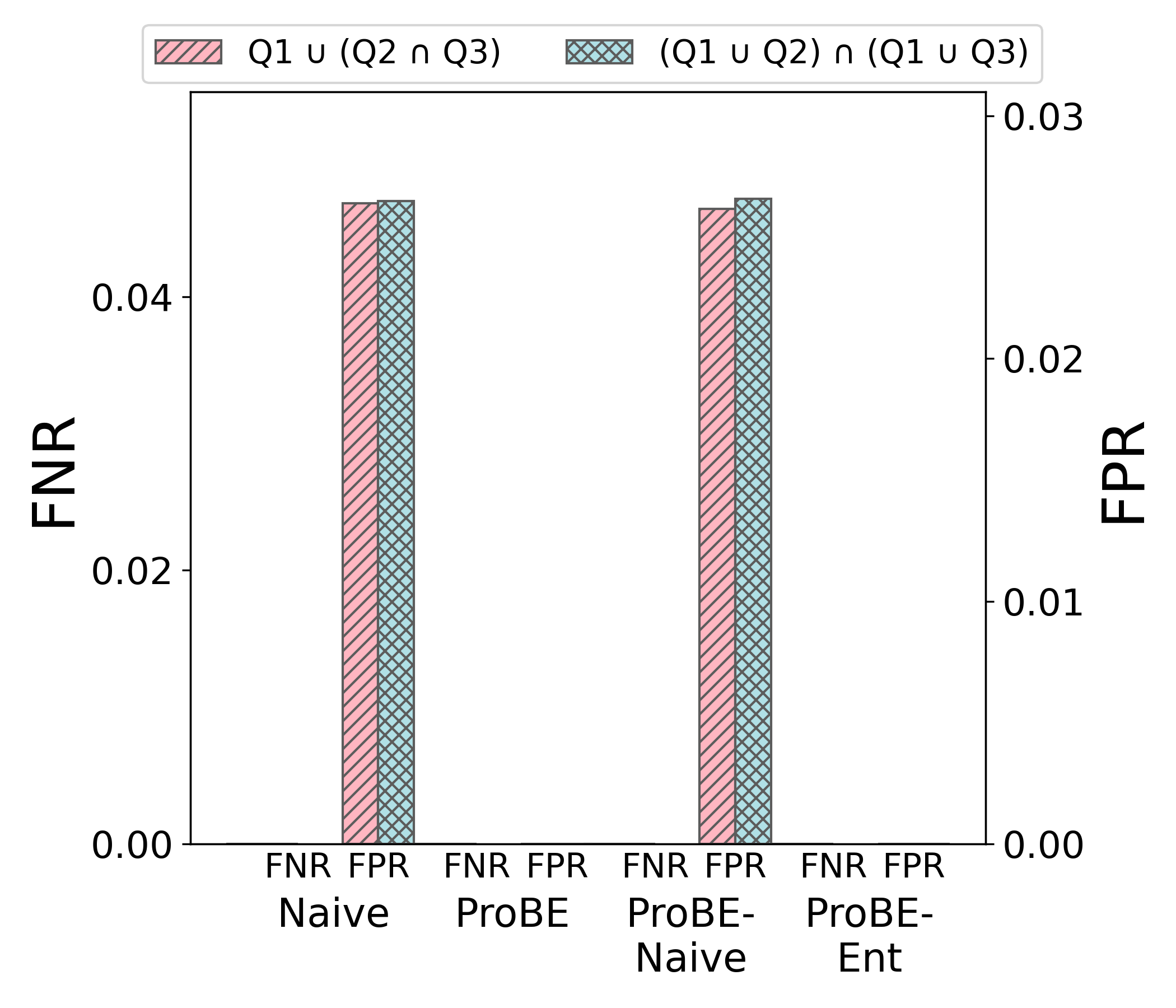}
    \vspace*{-6mm}
    \caption{FNR and FPR}
    \label{fig:distEnt}
    \end{subfigure}
   %  \begin{subfigure}[b]{.285\textwidth}
    
   %  \centering
   %  \includegraphics[width=\textwidth]{figures/experiments/varyingSALES/varyacc.png}
   % \caption{FNR and FPR}
   % \label{fig:varFNRFPR}
   %  \end{subfigure}
    \caption{Privacy ($\epsilon$) and Accuracy (FNR and FPR) at $\beta=0.05$ for varying $u$ (a-b) and for two equivalent query trees (c-d) on Sales data}
    \label{fig:varyingdistrib}
\end{figure*}

All algorithms achieve the guaranteed bound of $\beta=0.05$ as seen in Figure \ref{fig:privacyres} (columns 3-4), where the FNR is zero as the number of sub-queries increases for all datasets, hence the overlap of solid lines (FNR) for all algorithms.  For the Naive algorithm, the FPR mostly sees a steady increase both in the conjunction-only and combination queries, whereas ProBE-based algorithms successfully meet the bound on FPR $\alpha$ at a lower privacy loss due to the upper bound estimation and dynamic re-computation of the uncertain region parameter $u_{opt}$. The pattern, however, largely depends on the data distribution of the underlying dataset; e.g. for the Sales dataset we note that the FPR decreases as the number of sub-queries increases, which may be attributed to selectivity of the query along with the distance of the data points from their respective thresholds. 
\stitle{Varying Uncertain Region}. We vary the uncertain region parameter $u$ for the Naive algorithm  by setting its values to $\{5,10,...,30\}$\% of the data range on the Sales dataset in order to evaluate its performance compared to ProBE-based algorithms, which internally set their $u$ (first by setting its initial value to a conservative $30\%$ then by recomputing it in the second step).  We use the query $Q_1 \cup (Q_2 \cap Q_3)$ where $Q_1,Q_2,Q_3$ refer to the first three KPIs from Table \ref{tab:kpi}, and the thresholds are computed using the outlier method again. 
% \sm{is there a repetition of 5,10,... from earlier in the para. Also, where are 
% Q1, Q2 and Q3 descibed? You could start this setence as We use Sales dataset for the experiment and use the query q1 and q2 or q3
% to determine}
We fix the other parameters to $\beta = 0.05$ and $\alpha=0.1$. 
% \sm{ We vary the uncertain region parameter $u$ for the Naive algorithm settings its values to $\{5,10,...,30\}$\% 
% of the data range on the Sales dataset to explore how the naive algorithm with varying uncertainty region compares to the
% ProBE which sets its uncertainty interval by itself during
% phase 2 after starting with a very conservative value of $u$ 
% ($30\%$) in Phase 1. We use the  query $Q_1 \cup (Q_2 \cap Q_3)$. We fix the other parameter to $\beta = 0.05$ and $\alpha=0.1$.}

% \begin{table}[H]
%     \centering 
%     \begin{tabular}{|c|c|c|c|}
%     \hline
%     % \cline{2-4}
%     {\bf Correlation} & Negative& Low/None & Positive \\ \hline
%          {\bf FNR} & 0.0015 & 0.0 & 0.0  \\ \hline
%     \end{tabular}
%     \caption{False Negative Rate for Conjunction Query with Sub-Queries of Varying Correlation with $\beta=0.05$, $s_{min}=0.05$.}
%     \label{tab:corrQueries}
% \end{table}

Figure \ref{fig:varyingdistrib}(a-b) shows that as the uncertain region increases, privacy loss declines for all algorithms as expected, seen in Figure \ref{fig:varyingdistrib}\subref{fig:varEp}. We see that for the Naive algorithm, the FPR steadily increases as the uncertain region increases, whereas the privacy loss $\epsilon$ steadily decreases as the $u$ increases. This shows not only the direct correlation between false positives and $u$, but also the extent of the trade-off between privacy loss and the FPR. For our ProBE algorithms, since they do not take the parameter $u$ but rather internally set it in an optimal way, the privacy level and FPR are constant across plots. The FNR does not change for any algorithm due to the upheld $\beta$ bound guarantee.
% \nl{rerun exp}
% \noindent
% \textbf{Correlated Sub-Queries}. We evaluate how our approach performs when using the $\beta$-bound on FNR, and vary the correlation factor between sub-queries due to its direct effect on the $\beta$ bound as explained in \secref{independence} (i.e. our apportionment technique does not always guarantee a $\beta$-bound on FNR depending on the correlation factor between sub-queries). We compare three pairs of sub-queries on the Sales dataset with negative, low, and positive correlations using the Pearson correlation factor. We test each level on a 2-way conjunction query $Q = Q_1 \cap Q_2$ using the ProBE-SS algorithm with the default parameters of $\beta=0.05$ and $u=12\%$. Table \ref{tab:corrQueries} shows that for all levels of correlation, we successfully achieve the $\beta$-bound FNR. For negatively correlated sub-queries, the FNR is slightly higher, which is consistent with Eq. \eqref{eq:conjf}, but is still lower than $\beta$.

\stitle{Query Trees and Operator Distribution}. As discussed in \secref{treestructure}, the structure of a query tree has an effect on privacy loss. To evaluate the impact of operator distribution (i.e. distributing a conjunction over a disjunction and vice versa) we run ProBE-based algorithms and Naive algorithms for the two queries illustrated in Figure \ref{fig:q-tree} (c-d): $Q_I = Q_1 \cup (Q_2 \cap Q_3)$ and $Q_{II} = (Q_1 \cup Q_2) \cap (Q_1 \cup Q_3)$. We run these queries on the Sales dataset with the default parameters of $\beta = 0.05$, as well as a uncertain region parameter of $u_0=12\%$ for the Naive algorithm. 
Figure \ref{fig:varyingdistrib} shows that the distributed query $Q_{II}$ incurs a higher ex-post privacy cost than the grouped query $Q_{I}$ for all algorithms. Similarly, the FPR for the Naive algorithm sees a slightly higher value for query $Q_{II}$ as compared to the original query $Q_{I}$. Experiments ran on distributing the AND operator over or (i.e. $Q_1 \cap (Q_2 \cup Q_3)$ vs. $(Q_1 \cap Q_2) \cup (Q_1 \cap Q_3)$) showed similar results. This is attributed to the fact that $Q_1$ was allocated an additional privacy budget due to its second occurrence in $Q_{II}$, thus incurring a higher cost on the overall privacy loss due to the use of the Sequential Composition theorem.

\begin{figure*}
\captionsetup[subfigure]{justification=centering}
    \centering
    \begin{subfigure}[b]{.22\textwidth}
    \centering
    \includegraphics[width=\textwidth]{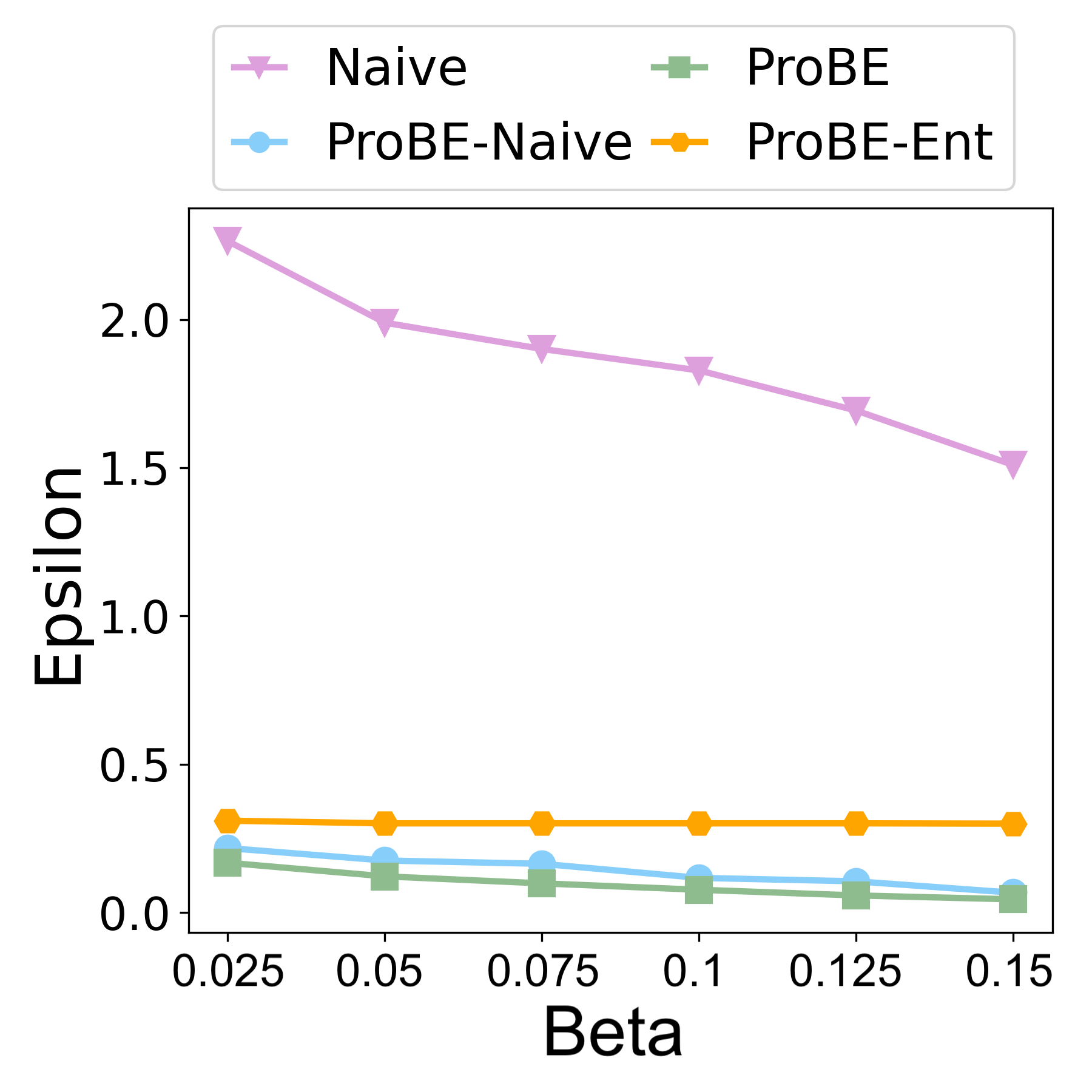}
    \caption{Ex-Post DP $\epsilon$}
    \label{fig:varBetaEp}
    \end{subfigure} 
    \begin{subfigure}[b]{.255\textwidth}
    \centering
    \includegraphics[width=\textwidth]{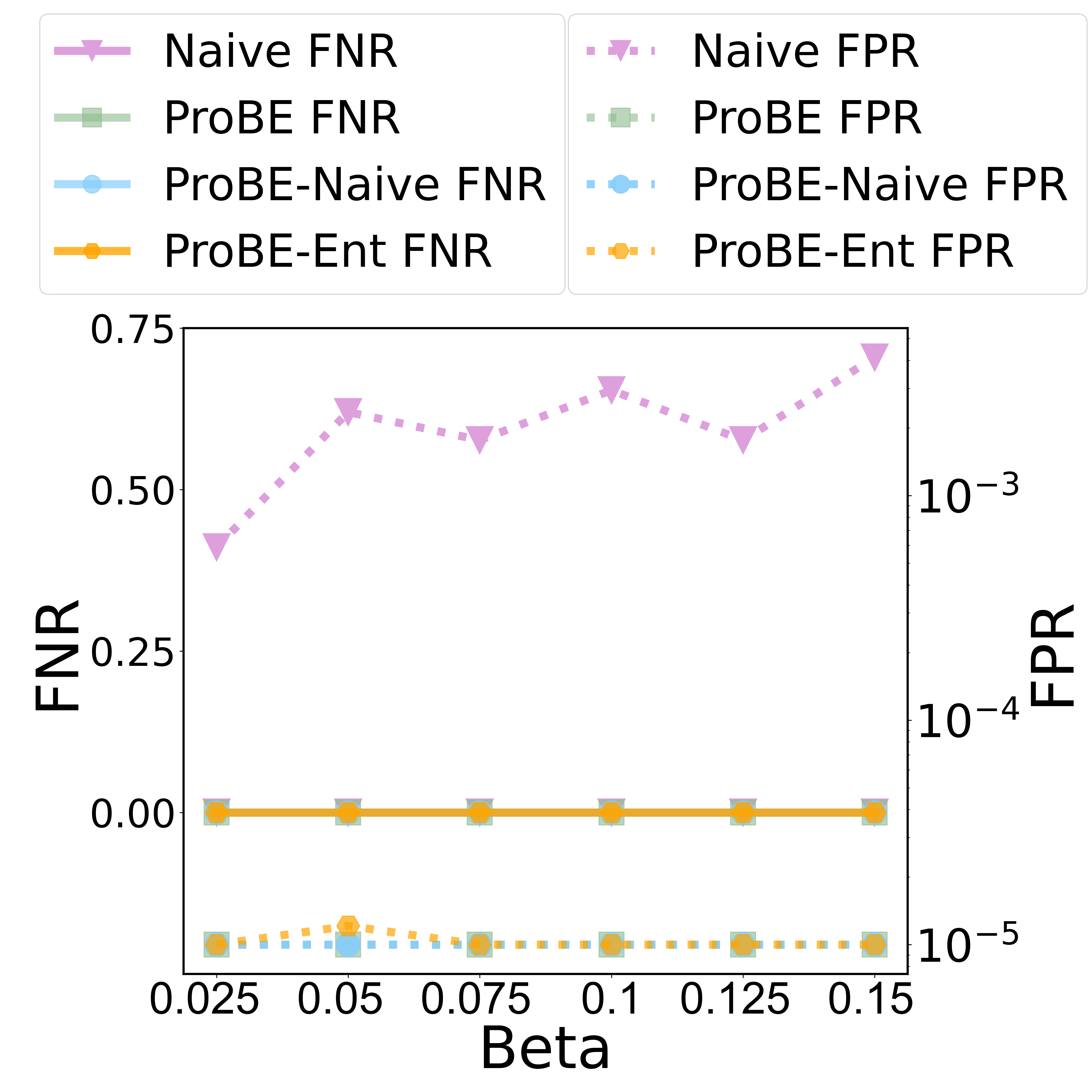}
   \caption{FNR and FPR}
   \label{fig:varBetaFNRFPR}
    \end{subfigure}
    \begin{subfigure}[b]{.22\textwidth}
    \centering
    \includegraphics[width=\textwidth]{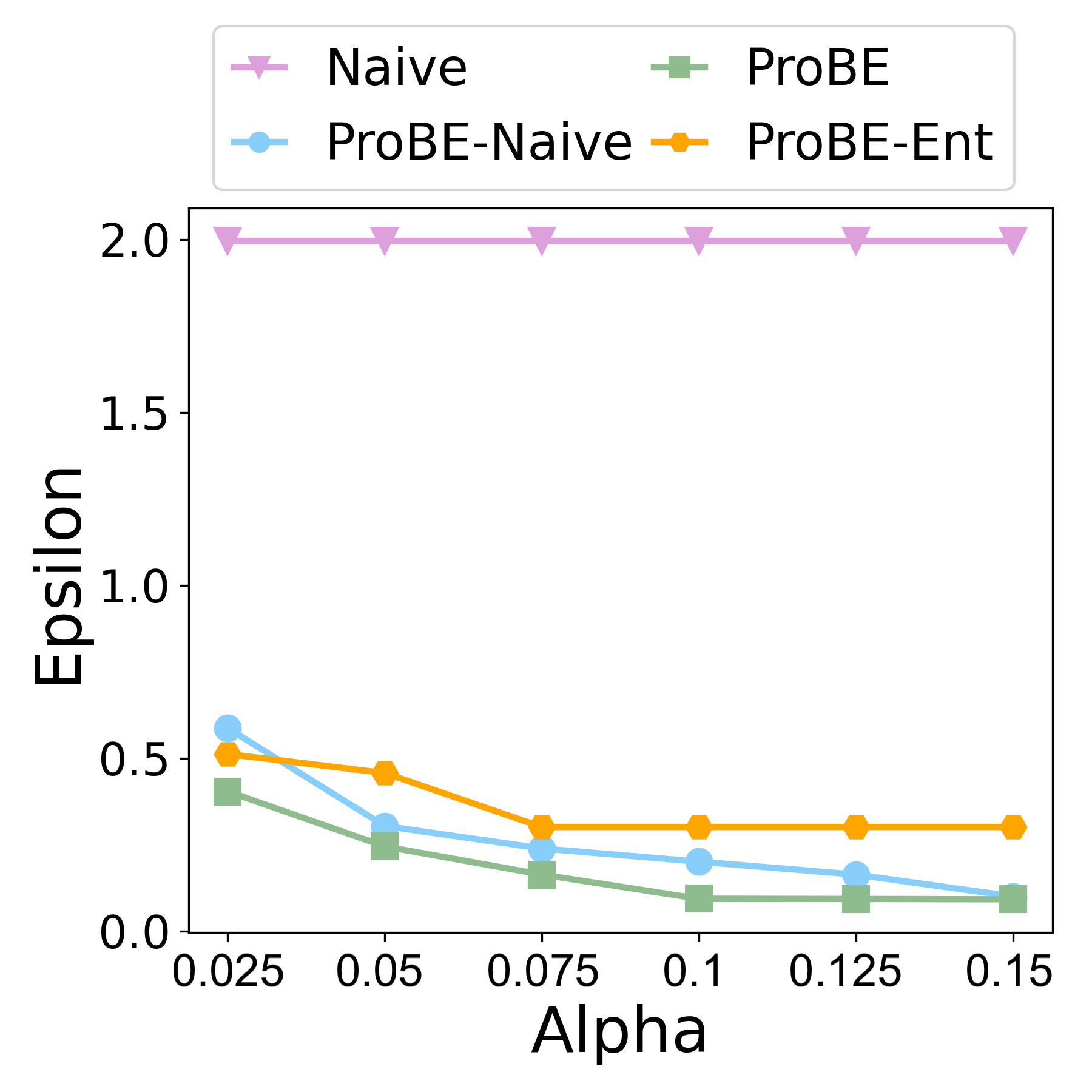}
    \caption{Ex-Post DP $\epsilon$}
    \label{fig:varBetaEp}
    \end{subfigure} 
    \begin{subfigure}[b]{.255\textwidth}
    \centering
    \includegraphics[width=\textwidth]{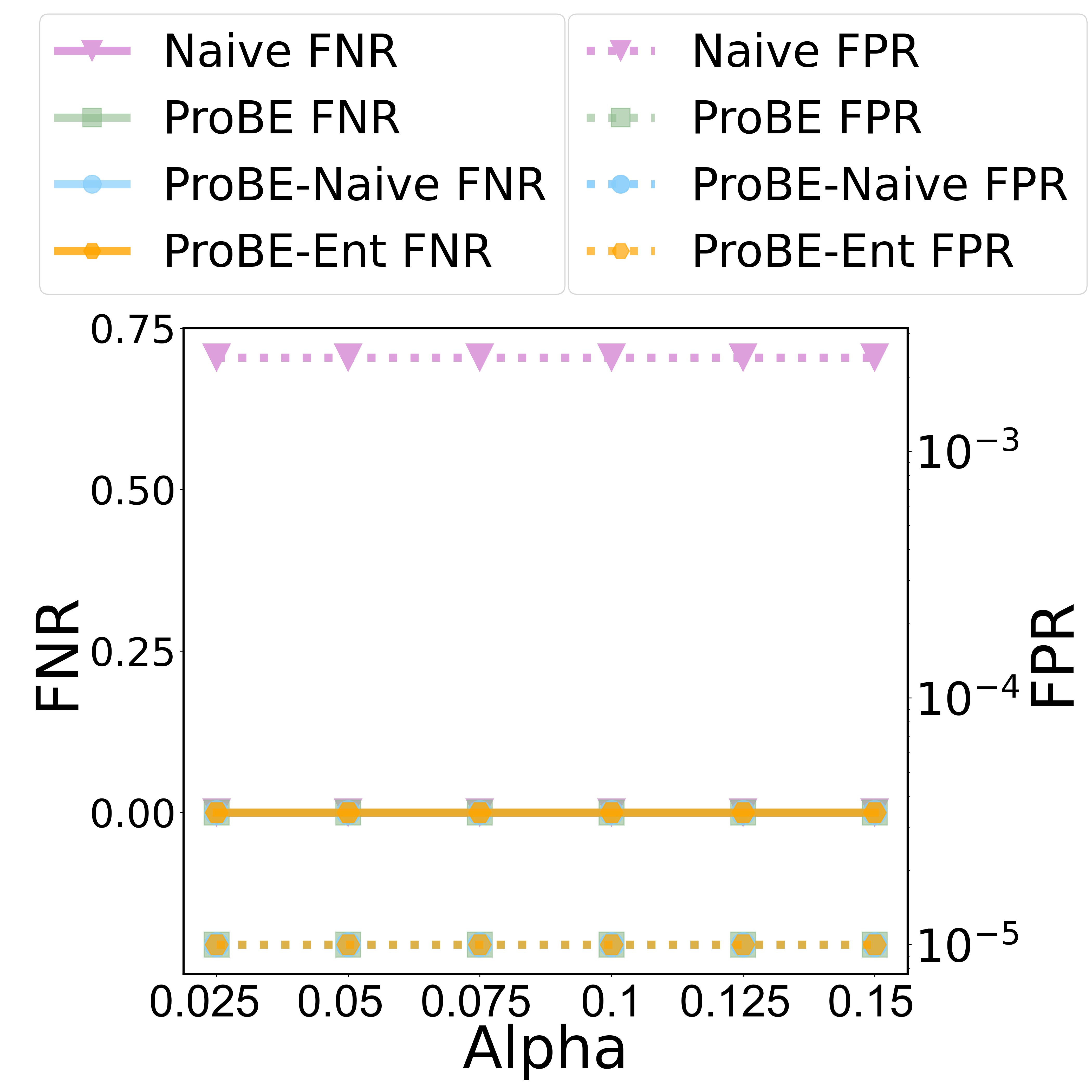}
   \caption{FNR and FPR}
   \label{fig:varBetaFNRFPR}
   \end{subfigure}
    \caption{Ex-Post DP ($\epsilon$) and Accuracy (FNR, FPR) Results for varying $\beta$ (a-b) and $\alpha$ (c-d) values on Taxi data.}
    \label{fig:varyAB}
    
\end{figure*}

\textbf{Varying $\beta$ and $\alpha$ Parameters}. We analyze the effect of selecting different values for the user-set FNR bound $\beta$ and FPR bound $\alpha$ on all four algorithms. We vary the two bounds by setting their values to $\{0.025, 0.05, 0.075, 0.1, 0.125, 0.15\}$ on the Taxi dataset with a 3-disjunction query (i.e. $Q_1 \cup Q_2 \cup Q_3$). Figure \ref{fig:varyAB} (a) shows that varying $\beta$ causes privacy loss $\epsilon$ to decrease as $\beta$ increases across all algorithms, which is to be expected due to the inverse correlation between $\epsilon$ and $\beta$. The accuracy measures depicted in Figure \ref{fig:varyAB} (b), i.e. the FNR and FPR, are not significantly impacted as their respective bounds are met by the ProBE-based algorithms. Conversely Figure \ref{fig:varyAB} (c-d) depicts the effect of varying the $\alpha$ bound on FPR. As the Naive algorithm does not support a mechanism to bound the FPR, the privacy and accuracy results remain constant.  On the other hand, we note that the privacy loss $\epsilon$ also decreases as the $\alpha$ bound is increased for ProBE-based algorithms. This is attributed to the fact that, as the $\alpha$ bound increases, the probability of Phase Two being run decreases as the tolerance for FP errors is higher, thus avoiding the additional privacy cost of the second step. However, this decrease is dependent upon the data distribution and how many elements are in the uncertain region; if the estimated FPR is lower than a smaller bound (e.g. $0.1$) then increasing $\alpha$ will have no additional on privacy loss, hence the somewhat constant FPR between $0.1-0.15$ for the ProBE-based algorithms. In terms of accuracy, the FNR and FPR similarly do not see a significant change, as their respective bounds are met regardless of their values. Setting a value for the $\beta$ and $\alpha$ bounds are thus entirely dependent on the use case of the decision support application built upon our ProBE algorithm and its purpose, as well as the underlying data distribution. Therefore, ProBE allows the user the flexibility of exploring the trade-off between privacy and accuracy in a way that meets their various requirements.

\section{RELATED WORK}
Differential Privacy \cite{Dwork:2014:AFD:2693052.2693053,dwork2} has become a well-studied standard for privacy-preserving data exploration and analysis \cite{relatedDP,relatedDP2,relatedDP3,relatedDP4}. Various work has been proposed to answer queries with DP, such as range queries \cite{rangequery1,rangequery2,rangequery3}, and linear counting queries \cite{countquery1,countquery2,matrix}. This body of work, however, is not applicable to the aggregate threshold queries that our solution considers. Join queries \cite{join1,join2,join3,join4} may be more applicable to such queries (specifically for conjunctions), but they do not encompass the full scope of the complex queries we tackle, nor does recent work apply a utility-first approach to their solutions.

Accuracy-constrained systems for differentially private data analysis have been proposed in recent years \cite{Apex:2019:RQP:1007568.1007642,dpella,cache,ex-postdp}, which allow data analysts to interactively specify accuracy requirements over their queries while providing a formal privacy guarantee. However, these solutions either do not specifically focus on decision support queries (CacheDP \cite{cache}, Ligett et al. \cite{ex-postdp}), or do not take into account their asymmetric utility accuracy requirements (APEx \cite{Apex:2019:RQP:1007568.1007642} and DPella \cite{dpella}).

The problem of incorporating differential privacy in decision support applications has been tackled in MIDE \cite{mide}, which we have previously described and extended in our work. Other solutions such as Fioretto et al. \cite{fairness_ds} also studies decision support on differentially private data, but does so from a fairness lens by proposing recommendations to mitigate bias resulting from making decisions on DP data. Detailed related work can be found in Appendix C.

\section{Conclusion and Future Work}
In this paper, we proposed ProBE, an optimization framework which enables the execution of complex decision support queries under utility requirements on the false positive and negative rates at a minimal privacy loss. A natural generalization of our framework is implementing it with different DP mechanisms. One such mechanism defined in \cite{mide} explores a more fine-grained definition of DP, where predicates have different privacy budgets, thus necessitating a new metric to quantify this privacy loss entitled Min-Entropy. We thus extend the entropy-based algorithm from \cite{mide} to answer complex queries with guarantees on both error rates as formalized in Appendix \ref{app:probeent}.
% In future work, we would like to extend ProBE to provide formal bounds on both the false negative and positive rates simultaneously. This problem is challenging in that the very nature of this algorithm relies on the trade-off between false positive and false negative rates, where bounding one incurs an unbounded cost on the other as seen in our experimental results. This means that simply shifting the threshold twice is not enough to bound both errors, specifically in the uncertain region where no guarantee on error can be made. Bounding both rates thus requires knowledge of the underlying data distribution in order to shift the threshold in such a way that both rates are within their respective bounds across the entire range of data. 
Another interesting direction would be to implement the ProBE framework with other widely used DP mechanisms such as the exponential mechanism \cite{Dwork:2014:AFD:2693052.2693053} or the matrix mechanism \cite{matrix} to compare their performance to the algorithms previously used.

\begin{acks}
This work was supported by the HPI Research Center in Machine Learning and Data Science at UC Irvine. It is also supported by NSF Grants No. 2032525, 1545071, 1527536, 1952247, 2008993,
2133391, and 2245372.
\end{acks}
\bibliographystyle{ACM-Reference-Format}
\balance
\bibliography{references}


\begin{thebibliography}{54}

%%% the Web address.

\ifx \showCODEN    \undefined \def \showCODEN     #1{\unskip}     \fi
\ifx \showDOI      \undefined \def \showDOI       #1{#1}\fi
\ifx \showISBNx    \undefined \def \showISBNx     #1{\unskip}     \fi
\ifx \showISBNxiii \undefined \def \showISBNxiii  #1{\unskip}     \fi
\ifx \showISSN     \undefined \def \showISSN      #1{\unskip}     \fi
\ifx \showLCCN     \undefined \def \showLCCN      #1{\unskip}     \fi
\ifx \shownote     \undefined \def \shownote      #1{#1}          \fi
\ifx \showarticletitle \undefined \def \showarticletitle #1{#1}   \fi
\ifx \showURL      \undefined \def \showURL       {\relax}        \fi
% invisible to TeX
\providecommand\bibfield[2]{#2}
\providecommand\bibinfo[2]{#2}
\providecommand\natexlab[1]{#1}
\providecommand\showeprint[2][]{arXiv:#2}

\bibitem[nyt(2020)]%
        {nytaxi}
 \bibinfo{year}{2020}\natexlab{}.
\newblock \bibinfo{title}{TLC Trip Record Data}.
\newblock \bibinfo{howpublished}{https://www1.nyc.gov/site/tlc/about/tlc-trip-record-data.page}.
\newblock
\newblock
\shownote{Accessed: 2021-12-31}.


\bibitem[tur(2020)]%
        {turkishsales}
 \bibinfo{year}{2020}\natexlab{}.
\newblock \bibinfo{title}{Turkish Market Sales Dataset}.
\newblock \bibinfo{howpublished}{https://www.kaggle.com/datasets/ omercolakoglu/turkish-market-sales-dataset-with-9000items}.
\newblock
\newblock
\shownote{Accessed = 2023-01-15}.


\bibitem[Banerjee et~al\mbox{.}(2009)]%
        {hypo1}
\bibfield{author}{\bibinfo{person}{Amitav Banerjee}, \bibinfo{person}{U.~B. Chitnis}, \bibinfo{person}{S.~L. Jadhav}, \bibinfo{person}{J.~S. Bhawalkar}, {and} \bibinfo{person}{S. Chaudhury}.} \bibinfo{year}{2009}\natexlab{}.
\newblock \showarticletitle{Hypothesis testing, type I and type II errors}.
\newblock \bibinfo{journal}{\emph{Industrial Psychiatry Journal}}  \bibinfo{volume}{18} (\bibinfo{year}{2009}).
\newblock
Issue 2.


\bibitem[Bertsekas(1982)]%
        {lagrange}
\bibfield{author}{\bibinfo{person}{Dimitri~P. Bertsekas}.} \bibinfo{year}{1982}\natexlab{}.
\newblock \bibinfo{booktitle}{\emph{Constrained Optimization and Lagrange Multiplier Methods}}.
\newblock \bibinfo{publisher}{Academic Press}.
\newblock
\urldef\tempurl%
\url{https://doi.org/10.1016/C2013-0-10366-2}
\showURL{%
\tempurl}


\bibitem[Chaudhuri and Dayal(1997)]%
        {olapDS}
\bibfield{author}{\bibinfo{person}{Surajit Chaudhuri} {and} \bibinfo{person}{Umeshwar Dayal}.} \bibinfo{year}{1997}\natexlab{}.
\newblock \showarticletitle{Data warehousing and OLAP for decision support}.
\newblock \bibinfo{journal}{\emph{Proceedings of the 1997 ACM SIGMOD International Conference on Management}} (\bibinfo{year}{1997}), \bibinfo{pages}{507--508}.
\newblock
\urldef\tempurl%
\url{https://doi.org/10.1145/253260.253373}
\showURL{%
\tempurl}


\bibitem[Chaudhuri et~al\mbox{.}(2017)]%
        {aqp}
\bibfield{author}{\bibinfo{person}{Surajit Chaudhuri}, \bibinfo{person}{Bolin Ding}, {and} \bibinfo{person}{Srikanth Kandula}.} \bibinfo{year}{2017}\natexlab{}.
\newblock \showarticletitle{Approximate Query Processing: No Silver Bullet}.
\newblock \bibinfo{journal}{\emph{SIGMOD '17}} (\bibinfo{year}{2017}), \bibinfo{pages}{511--519}.
\newblock
\urldef\tempurl%
\url{https://doi.org/10.1145/3035918.3056097}
\showURL{%
\tempurl}


\bibitem[Dellino et~al\mbox{.}(2018)]%
        {supply2}
\bibfield{author}{\bibinfo{person}{Gabriella Dellino}, \bibinfo{person}{Teresa Laudadio}, \bibinfo{person}{Renato Mari}, \bibinfo{person}{Nicola Mastronardi}, {and} \bibinfo{person}{Carlo Meloni}.} \bibinfo{year}{2018}\natexlab{}.
\newblock \showarticletitle{A reliable decision support system for fresh food supply chain management}.
\newblock \bibinfo{journal}{\emph{International Journal of Production Research}}  \bibinfo{volume}{56} (\bibinfo{year}{2018}), \bibinfo{pages}{1458--1485}.
\newblock
Issue 4.


\bibitem[Dinur and Nissim(2003)]%
        {recons_attack:dinur2003revealing}
\bibfield{author}{\bibinfo{person}{Irit Dinur} {and} \bibinfo{person}{Kobbi Nissim}.} \bibinfo{year}{2003}\natexlab{}.
\newblock \showarticletitle{Revealing information while preserving privacy}. In \bibinfo{booktitle}{\emph{Proceedings of the twenty-second ACM SIGMOD-SIGACT-SIGART symposium on Principles of database systems}}. \bibinfo{pages}{202--210}.
\newblock


\bibitem[Dombrowski et~al\mbox{.}(2013)]%
        {kpi}
\bibfield{author}{\bibinfo{person}{Uwe Dombrowski}, \bibinfo{person}{David Ebentreich}, {and} \bibinfo{person}{K. Schmidtchen}.} \bibinfo{year}{2013}\natexlab{}.
\newblock \showarticletitle{Balanced Key Performance Indicators in Product Development}.
\newblock \bibinfo{journal}{\emph{International Journal of Materials, Mechanics and Manufacturing}} \bibinfo{volume}{1}, \bibinfo{number}{1} (\bibinfo{year}{2013}), \bibinfo{pages}{27--31}.
\newblock
\urldef\tempurl%
\url{doi.org/10.7763/IJMMM.2013.V1.6}
\showURL{%
\tempurl}


\bibitem[Dong et~al\mbox{.}(2022)]%
        {fdp}
\bibfield{author}{\bibinfo{person}{Jinshuo Dong}, \bibinfo{person}{Aaron Roth}, {and} \bibinfo{person}{Weijie~J. Su}.} \bibinfo{year}{2022}\natexlab{}.
\newblock \showarticletitle{Gaussian Differential Privacy}.
\newblock \bibinfo{journal}{\emph{Journal of the Royal Statistical Society Series B: Statistical Methodology}}  \bibinfo{volume}{84} (\bibinfo{year}{2022}), \bibinfo{pages}{3--37}.
\newblock
Issue 1.
\urldef\tempurl%
\url{https://doi.org/10.1111/rssb.12454}
\showURL{%
\tempurl}


\bibitem[Dong et~al\mbox{.}(2023)]%
        {join2}
\bibfield{author}{\bibinfo{person}{Wei Dong}, \bibinfo{person}{Juanru Fang}, \bibinfo{person}{Ke Yi}, \bibinfo{person}{Yuchao Tao}, {and} \bibinfo{person}{Ashwin Machanavajjhala}.} \bibinfo{year}{2023}\natexlab{}.
\newblock \showarticletitle{R2T: Instance-optimal Truncation for Differentially Private Query Evaluation with Foreign Keys}.
\newblock \bibinfo{journal}{\emph{ACM SIGMOD Record}}  \bibinfo{volume}{52} (\bibinfo{year}{2023}), \bibinfo{pages}{115--123}.
\newblock
Issue 1.
\urldef\tempurl%
\url{https://doi.org/10.1145/3604437.3604462}
\showURL{%
\tempurl}


\bibitem[Dong and Yi(2021a)]%
        {relatedDP4}
\bibfield{author}{\bibinfo{person}{Wei Dong} {and} \bibinfo{person}{Ke Yi}.} \bibinfo{year}{2021}\natexlab{a}.
\newblock \showarticletitle{Residual Sensitivity for Differentially Private Multi-Way Joins}.
\newblock \bibinfo{journal}{\emph{SIGMOD '21}} (\bibinfo{year}{2021}), \bibinfo{pages}{432–444}.
\newblock
\urldef\tempurl%
\url{https://doi.org/10.1145/3448016.3452813}
\showURL{%
\tempurl}


\bibitem[Dong and Yi(2021b)]%
        {join3}
\bibfield{author}{\bibinfo{person}{Wei Dong} {and} \bibinfo{person}{Ke Yi}.} \bibinfo{year}{2021}\natexlab{b}.
\newblock \showarticletitle{Residual Sensitivity for Differentially Private Multi-Way Joins}.
\newblock \bibinfo{journal}{\emph{SIGMOD '21}} (\bibinfo{year}{2021}), \bibinfo{pages}{432--444}.
\newblock
\urldef\tempurl%
\url{https://doi.org/10.1145/3448016.3452813}
\showURL{%
\tempurl}


\bibitem[Doukas et~al\mbox{.}(2007)]%
        {bmsds2}
\bibfield{author}{\bibinfo{person}{Haris Doukas}, \bibinfo{person}{Konstantinos~D. Patlitzianas}, \bibinfo{person}{Konstantinos Iatropoulos}, {and} \bibinfo{person}{John Psarras}.} \bibinfo{year}{2007}\natexlab{}.
\newblock \showarticletitle{Intelligent building energy management system using rule sets}.
\newblock \bibinfo{journal}{\emph{Building and Environment}}  \bibinfo{volume}{42} (\bibinfo{year}{2007}), \bibinfo{pages}{3562--3569}.
\newblock
Issue 10.


\bibitem[Dowdy et~al\mbox{.}(2011)]%
        {hypo}
\bibfield{author}{\bibinfo{person}{Shirley Dowdy}, \bibinfo{person}{Stanley Wearden}, {and} \bibinfo{person}{Daniel Chilko}.} \bibinfo{year}{2011}\natexlab{}.
\newblock \bibinfo{booktitle}{\emph{Statistics for research}}.
\newblock \bibinfo{publisher}{John Wiley \& Sons,}.
\newblock


\bibitem[Du et~al\mbox{.}(2021)]%
        {rangequery2}
\bibfield{author}{\bibinfo{person}{Linkang Du}, \bibinfo{person}{Zhikun Zhang}, \bibinfo{person}{Shaojie Bai}, \bibinfo{person}{Changchang~Liu Liu}, \bibinfo{person}{Shouling Ji}, \bibinfo{person}{Peng Cheng}, {and} \bibinfo{person}{Jiming Chen}.} \bibinfo{year}{2021}\natexlab{}.
\newblock \showarticletitle{AHEAD: Adaptive Hierarchical Decomposition for Range Query under Local Differential Privacy}.
\newblock \bibinfo{journal}{\emph{CCS '21}} (\bibinfo{year}{2021}), \bibinfo{pages}{1266--1288}.
\newblock
\urldef\tempurl%
\url{https://doi.org/10.1145/3460120.3485668}
\showURL{%
\tempurl}


\bibitem[Dweiri et~al\mbox{.}(2016)]%
        {carDS}
\bibfield{author}{\bibinfo{person}{Fikri Dweiri}, \bibinfo{person}{Sameer Kumar}, \bibinfo{person}{Sharfuddin~Ahmed Khan}, {and} \bibinfo{person}{Vipul Jain}.} \bibinfo{year}{2016}\natexlab{}.
\newblock \showarticletitle{Designing an integrated AHP based decision support system for supplier selection in automotive industry}.
\newblock \bibinfo{journal}{\emph{Expert Systems with Applications: An International Journal}}  \bibinfo{volume}{62} (\bibinfo{year}{2016}), \bibinfo{pages}{273--283}.
\newblock
Issue C.
\urldef\tempurl%
\url{https://doi.org/10.1016/j.eswa.2016.06.030}
\showURL{%
\tempurl}


\bibitem[Dwork et~al\mbox{.}(2006)]%
        {dwork2}
\bibfield{author}{\bibinfo{person}{Cynthia Dwork}, \bibinfo{person}{Frank McSherry}, \bibinfo{person}{Kobbi Nissim}, {and} \bibinfo{person}{Adam Smith}.} \bibinfo{year}{2006}\natexlab{}.
\newblock \showarticletitle{Calibrating noise to sensitivity in private data analysis}.
\newblock \bibinfo{journal}{\emph{Proceedings of the Third Conference on Theory of Cryptography}} (\bibinfo{year}{2006}), \bibinfo{pages}{265--284}.
\newblock
\urldef\tempurl%
\url{https://doi.org/10.1007/11681878_14}
\showURL{%
\tempurl}


\bibitem[Dwork and Roth(2014)]%
        {Dwork:2014:AFD:2693052.2693053}
\bibfield{author}{\bibinfo{person}{Cynthia Dwork} {and} \bibinfo{person}{Aaron Roth}.} \bibinfo{year}{2014}\natexlab{}.
\newblock \showarticletitle{The Algorithmic Foundations of Differential Privacy}.
\newblock \bibinfo{journal}{\emph{Foundation and Trends in Theoretical Computer Science}} \bibinfo{volume}{9}, \bibinfo{number}{3--4} (\bibinfo{date}{Aug.} \bibinfo{year}{2014}), \bibinfo{pages}{211--407}.
\newblock
\showISSN{1551-305X}
\urldef\tempurl%
\url{https://doi.org/10.1561/0400000042}
\showDOI{\tempurl}


\bibitem[Dwork and Yekhanin(2008)]%
        {recons_attack:dwork2008new}
\bibfield{author}{\bibinfo{person}{Cynthia Dwork} {and} \bibinfo{person}{Sergey Yekhanin}.} \bibinfo{year}{2008}\natexlab{}.
\newblock \showarticletitle{New efficient attacks on statistical disclosure control mechanisms}. In \bibinfo{booktitle}{\emph{Annual International Cryptology Conference}}. Springer, \bibinfo{pages}{469--480}.
\newblock


\bibitem[Fanti et~al\mbox{.}(2016)]%
        {google}
\bibfield{author}{\bibinfo{person}{Giulia Fanti}, \bibinfo{person}{Vasyl Pihur}, {and} \bibinfo{person}{Úlfar Erlingsson}.} \bibinfo{year}{2016}\natexlab{}.
\newblock \showarticletitle{Building a rappor with the unknown: Privacy-preserving learning of associations and data dictionaries}.
\newblock \bibinfo{journal}{\emph{Proceedings on Privacy Enhancing Technologies}}  \bibinfo{volume}{3} (\bibinfo{year}{2016}).
\newblock


\bibitem[Fioretto et~al\mbox{.}(2021)]%
        {fairness_ds}
\bibfield{author}{\bibinfo{person}{Ferdinando Fioretto}, \bibinfo{person}{Cuong Tran}, \bibinfo{person}{Pascal Van~Hentenryck}, {and} \bibinfo{person}{Zhiyan Yao}.} \bibinfo{year}{2021}\natexlab{}.
\newblock \showarticletitle{Decision Making with Differential Privacy under a Fairness Lens}.
\newblock
\urldef\tempurl%
\url{https://doi.org/10.24963/ijcai.2021/78}
\showDOI{\tempurl}


\bibitem[Ge et~al\mbox{.}(2019)]%
        {Apex:2019:RQP:1007568.1007642}
\bibfield{author}{\bibinfo{person}{Chang Ge}, \bibinfo{person}{Xi He}, \bibinfo{person}{Ihab~F Ilyas}, {and} \bibinfo{person}{Ashwin Machanavajjhala}.} \bibinfo{year}{2019}\natexlab{}.
\newblock \showarticletitle{APEx: Accuracy-Aware Differentially Private Data Exploration} \emph{(\bibinfo{series}{SIGMOD})}.
\newblock


\bibitem[Ghayyur et~al\mbox{.}(2022)]%
        {mide}
\bibfield{author}{\bibinfo{person}{Sameera Ghayyur}, \bibinfo{person}{Dhrubajyoti Ghosh}, \bibinfo{person}{Xi He}, {and} \bibinfo{person}{Sharad Mehrotra}.} \bibinfo{year}{2022}\natexlab{}.
\newblock \showarticletitle{MIDE: Accuracy Aware Minimally Invasive Data Exploration For Decision Support}.
\newblock \bibinfo{journal}{\emph{Proceedings of the VLDB Endowment}} \bibinfo{volume}{15}, \bibinfo{number}{11} (\bibinfo{year}{2022}), \bibinfo{pages}{2653--2665}.
\newblock
\urldef\tempurl%
\url{https://doi.org/10.14778/3551793.3551821}
\showURL{%
\tempurl}


\bibitem[Govindan et~al\mbox{.}(2020)]%
        {healthDS}
\bibfield{author}{\bibinfo{person}{Kannan Govindan}, \bibinfo{person}{Hassan Mina}, {and} \bibinfo{person}{Behrouz Alavi}.} \bibinfo{year}{2020}\natexlab{}.
\newblock \showarticletitle{A decision support system for demand management in healthcare supply chains considering the epidemic outbreak: A case study of coronavirus disease 2019}.
\newblock \bibinfo{journal}{\emph{Transportation Research Part E: Logistics and Transportation Review}}  \bibinfo{volume}{138} (\bibinfo{year}{2020}).
\newblock
\urldef\tempurl%
\url{https://doi.org/10.1016/j.tre.2020.101967}
\showURL{%
\tempurl}


\bibitem[Haney et~al\mbox{.}(2017)]%
        {census}
\bibfield{author}{\bibinfo{person}{Samuel Haney}, \bibinfo{person}{Ashwin Machanavajjhala}, \bibinfo{person}{John~M. Abowd}, \bibinfo{person}{Matthew Graham}, \bibinfo{person}{Mark Kutzbach}, {and} \bibinfo{person}{Lars Vilhuber}.} \bibinfo{year}{2017}\natexlab{}.
\newblock \showarticletitle{Utility cost of formal privacy for releasing national employer-employee statistics}.
\newblock \bibinfo{journal}{\emph{SIGMOD '17}} (\bibinfo{year}{2017}), \bibinfo{pages}{1339–1354}.
\newblock
\urldef\tempurl%
\url{https://doi.org/10.1145/3035918.3035940}
\showURL{%
\tempurl}


\bibitem[Kato et~al\mbox{.}(2022)]%
        {relatedDP3}
\bibfield{author}{\bibinfo{person}{Fumiyuki Kato}, \bibinfo{person}{Tsubasa Takahashi}, \bibinfo{person}{Shun Takagi}, \bibinfo{person}{Yang Cao}, \bibinfo{person}{Seng~Pei Liew}, {and} \bibinfo{person}{Masatoshi Yoshikawa}.} \bibinfo{year}{2022}\natexlab{}.
\newblock \showarticletitle{HDPView: Differentially Private Materialized View for Exploring High Dimensional Relational Data}.
\newblock \bibinfo{journal}{\emph{Proceedings of the VLDB Endowment}} \bibinfo{volume}{15}, \bibinfo{number}{9} (\bibinfo{year}{2022}).
\newblock


\bibitem[Kotsogiannis et~al\mbox{.}(2019)]%
        {join1}
\bibfield{author}{\bibinfo{person}{Ios Kotsogiannis}, \bibinfo{person}{Yuchao Tao}, \bibinfo{person}{Xi He}, \bibinfo{person}{Maryam Fanaeepour}, \bibinfo{person}{Ashwin Machanavajjhala}, \bibinfo{person}{Michael~Hay Hay}, {and} \bibinfo{person}{Gerome Miklau}.} \bibinfo{year}{2019}\natexlab{}.
\newblock \showarticletitle{PrivateSQL: A Differentially Private SQL Query Engine}.
\newblock \bibinfo{journal}{\emph{Proceedings of the VLDB Endowment}}  \bibinfo{volume}{12} (\bibinfo{year}{2019}), \bibinfo{pages}{1371--1384}.
\newblock
Issue 11.
\urldef\tempurl%
\url{https://doi.org/10.14778/3342263.3342274}
\showURL{%
\tempurl}


\bibitem[Lallich et~al\mbox{.}(2006)]%
        {statinf1}
\bibfield{author}{\bibinfo{person}{Stéphane Lallich}, \bibinfo{person}{Olivier Teytaud}, {and} \bibinfo{person}{Elie Prudhomme}.} \bibinfo{year}{2006}\natexlab{}.
\newblock \showarticletitle{Statistical inference and data mining: false discoveries control}.
\newblock \bibinfo{journal}{\emph{Compstat 2006 - Proceedings in Computational Statistics}} (\bibinfo{year}{2006}), \bibinfo{pages}{325--336}.
\newblock


\bibitem[Li et~al\mbox{.}(2014a)]%
        {rangequery1}
\bibfield{author}{\bibinfo{person}{Chao Li}, \bibinfo{person}{Michael Hay}, \bibinfo{person}{Gerome Miklau}, {and} \bibinfo{person}{Yue Wang}.} \bibinfo{year}{2014}\natexlab{a}.
\newblock \showarticletitle{Data- and Workload-Aware Algorithm for Range Queries Under Differential Privacy}.
\newblock \bibinfo{journal}{\emph{Proceedings of the VLDB Endowment}}  \bibinfo{volume}{7} (\bibinfo{year}{2014}), \bibinfo{pages}{341--352}.
\newblock
Issue 5.
\urldef\tempurl%
\url{https://doi.org/10.14778/2732269.2732271}
\showURL{%
\tempurl}


\bibitem[Li et~al\mbox{.}(2014b)]%
        {dawa}
\bibfield{author}{\bibinfo{person}{Chao Li}, \bibinfo{person}{Michael Hay}, \bibinfo{person}{Gerome Miklau}, {and} \bibinfo{person}{Yue Wang}.} \bibinfo{year}{2014}\natexlab{b}.
\newblock \showarticletitle{A data-and workload-aware algorithm for range queries under differential privacy}.
\newblock \bibinfo{journal}{\emph{Proceedings of the VLDB Endowment}} \bibinfo{volume}{7}, \bibinfo{number}{5} (\bibinfo{year}{2014}), \bibinfo{pages}{341--352}.
\newblock
\urldef\tempurl%
\url{https://arxiv.org/abs/1410.0265}
\showURL{%
\tempurl}


\bibitem[Li et~al\mbox{.}(2015)]%
        {matrix}
\bibfield{author}{\bibinfo{person}{Chao Li}, \bibinfo{person}{Gerome Miklau}, \bibinfo{person}{Michael Hay}, \bibinfo{person}{Andrew McGregor}, {and} \bibinfo{person}{Vibhor Rastogi}.} \bibinfo{year}{2015}\natexlab{}.
\newblock \showarticletitle{The matrix mechanism: optimizing linear counting queries under differential privacy}.
\newblock \bibinfo{journal}{\emph{VLDB Journal}} \bibinfo{volume}{24}, \bibinfo{number}{6} (\bibinfo{year}{2015}).
\newblock
\urldef\tempurl%
\url{https://dl.acm.org/doi/10.1007/s00778-015-0398-x}
\showURL{%
\tempurl}


\bibitem[Ligett et~al\mbox{.}(2017)]%
        {ex-postdp}
\bibfield{author}{\bibinfo{person}{Katrina Ligett}, \bibinfo{person}{Seth Neel}, \bibinfo{person}{Aaron Roth}, \bibinfo{person}{Bo Waggoner}, {and} \bibinfo{person}{Zhiwei Wu}.} \bibinfo{year}{2017}\natexlab{}.
\newblock \showarticletitle{Accuracy First: Selecting a Differential Privacy Level for Accuracy-Constrained ERM}.
\newblock \bibinfo{journal}{\emph{Journal of Privacy and Confidentiality}}  \bibinfo{volume}{9} (\bibinfo{date}{05} \bibinfo{year}{2017}).
\newblock
\urldef\tempurl%
\url{https://doi.org/10.29012/jpc.682}
\showDOI{\tempurl}


\bibitem[Lobo-Vesga et~al\mbox{.}(2020)]%
        {dpella}
\bibfield{author}{\bibinfo{person}{E. Lobo-Vesga}, \bibinfo{person}{A. Russo}, {and} \bibinfo{person}{M. Gaboardi}.} \bibinfo{year}{2020}\natexlab{}.
\newblock \showarticletitle{A Programming Framework for Differential Privacy with Accuracy Concentration Bounds}. In \bibinfo{booktitle}{\emph{2020 IEEE Symposium on Security and Privacy (SP)}}. \bibinfo{publisher}{IEEE Computer Society}, \bibinfo{address}{Los Alamitos, CA, USA}, \bibinfo{pages}{411--428}.
\newblock
\urldef\tempurl%
\url{https://doi.org/10.1109/SP40000.2020.00086}
\showDOI{\tempurl}


\bibitem[Mansouri et~al\mbox{.}(2012)]%
        {supply1}
\bibfield{author}{\bibinfo{person}{S.~Afshin Mansouri}, \bibinfo{person}{David Gallear}, {and} \bibinfo{person}{Mohammad~H. Askariazad}.} \bibinfo{year}{2012}\natexlab{}.
\newblock \showarticletitle{Decision support for build-to-order supply chain management through multiobjective optimization}.
\newblock \bibinfo{journal}{\emph{International Journal of Production Economics}}  \bibinfo{volume}{135} (\bibinfo{year}{2012}), \bibinfo{pages}{24--36}.
\newblock
Issue 1.


\bibitem[Mazmudar et~al\mbox{.}(2022)]%
        {cache}
\bibfield{author}{\bibinfo{person}{Miti Mazmudar}, \bibinfo{person}{Thomas Humphries}, \bibinfo{person}{Jiaxiang Liu}, \bibinfo{person}{Matthew Rafuse}, {and} \bibinfo{person}{Xi He}.} \bibinfo{year}{2022}\natexlab{}.
\newblock \showarticletitle{Cache Me If You Can: Accuracy-Aware Inference Engine for Diferentially Private Data Exploration}.
\newblock \bibinfo{journal}{\emph{Proceedings of the VLDB Endowment}}  \bibinfo{volume}{16} (\bibinfo{year}{2022}), \bibinfo{pages}{574–586}.
\newblock
Issue 4.
\urldef\tempurl%
\url{https://doi.org/10.14778/3574245.3574246}
\showURL{%
\tempurl}


\bibitem[McCluskey(1956)]%
        {mccluskey}
\bibfield{author}{\bibinfo{person}{E.~J. McCluskey}.} \bibinfo{year}{1956}\natexlab{}.
\newblock \showarticletitle{Minimization of Boolean functions}.
\newblock \bibinfo{journal}{\emph{The Bell System Technical Journal}} \bibinfo{volume}{35}, \bibinfo{number}{6} (\bibinfo{year}{1956}).
\newblock


\bibitem[McKenna et~al\mbox{.}(2020)]%
        {countquery1}
\bibfield{author}{\bibinfo{person}{Ryan McKenna}, \bibinfo{person}{Raj Kumar~Maity}, \bibinfo{person}{Arya Mazumdar}, {and} \bibinfo{person}{Gerome Miklau}.} \bibinfo{year}{2020}\natexlab{}.
\newblock \showarticletitle{A workload-adaptive mechanism for linear queries under local differential privacy}.
\newblock \bibinfo{journal}{\emph{Proceedings of the VLDB Endowment}} \bibinfo{volume}{13}, \bibinfo{number}{12} (\bibinfo{year}{2020}), \bibinfo{pages}{1905--1918}.
\newblock
\urldef\tempurl%
\url{https://doi.org/10.14778/3407790.3407798}
\showURL{%
\tempurl}


\bibitem[McKenna et~al\mbox{.}(2018)]%
        {countquery2}
\bibfield{author}{\bibinfo{person}{Ryan McKenna}, \bibinfo{person}{Gerome Miklau}, \bibinfo{person}{Michael Hay}, {and} \bibinfo{person}{Ashwin Machanavajjhala}.} \bibinfo{year}{2018}\natexlab{}.
\newblock \showarticletitle{Optimizing error of high-dimensional statistical queries under differential privacy}.
\newblock \bibinfo{journal}{\emph{Proceedings of the VLDB Endowment}} \bibinfo{volume}{11}, \bibinfo{number}{10} (\bibinfo{year}{2018}), \bibinfo{pages}{1206--1219}.
\newblock
\urldef\tempurl%
\url{https://doi.org/10.14778/3231751.3231769}
\showURL{%
\tempurl}


\bibitem[Mehrotra et~al\mbox{.}(2016)]%
        {tippers}
\bibfield{author}{\bibinfo{person}{Sharad Mehrotra}, \bibinfo{person}{Kobsa Alfred}, \bibinfo{person}{Nalini Venkatasubramanian}, {and} \bibinfo{person}{Siva~Raj Rajagopalan}.} \bibinfo{year}{2016}\natexlab{}.
\newblock \showarticletitle{TIPPERS: A privacy cognizant IoT environment}. In \bibinfo{booktitle}{\emph{2016 IEEE PerCom Workshops}}.
\newblock


\bibitem[Mironov(2017)]%
        {renyi}
\bibfield{author}{\bibinfo{person}{Ilya Mironov}.} \bibinfo{year}{2017}\natexlab{}.
\newblock \showarticletitle{Rényi Differential Privacy}.
\newblock \bibinfo{journal}{\emph{2017 IEEE 30th Computer Security Foundations Symposium (CSF)}} (\bibinfo{year}{2017}), \bibinfo{pages}{263--275}.
\newblock
\urldef\tempurl%
\url{https://doi.org/10.1109/CSF.2017.11}
\showURL{%
\tempurl}


\bibitem[Mizrahi(2020)]%
        {hypo3}
\bibfield{author}{\bibinfo{person}{Moti Mizrahi}.} \bibinfo{year}{2020}\natexlab{}.
\newblock \showarticletitle{Hypothesis Testing in Scientific Practice: An Empirical Study}.
\newblock \bibinfo{journal}{\emph{International Studies in the Philosophy of Science}}  \bibinfo{volume}{33} (\bibinfo{year}{2020}).
\newblock
Issue 1.


\bibitem[Mohan et~al\mbox{.}(2012)]%
        {gupt}
\bibfield{author}{\bibinfo{person}{Prashanth Mohan}, \bibinfo{person}{Abhradeep Thakurta}, \bibinfo{person}{Elaine Shi}, \bibinfo{person}{Dawn Song}, {and} \bibinfo{person}{David Culler}.} \bibinfo{year}{2012}\natexlab{}.
\newblock \showarticletitle{GUPT: privacy preserving data analysis made easy}. In \bibinfo{booktitle}{\emph{Proceedings of the 2012 ACM SIGMOD International Conference on Management of Data}}. \bibinfo{pages}{349--360}.
\newblock


\bibitem[Musen et~al\mbox{.}(2021)]%
        {med3}
\bibfield{author}{\bibinfo{person}{Mark~A. Musen}, \bibinfo{person}{Blackford Middleton}, {and} \bibinfo{person}{Robert~A. Greenes}.} \bibinfo{year}{2021}\natexlab{}.
\newblock \showarticletitle{Clinical Decision-Support Systems}.
\newblock In \bibinfo{booktitle}{\emph{Biomedical Informatics}}. \bibinfo{publisher}{Springer}, \bibinfo{pages}{795--840}.
\newblock


\bibitem[Newey and McFadden(1994)]%
        {hypo2}
\bibfield{author}{\bibinfo{person}{Whitney~K. Newey} {and} \bibinfo{person}{Daniel McFadden}.} \bibinfo{year}{1994}\natexlab{}.
\newblock \showarticletitle{Large sample estimation and hypothesis testing}.
\newblock In \bibinfo{booktitle}{\emph{Handbook of Econometrics}}. \bibinfo{publisher}{Elsevier B.V.}, \bibinfo{pages}{2111--2245}.
\newblock


\bibitem[Peiffer-Smadja et~al\mbox{.}(2020)]%
        {med2}
\bibfield{author}{\bibinfo{person}{N. Peiffer-Smadja}, \bibinfo{person}{T.M.~Rawson Rawson}, \bibinfo{person}{R. Ahmad}, \bibinfo{person}{A. Buchard}, \bibinfo{person}{P. Georgiou}, \bibinfo{person}{F.-X.~Lescure Lescure}, \bibinfo{person}{G. Birgand}, {and} \bibinfo{person}{A.H. Holmes}.} \bibinfo{year}{2020}\natexlab{}.
\newblock \showarticletitle{Machine learning for clinical decision support in infectious diseases: a narrative review of current applications}.
\newblock \bibinfo{journal}{\emph{Clinical Microbiology and Infection}}  \bibinfo{volume}{26} (\bibinfo{year}{2020}).
\newblock
Issue 5.


\bibitem[Poess et~al\mbox{.}(2007)]%
        {tpcds}
\bibfield{author}{\bibinfo{person}{Meikel Poess}, \bibinfo{person}{R.O. Nambiar}, {and} \bibinfo{person}{David Walrath}.} \bibinfo{year}{2007}\natexlab{}.
\newblock \showarticletitle{Why you should run TPC-DS: a workload analysis}.
\newblock \bibinfo{journal}{\emph{Proceedings of the 33rd International Conference on Very Large Databases}} (\bibinfo{year}{2007}), \bibinfo{pages}{1138--1149}.
\newblock
\urldef\tempurl%
\url{https://dl.acm.org/doi/10.5555/1325851.1325979}
\showURL{%
\tempurl}


\bibitem[Pujol et~al\mbox{.}(2022)]%
        {relatedDP2}
\bibfield{author}{\bibinfo{person}{David Pujol}, \bibinfo{person}{Albert Sun}, \bibinfo{person}{Brandon Fain}, {and} \bibinfo{person}{Ashwin Machanavajjhala}.} \bibinfo{year}{2022}\natexlab{}.
\newblock \showarticletitle{Multi-Analyst Differential Privacy for Online Query Answering}.
\newblock \bibinfo{journal}{\emph{Proceedings of the VLDB Endowment}} \bibinfo{volume}{16}, \bibinfo{number}{4} (\bibinfo{year}{2022}).
\newblock


\bibitem[Rogers et~al\mbox{.}(2016)]%
        {expostcomposition}
\bibfield{author}{\bibinfo{person}{Ryan Rogers}, \bibinfo{person}{Aaron Roth}, \bibinfo{person}{Jonathan Ullman}, {and} \bibinfo{person}{Salil Vadhan}.} \bibinfo{year}{2016}\natexlab{}.
\newblock \showarticletitle{Privacy Odometers and Filters: Pay-as-you-Go Composition}.
\newblock \bibinfo{journal}{\emph{NIPS'16}} (\bibinfo{year}{2016}), \bibinfo{pages}{1929–1937}.
\newblock


\bibitem[Sutton et~al\mbox{.}(2020)]%
        {med1}
\bibfield{author}{\bibinfo{person}{Reed~T. Sutton}, \bibinfo{person}{David Pincock}, \bibinfo{person}{Daniel~C. Baumgart}, \bibinfo{person}{Daniel~C. Sadowski}, \bibinfo{person}{Richard~N. Fedorak}, {and} \bibinfo{person}{Karen~I. Kroeker}.} \bibinfo{year}{2020}\natexlab{}.
\newblock \showarticletitle{An overview of clinical decision support systems: benefits, risks, and strategies for success}.
\newblock \bibinfo{journal}{\emph{npj Digit. Med}}  \bibinfo{volume}{3} (\bibinfo{year}{2020}).
\newblock
Issue 17.


\bibitem[Tao et~al\mbox{.}(2022)]%
        {relatedDP}
\bibfield{author}{\bibinfo{person}{Yuchao Tao}, \bibinfo{person}{Amir Gilad}, \bibinfo{person}{Ashwin Machanavajjhala}, {and} \bibinfo{person}{Sudeepa Roy}.} \bibinfo{year}{2022}\natexlab{}.
\newblock \showarticletitle{DPXPlain: Privately Explaining Aggregate Query Answers}.
\newblock \bibinfo{journal}{\emph{Proceedings of the VLDB Endowment}} \bibinfo{volume}{16}, \bibinfo{number}{1} (\bibinfo{year}{2022}).
\newblock


\bibitem[Tao et~al\mbox{.}(2020)]%
        {join4}
\bibfield{author}{\bibinfo{person}{Yuchao Tao}, \bibinfo{person}{Xi He}, \bibinfo{person}{Ashwin Machanavajjhala}, {and} \bibinfo{person}{Sudeepa Roy}.} \bibinfo{year}{2020}\natexlab{}.
\newblock \showarticletitle{Computing Local Sensitivities of Counting Queries with Joins}.
\newblock \bibinfo{journal}{\emph{SIGMOD '20}} (\bibinfo{year}{2020}), \bibinfo{pages}{479--494}.
\newblock
\urldef\tempurl%
\url{https://doi.org/10.1145/3318464.3389762}
\showURL{%
\tempurl}


\bibitem[Thompson and Bank(2010)]%
        {bmsds1}
\bibfield{author}{\bibinfo{person}{Benjamin~P. Thompson} {and} \bibinfo{person}{Lawrence~C. Bank}.} \bibinfo{year}{2010}\natexlab{}.
\newblock \showarticletitle{Use of system dynamics as a decision-making tool in building design and operation}.
\newblock \bibinfo{journal}{\emph{Building and Environment}}  \bibinfo{volume}{45} (\bibinfo{year}{2010}), \bibinfo{pages}{1006--1015}.
\newblock
Issue 4.


\bibitem[Yang et~al\mbox{.}(2020)]%
        {rangequery3}
\bibfield{author}{\bibinfo{person}{Jianyu Yang}, \bibinfo{person}{Tianhao Wang}, \bibinfo{person}{Ninghui Li}, \bibinfo{person}{Xiang Cheng}, {and} \bibinfo{person}{Sen Su}.} \bibinfo{year}{2020}\natexlab{}.
\newblock \showarticletitle{Answering Multi-Dimensional Range Queries under Local Differential Privacy}.
\newblock \bibinfo{journal}{\emph{Proceedings of the VLDB Endowment}} \bibinfo{volume}{14}, \bibinfo{number}{3} (\bibinfo{year}{2020}), \bibinfo{pages}{378--390}.
\newblock
\urldef\tempurl%
\url{https://doi.org/10.14778/3430915.3430927}
\showURL{%
\tempurl}


\end{thebibliography}
\appendix
\section{PROOFS}

\subsection{Proofs for Single Operator Queries (FNR)} \label{app:fnrbound}
\textbf{1. Proof For 2-Query Conjunction}.
 Consider a complex decision support query $Q$ composed of two atomic aggregate threshold queries $Q=Q_1 \cap Q_2$. $Q_1$ and $Q_2$ are answered by TSLM $M_1$, $M_2$ respectively, and mechanism $M = M_1 \cap M_2$ is used to answer query $Q$. If $M_1$ and $M_2$ guarantee an FNR bound of $\beta_1$ and $\beta_2$ respectively, then $M$ has an FNR bound of $\beta_1 + \beta_2$.

\noindent
\begin{proof}
$\forall \lambda_i \in \Lambda,$ \\
Let $A$ be the event that $\lambda_i \in M_1(D)$, $B$ be the event that $\lambda_i \in M_2(D)$, $C$ be the event that $\lambda_i \in Q_1(D)$ and $D$ be the event that $\lambda_i \in Q_2(D)$. The probability of false negatives can be derived as:
\begin{eqnarray}
FNR &=& P[\lambda_i \not \in M_1(D) \cap M_2(D)| \lambda_i  \in Q_1(D)\cap Q_2(D)]\nonumber\\
&=&P[\overline{AB}| CD]\nonumber=P[\overline{A}\vee \overline{B}|CD]\nonumber=P[\overline{A}B \vee A\overline{B} \vee \overline{AB}|CD] \nonumber \\
&=&\frac{P[(\overline{A}B \vee A\overline{B} \vee \overline{AB})\wedge CD]}{P[CD]}\nonumber\\
&=& \frac{P[(\overline{A}B \wedge CD) \vee (A\overline{B} \wedge CD) \vee( \overline{AB}\wedge CD)]}{P[CD]}
\end{eqnarray}
Given that the three clauses are mutually exclusive events
\begin{eqnarray*}
&=&\frac{P[\overline{A}B \wedge CD ]+P[A\overline{B} \wedge CD] +P[ \overline{AB}\wedge CD]}{P[CD]}\\
&=& P[\overline{A}B | CD ]+P[A\overline{B} | CD] +P[ \overline{AB}| CD]
\end{eqnarray*}
We know that $M_1$ and $M_2$ are mechanisms of independent randomness, i.e. add random noise. This means that $A$ is independent of $B$ given $CD$ and vice versa. Therefore we can rewrite $P[\overline{A}B|CD] = P[\overline{A}|CD] P[B|CD]$. We obtain:
\begin{eqnarray*}
    &=& P[\overline{A}|CD] P[B|CD] + P[A|CD]P[\overline{B}|CD] + P[\overline{A}|CD]P[\overline{B}|CD]
\end{eqnarray*}
There is a conditional independence between $A$ and $D$ given $C$ due to the nature of the randomized mechanisms. Similarly there is a conditional independence between $B$ and $C$ given $D$. Thus, we can rewrite $P[\overline{A}|CD] P[B|CD] = P[\overline{A}|C] P[B|D] $. Therefore,
\begin{eqnarray*}
&=& P[\overline{A}|C]P[B|D] + P[A|C]P[\overline{B}|D] + P[\overline{A}|C]P[\overline{B}|D]
\end{eqnarray*}
 Knowing that  $FNR_1 = P[\overline{A}|C]$, $FNR_2 = P[\overline{B}|D]$, $TPR_1 = P[A|C]$ and $TPR_2 = P[B|D]$, we substitute:
 \begin{eqnarray}
     &=& FNR_1 \cdot TPR_2 + TPR_1 \cdot FNR_2 + FNR_1 \cdot FNR_2 \nonumber \\
     &=& FNR_1(1-FNR_2) + (1-FNR_1)FNR_2 + FNR_1 FNR_2 \nonumber\\
     &=& FNR_1 + FNR_2 - FNR_1 FNR_2  \nonumber \\
     &\leq& FNR_1 + FNR_2 \leq \beta_1 + \beta_2
 \end{eqnarray}
\end{proof}

\noindent
\textbf{2. Proof For 2-Query Disjunction}.
 Consider a complex decision support query $Q$ composed of two atomic aggregate threshold queries $Q=Q_1 \cup Q_2$. $Q_1$ and $Q_2$ are answered by TSLM $M_1$, $M_2$ respectively, and mechanism $M = M_1 \cup M_2$ is used to answer query $Q$. If $M_1$ and $M_2$ guarantee an FNR bound of $\beta_1$ and $\beta_2$ respectively, then $M$ has an FNR bound of $\beta_1 + \beta_2$.

\noindent
\begin{proof}
$\forall \lambda_i \in \Lambda,$ \\
Let $A$ be the event that $\lambda_i \in M_1(D)$, $B$ be the event that $\lambda_i \in M_2(D)$, $C$ be the event that $\lambda_i \in Q_1(D)$ and $D$ be the event that $\lambda_i \in Q_2(D)$. The probability of false negatives can be derived as:
\begin{eqnarray}
    FNR &=& P[\lambda_i \not \in M_1(D) \cap M_2(D)| \lambda_i  \in Q_1(D)\cap Q_2(D)]\nonumber\\  
    &=& P[\overline{A\vee B}| C\vee D]\nonumber\\
    &=& \frac{P[(\overline{AB}) \wedge (\bar{C}D \vee C\bar{D}\vee CD)]}{ P[C\vee D]} \nonumber
\end{eqnarray}
Given that the three clauses are mutually exclusive events
\begin{eqnarray*}
    &=& \frac{P[(\overline{AB}) \wedge \bar{C}D]
 + P[(\overline{AB}) \wedge C\bar{D}]
 + P[(\overline{AB}) \wedge CD]}{ P[C\vee D]} \\
 &=& \frac{P[(\overline{AB}) | \bar{C}D] P[\bar{C}D] }{ P[C\vee D]}
 + \frac{P[(\overline{AB}) | C\bar{D}] P[C\bar{D}]}{ P[C\vee D]} \\
 &+& \frac{P[(\overline{AB})| CD] P[CD]}{ P[C\vee D]} \\
 &\leq& P[\overline{AB} | \bar{C}D] 
 + P[\overline{AB} | C\bar{D}] 
 + P[\overline{AB}| CD] \\
 \end{eqnarray*}
 We know that $M_1$ and $M_2$ are mechanisms of independent randomness, i.e. add random noise. This means that $A$ is independent of $B$ given $CD$ and vice versa. Therefore we can rewrite $P[\overline{AB}|\overline{C}D] = P[\overline{A}|\overline{C}D] P[\overline{B}|\overline{C}D]$. We obtain:
 \begin{eqnarray*}
     &=& P[\overline{A}|\overline{C}D] P[\overline{B}|\overline{C}D] + P[\overline{A}|C\overline{D}]P[\overline{B}|C\overline{D}] + P[\overline{A}|CD]P[\overline{B}|CD]
 \end{eqnarray*}
 There is a conditional independence between $A$ and $D$ given $C$ due to the nature of the randomized mechanisms. Similarly there is a conditional independence between $B$ and $C$ given $D$. Thus, we can rewrite $ P[\overline{A}|\overline{C}D] P[\overline{B}|\overline{C}D] = P[\overline{A}|\overline{C}] P[\overline{B}|D] $. Therefore,
 \begin{eqnarray*}
 &=& P[\overline{A}|\overline{C}] P[\overline{B}|D] + P[\overline{A}|C]P[\overline{B}|\overline{D}] + P[\overline{A}|C]P[\overline{B}|D]
\end{eqnarray*}
 Knowing that  $FNR_1 = P[\overline{A}|C]$, $FNR_2 = P[\overline{B}|D]$, $TNR_1 = P[\overline{A}|\overline{C}]$ and $TNR_2 = P[\overline{B}|\overline{D}]$, we substitute:
 \begin{eqnarray}
    &=& TNR_1 \cdot FNR_2 + FNR_1 \cdot TNR_2 + FNR_1 \cdot FNR_2 \nonumber \\
    &=& TNR_1 \cdot FNR_2 + FNR_1 ( TNR_2 + FNR_2) \nonumber \\
    &\leq& FNR_2 + FNR_1 \leq \beta_1 + \beta_2
\end{eqnarray}

\end{proof}

\noindent
\textbf{3. Proof for Theorem \ref{def:fnr_conj_theorem}}. Given a disjunction/conjunction query $Q$ answered by a disjunction/conjunction mechanism $M(D)$ where $M_i$ is a Threshold Shift Laplace mechanism and $\epsilon = \epsilon_1 + \epsilon_2$, we achieve minimum privacy loss $\epsilon$ and a $\beta$-bound on FNR by budgeting $\beta$ as:
\begin{eqnarray}
    \beta_1=\frac{u_2\Delta g_1\beta}{u_1\Delta g_2+u_2\Delta g_1},
    \beta_2=\frac{u_1\Delta g_2\beta}{u_1\Delta g_2+u_2\Delta g_1}
\end{eqnarray}
\begin{proof}
Consider a complex decision support query $Q$ composed of two atomic aggregate threshold queries connected by the disjunction or conjunction operator. For both conjunction and disjunction mechanisms, we obtained the optimization problem of:
\begin{maxi}|l|[0]
{}{ (\beta_1)^{\frac{\Delta g_1}{u_1}} (\beta_2)^{\frac{\Delta g_2}{u_2}}}
{}{}
\addConstraint{ \beta_1 + \beta_2 \leq \beta}
\end{maxi}
from Eq.  \eqref{eq:beta12conj}, \eqref{eq:maxconj} for conjunction and Eq. \eqref{eq:final_fnr_disj}, \eqref{eq:ep_min_function} for disjunction. We use the Lagrange Multipliers method to solve this constrained optimization problem, which consists of solving the Lagrangian function that sets the gradient of the function equal to the gradient of the constraint multiplied by the Lagrange Multiplier $\lambda$. We solve for $\lambda$ below:
\begin{eqnarray}
  \frac{d}{d \beta_2} ((\beta_1)^{\frac{\Delta g_1}{u_1}} (\beta_2)^{\frac{\Delta g_2}{u_2}}) &=& \lambda \frac{d}{d \beta_2} (\beta_1+\beta_2 \nonumber)\\
   (\beta_1)^{\frac{\Delta g_1}{u_1}} (\beta_2)^{\frac{\Delta g_2}{u_2}-1}\cdot \frac{\Delta g_2}{u_2} &=& \lambda  \label{eq:fnr_conj_l_1}
   \end{eqnarray}
Similarly,
\begin{eqnarray}
 \frac{d}{d \beta_1} ((\beta_1)^{\frac{\Delta g_1}{u_1}} (\beta_2)^{\frac{\Delta g_2}{u_2}})&=& \lambda \frac{d}{d \beta_1} (\beta_1+\beta_2 \nonumber)\\
 (\beta_1)^{\frac{\Delta g_1}{u_1}-1} (\beta_2)^{\frac{\Delta g_2}{u_2}}\cdot \frac{\Delta g_1}{u_1}&=& \lambda  \label{eq:fnr_conj_l_2}
 \end{eqnarray}
 From setting Equations \eqref{eq:fnr_conj_l_1} and \eqref{eq:fnr_conj_l_2} equal we obtain,
 \begin{eqnarray}
 &&\beta_1=\frac{u_2\beta_2\Delta g_1}{u_1 \Delta g_2}, \beta_2=\frac{u_1\beta_1\Delta g_2}{u_2\Delta g_1}\nonumber
 \end{eqnarray}
 Substituting in the original optimization inequality constraint,
 \begin{eqnarray}
 \beta_2+(\frac{u_2\beta_2\Delta g_1}{u_1 \Delta g_2})&=&\beta\nonumber\\
    \beta_2&=&\frac{u_1\Delta g_2\beta}{u_1\Delta g_2+u_2\Delta g_1}\nonumber
\end{eqnarray}
Similarly,
\begin{eqnarray}
 \beta_1+(\frac{u_1\beta_1\Delta g_2}{u_2\Delta g_1})&=&\beta\nonumber\\
 \beta_1&=&\frac{u_2\Delta g_1\beta}{u_1\Delta g_2+u_2\Delta g_1} \nonumber
\end{eqnarray}

\end{proof}

\subsection{Proofs of PROBE Generalization for $n$-Queries}
\label{app:genbound}
\textbf{Proof for Theorem \ref{def:generalbeta}}. Given a complex decision support query $Q^{\Lambda,F}$ composed of $n$ aggregate threshold queries connected by $n-1$ conjunctions/disjunctions, we achieve minimum privacy loss $\epsilon$ by budgeting the false negative bound $\beta$ as:
    \begin{eqnarray}
    \beta_j =\frac{\Delta g_j \beta  \prod_{x=1}^{n,x\neq j } (u_x )}{\sum_{y=1}^{n}\prod_{x=1}^{n,x\neq y } (u_x \Delta g_y )},  \forall j=\{1,2,...,n\}
    \label{eq:gen_beta_app}
    \end{eqnarray}
\begin{proof}
We will consider exclusive conjunction for this proof because the false negative rate equation is identical for both 2-way conjunction and disjunction. Consider the conjunction of $n$ aggregate threshold queries $Q_1, Q_2,..., Q_n$, where $Q_j$ is defined as $Q_{\mathsf{g_j(.)}>C_j}^{\Lambda,f_j}(D) =\{\lambda_i\!\in\!\Lambda\!\mid\! \mathsf{g_j}(D^{f_j}_{\lambda_i})\!\!>\!\!{c_j}_i\}$.
All $Q_j$ have the same predicates $\Lambda={\lambda_i,\lambda_2,...,\lambda_k}$ but different filters $f_j$, aggregate function $g_j(.)$  and threshold $C_j={{c_j}_1,...,{c_j}_k}$. Let mechanism $M_i:\mathcal{D} \to O_i$ satisfy a $\beta_i$-bound on FNR for aggregate threshold query $Q_i$. We can answer query $Q$ which is a conjunction of $n$ aggregate threshold queries using mechanism $M$ where $M(D)=M_1(D)\cap M_2(D)\cap...\cap M_n(D)$. First, we formally define $M$'s overall false negative rate bound for exclusive conjunction.

We can evaluate mechanism $M$ by sequentially evaluating the 2-way conjunction of each pair of sub-mechanisms $(M_i,M_{i+1})$. Let $M_{i \ i+1}$ be the resulting conjunction mechanism. For an even number of queries $n$ we obtain:
\begin{equation*}
    M(D) = M_{12}(D)\cap M_{34}(D)\cap...\cap M_{n-1\,n}
\end{equation*}
Each mechanism $M_{i\,i+1}$ has the resulting FNR bound of $\beta_i +\beta_{i+1}$ according to Theorem \ref{def:fnr_conj_theorem}, e.g. $M_{1,2}$ has an FNR bound of $\beta_{12}=\beta_1+\beta_2$. We subsequently run the 2-way conjunction mechanism on every pair $(M_{i\,i+1}, M_{i+2\,i+3})$. We obtain:
\begin{equation*}
    M(D) = M_{1234}(D)\cap...\cap M_{n-3 \, n-2 \, n-1 \, n}(D)
\end{equation*}
Where each mechanism $M_{i\,i+1\,i+2\,i+3}$ will have the FNR bound $\beta_{i\,i+1} +\beta_{i+2\,i+3}$ e.g. $\beta_{1234} = \beta_{12}+\beta_{34} = \beta_1+\beta_2+\beta_3+\beta_4$. We can thus recursively run the 2-way conjunction mechanism in this subsequent manner to obtain:
\begin{equation}
 \beta_1 + \beta_2 + ... + \beta_n \leq \beta
\end{equation}
For an odd number of atomic queries $n$ we similarly recursively group sub-mechanisms in pairs except for the last sub-mechanism $M_n$, which will be paired last with the mechanism composed of $n_1$ sub-mechanisms. We thus obtain the same result.

Second, we define mechanism $M$'s overall privacy loss $\epsilon$ in terms of $n$-bounds $\beta_1,\beta_2,...\beta_n$. For conjunction and disjunction alike, we use the sequential composition theorem (Def. \ref{def:seq_comp}) to compute the final privacy loss. Formally,
\begin{eqnarray}
\epsilon &=& \epsilon_1 +\epsilon_2 +...+ \epsilon_n  \nonumber\\ 
 &=& \Delta g_1\frac{\ln(1/(2\beta_1))}{u_1} +\Delta g_2\frac{\ln(1/(2\beta_2))}{u_2}+...+\Delta g_n\frac{\ln(1/(2\beta_n))}{u_n}\nonumber\\ 
 &=&  \ln(1/2\beta_1)^{\frac{\Delta g_1}{u_1}} +\ln(1/2\beta_2)^{\frac{\Delta g_2}{u_2}} +...+\ln(1/2\beta_n)^{\frac{\Delta g_n}{u_n}}   \nonumber\\ 
 &=&  -\ln(2\beta_1)^{\frac{\Delta g_1}{u_1}} -\ln(2\beta_2)^{\frac{\Delta g_n}{u_2}} -...-\ln(2\beta_n)^{\frac{\Delta g_n}{u_n}}   \nonumber\\
 &=& - \ln((2\beta_1)^{\frac{\Delta g_1}{u_1}}(2\beta_2)^{\frac{\Delta g_2}{u_2}}... \ (2\beta_n)^{\frac{\Delta g_n}{u_n}}) \nonumber
 \end{eqnarray}
 To minimize $\epsilon$ we thus need to maximize:
 \begin{eqnarray}
    f_\beta(\beta_1,\beta_2,...,\beta_n)&=&(\beta_1)^{\frac{\Delta g_1}{u_1}} (\beta_2)^{\frac{\Delta g_2}{u_2}} ... \ (\beta_n)^{\frac{\Delta g_n}{u_n}}
\label{eq:apdxbeta}
\end{eqnarray}
We thus use the Lagrange method to solve the following optimization problem:
\begin{maxi}|l|[0]
{}{f_\beta(\beta_1,\beta_2,...,\beta_n) = (\beta_1)^{\frac{\Delta g_1}{u_1}} (\beta_2)^{\frac{\Delta g_2}{u_2}} ... \ (\beta_n)^{\frac{\Delta g_n}{u_n}} }
{}{}
\addConstraint{ \beta_1 + \beta_2 + ... + \beta_n \leq \beta}
\end{maxi}
Applying the Lagrangian function to our optimization problem,

we obtain $n$-functions:
\begin{eqnarray}
  \frac{d}{d \beta_1} ((\beta_1)^{\frac{\Delta g_1}{u_1}} (\beta_2)^{\frac{\Delta g_2}{u_2}} ... \ (\beta_n)^{\frac{\Delta g_n}{u_n}} )= \lambda \frac{d}{d \beta_1} (\beta_1 + \beta_2 + ... + \beta_n)  \nonumber\\
   \frac{d}{d \beta_2} ((\beta_1)^{\frac{\Delta g_1}{u_1}} (\beta_2)^{\frac{\Delta g_2}{u_2}} ... \ (\beta_n)^{\frac{\Delta g_n}{u_n}} )= \lambda \frac{d}{d \beta_2} (\beta_1 + \beta_2 + ... + \beta_n)  \nonumber\\
  \vdots \ \ \ \ \ \ \ \ \ \ \ \ \ \ \ \ \ \ \ \ \ \ \ \ \ \ \ \ \ \ \ \ \ \ \ \ \ \ \ \ \ \ \ \ \ \ \ \ \ \nonumber \\
 \frac{d}{d \beta_n} ((\beta_1)^{\frac{\Delta g_1}{u_1}} (\beta_2)^{\frac{\Delta g_2}{u_2}} ... \ (\beta_n)^{\frac{\Delta g_n}{u_n}} )= \lambda \frac{d}{d \beta_n} (\beta_1 + \beta_2 + ... + \beta_n)  \nonumber
\end{eqnarray}
Solving the equations results in,
\begin{eqnarray}
     \frac{\Delta g_1}{u_1}(\beta_1)^{\frac{\Delta g_1}{u_1}-1} (\beta_2)^{\frac{\Delta g_2}{u_2}}... (\beta_n)^{\frac{\Delta g_n}{u_n}}&=& \lambda \nonumber\\
     \frac{\Delta g_2}{u_2}(\beta_1)^{\frac{\Delta g_1}{u_1}} (\beta_2)^{\frac{\Delta g_2}{u_2}-1}... (\beta_n)^{\frac{\Delta g_n}{u_n}}&=& \lambda \nonumber\\
   \vdots \ \ \ \ \ \ \ \ \ \ \ \ \ \ \ \ \ \ \ \ \ \ \ \ \ \ \nonumber \\
   \frac{\Delta g_n}{u_n}(\beta_1)^{\frac{\Delta g_1}{u_1}} (\beta_2)^{\frac{\Delta g_2}{u_2}}... (\beta_n)^{\frac{\Delta g_n}{u_n}-1}&=& \lambda \nonumber
   \end{eqnarray}
 By setting every combination of equations equal, we obtain:
 \begin{eqnarray*}
   \frac{\Delta g_1}{\beta_1 u_1}&=&\frac{\Delta g_2}{\beta_2 u_2}=...=\frac{\Delta g_n}{\beta_n u_n}
\end{eqnarray*}
Solving for individual $\beta_i$,
 \begin{eqnarray}
   \beta_1&=&\frac{\beta_2 u_2 \Delta g_1}{u_1 \Delta g_2}=\frac{\beta_3 u_3 \Delta g_1}{u_1 \Delta g_3}=...=\frac{\beta_n u_n \Delta g_1}{u_1 \Delta g_n} \nonumber \\
      \beta_2&=&\frac{\beta_1 u_1 \Delta g_2}{u_2 \Delta g_1}=\frac{\beta_3 u_3 \Delta g_2}{u_2 \Delta g_3}=...=\frac{\beta_n u_n \Delta g_2}{u_2 \Delta g_n} \nonumber \\
        && \ \ \ \ \ \ \ \vdots \ \ \ \ \ \ \ \ \ \ \ \ \ \ \ \ \ \ \ \ \ \ \ \ \ \ \nonumber \\
   \beta_n&=&\frac{\beta_1 u_1 \Delta g_n}{u_n \Delta g_1}=\frac{\beta_2 u_2 \Delta g_n}{u_n \Delta g_2}=...=\frac{\beta_{n-1} u_{n-1} \Delta g_n}{u_n \Delta g_{n-1}} \nonumber 
 \end{eqnarray}
 We can now substitute $\beta_{j \not = i}$ in the original constraint to solve for individual $\beta_i$:
 \begin{eqnarray}
   \beta &=& \beta_1 + \frac{\beta_1 u_1 \Delta g_2}{u_2 \Delta g_1} + ... + \frac{\beta_1 u_1 \Delta g_n}{u_n \Delta g_1}  \nonumber
 \end{eqnarray}
 For $\beta_1$ we obtain:
  \begin{eqnarray}
   \beta_1  &=&\frac{\beta}{(1+ \frac{u_1}{\Delta g_1}(\frac{\Delta g_2}{u_2}+...+\frac{\Delta g_n}{u_n}))}\nonumber\\
   &=&\frac{\beta}{(1+ \frac{u_1}{\Delta g_1}(\frac{\Delta g_2 u_3u_4...u_n+\Delta g_3 u_2u_4...u_n+...+ \Delta g_n u_2u_3...u_{n-1}}{u_2u_3...u_n}))} \nonumber \\
   &=& \frac{\beta}{(1+ \frac{u_1}{\Delta g_1}(\frac{\sum_{i=1}^{n,i \not = 1}\prod_{j=1}^{n,j \not = i,1}u_j \Delta g_i}{\prod_{i=1}^{n,i \not = 1}u_i}))} \nonumber\\
   &=& \frac{\Delta g_1\beta\prod_{i=1}^{n,i \not = 1}u_i}{\Delta g_1\prod_{i=1}^{n,i \not = 1}u_i+u_1\sum_{i=1}^{n,i \not = 1}\prod_{j=1}^{n,j \not = i,1}u_j\Delta g_i}\nonumber \\
   &=&\frac{\Delta g_1\beta\prod_{i=1}^{n,i \not = 1}u_i}{\sum_{i=1}^{n}\prod_{j=1}^{n,j \not = i}u_j\Delta g_i}\nonumber
   \end{eqnarray}
   Similarly,
   \begin{eqnarray}
   \beta_2&=&\frac{\Delta g_2\beta\prod_{i=1}^{n,i \not = 2}u_i}{\sum_{i=1}^{n}\prod_{j=1}^{n,j \not = i}u_j \Delta g_i},... \ ,\beta_n=\frac{\Delta g_n\beta\prod_{i=1}^{n,i \not = n}u_i}{\sum_{i=1}^{n}\prod_{j=1}^{n,j \not = i}u_j\Delta g_i}\nonumber
    \end{eqnarray}
   Thus, we can generalize the expression for $i \in {1,2,...,n}$
   \begin{equation}
         \beta_i =\frac{\Delta g_i\beta  \prod_{x=1}^{n,x\neq i } (u_x )}{\sum_{y=1}^{n}\prod_{x=1}^{n,x\neq y } (u_x \Delta g_y )},  \forall i=\{1,2,...,n\}
   \end{equation}
\end{proof}

\noindent
\textbf{Proof for Theorem \ref{def:beta_occ}}. 
Given a complex decision support query $Q^{\Lambda,F}$ with query tree $T$ composed of $n$ aggregate threshold queries with an associated $o_i$ number of occurrences within the tree,  we achieve minimum privacy loss $\epsilon$ by budgeting the $\beta$-bound on FNR as:
    \begin{eqnarray}
    \beta_i =\frac{\Delta g_i \beta  \prod_{x=1}^{n,x\neq i } (u_x )}{\sum_{y=1}^{n}\prod_{x=1}^{n,x\neq y } (u_x o_y\Delta g_y )},  \forall i=\{1,2,...,n\}
    \label{eq:gen_beta_occ_app}
    \end{eqnarray}
        % \xh{change the subscript for $\beta$ from $j$ to $i$}
\begin{proof}
We first prove that the apportioning function $f_\beta$ that optimally apportions the FNR bound $\beta$ over the sub-queries given a query tree $T_c$ is:
\begin{equation}
    f_\beta( \beta_1, \beta_2, ...,\beta_n) = \sum_{i=1}^n o_i\beta_i \leq \beta
\label{eq:provebeta}
\end{equation}
We prove this by induction as follows:

\noindent
\underline{Base case:} \textit{Query tree with two nodes.} consider a complex query $Q^{\Lambda,F}$ with compact query tree $T$ composed of $n=2$ leaf nodes containing two sub-queries $Q_1, Q_{2}$ with occurrences $o_1 = o_{2} = 1$ connected by an operator $\cap$ or $\cup$. It follows from Appendix A.2 that the apportioning of the FNR in terms of $\beta_i$ is:
\begin{eqnarray*}
    f_\beta(\beta_1,\beta_2) &=& \beta_{1} + \beta_{2} \leq \beta \\
    f_{\beta_1,\beta_2} &=& o_1\beta_{1} + o_{2}\beta_{2} \leq \beta
\end{eqnarray*}
Therefore condition \eqref{eq:provebeta} holds for the base case of $n=2$. 

\noindent
\underline{Induction step:} Suppose that for a query tree $T$ of size $n=k$ nodes containing $m$ sub-queries $Q_1, Q_{2},...,Q_m $ with occurrences $o_1,o_2,...,o_m$ ,  the FNR is apportioned in terms of $\beta_i$ and bound by $\beta$ as:
\begin{equation}
     f_\beta( \beta_1, \beta_2, ...,\beta_m) = \sum_{i=1}^m o_i\beta_i \leq \beta
     \label{eq:condition}
\end{equation}
Now we show that the condition holds for $n=k+1$. 
\noindent
Let $T$ be a compact query tree composed of $n=k+1$ nodes containing $m$ sub-queries $Q_1, Q_{2},...,Q_m $ with occurrences $o_1,o_2,...,o_m$ connected by an operator $\cap$ or $\cup$. We refer to the right sub-tree as $T_R$ and the left sub-tree as $T_L$. Each sub-tree contains the sub-queries $Q_1, Q_{2},...,Q_m $ with local occurrences $T_L.o_i$ or $T_R.o_i$ which add up to the overall occurrence number in the tree $T.o_i$, i.e. $\forall i \in m$
\begin{equation}
     T.o_i = T_L.o_i + T_R.o_i
     \label{eq:overalltree}
\end{equation}
For the right sub-tree $T_R$, we know that the condition \eqref{eq:condition} holds for $n=k$. Therefore the FNR for $T_R$ in terms of $\beta_i$ is:
\begin{equation}
     \beta_{T_R} = \sum_{i=1}^m T_R.o_i\beta_i 
     \label{eq:condright}
\end{equation}
Similarly for the left sub-tree $T_L$, we obtain the FNR,
\begin{equation}
    \beta_{T_L} = \sum_{i=1}^m T_L.o_i\beta_i
     \label{eq:condleft}
\end{equation}
For overall tree $T$, we know from 2-way conjunction/disjunction that:
\begin{equation*}
    f_{\beta_{T}}( \beta_1, \beta_2, ...,\beta_m) = \beta_{T_L}+ \beta_{T_R} \leq \beta
\end{equation*}
Substituting in Eq. \eqref{eq:condright} and \eqref{eq:condleft} we obtain:
\begin{eqnarray*}
    f_{\beta_{T}}( \beta_1, \beta_2, ...,\beta_m) &=&\sum_{i=1}^m T_L.o_i\beta_i + \sum_{i=1}^m T_R.o_i\beta_i  \leq \beta \\
    &=&\sum_{i=1}^m T_L.o_i\beta_i + T_R.o_i\beta_i  \leq \beta \\
\end{eqnarray*}
From Eq. \eqref{eq:overalltree} we obtain
\begin{eqnarray*}
    f_{\beta{T}}(\beta_1, \beta_2, ...,\beta_m) &=&\sum_{i=1}^m T.o_i\beta_i  \leq \beta
\end{eqnarray*}
Therefore Eq. \eqref{eq:condition} holds for all query trees $T$ with $m$ atomic aggregate threshold queries.

Second, we derive the apportionment of $\beta_i$ from Eq. \eqref{eq:gen_beta_occ_app}. To do so, we derive our optimization problem which minimizes $\epsilon$ given the constraint from Eq. \eqref{eq:provebeta}. Given that we run each sub-query $Q_i$ using $M_i$ based on its number of occurrences $o_i$ in query tree $T$, its privacy budget will be proportional to the number of occurrences, i.e. the loss will be $o_i\epsilon_i$ for each $Q_i$. This means that the overall privacy loss, when using the Sequential Composition theorem, will be $\epsilon = \sum_{i=1}^n o_i\epsilon_i$. As we use TSLM for the sub-mechanisms $M_i$, we can use the formulation of $\epsilon_i = \Delta g_i\frac{\ln(1/(2\beta_i))}{u_i}$. Substituting in:
\begin{eqnarray}
\epsilon &=& \sum_{i=1}^n o_i\epsilon_i \nonumber\\ 
 &=& \sum_{i=1}^n o_i \Delta g_i\frac{\ln(1/(2\beta_i))}{u_i} \nonumber\\ 
 &=& -\sum_{i=1}^n\ln(2\beta_i)^{\frac{o_i\Delta g_i}{u_i}} \nonumber\\
 &=& - \ln((2\beta_1)^{\frac{o_1\Delta g_1}{u_1}}(2\beta_2)^{\frac{o_2\Delta g_2}{u_2}}... \ (2\beta_n)^{\frac{o_n\Delta g_n}{u_n}}) \nonumber
 \end{eqnarray}
 To minimize $\epsilon$ we thus need to maximize:
 \begin{eqnarray}
    f_\beta(\beta_1,\beta_2,...,\beta_n)&=&\prod_{i=1}^n(\beta_i)^{\frac{o_i\Delta g_i}{u_i}}
\label{eq:apdxgenbeta}
\end{eqnarray}
So our optimization problem is updated to:
\begin{maxi}|l|[0]
{}{\prod_{i=1}^n(\beta_i)^{\frac{o_i\Delta g_i}{u_i}}}
{}{}
\addConstraint{ \sum_{i=1}^n o_i\beta_i  \leq \beta}
\end{maxi}
We thus apply the Lagrange Multipliers Method as with the previous proof to obtain: 
 \begin{eqnarray}
    \beta_i =\frac{\Delta g_i \beta  \prod_{x=1}^{n,x\neq i } (u_x )}{\sum_{y=1}^{n}\prod_{x=1}^{n,x\neq y } (u_x o_y\Delta g_y )},  \forall i=\{1,2,...,n\}
    \end{eqnarray}
\end{proof}
\subsection{Proofs For FPR Bounds}
\label{app:fprbound}
\textbf{1. Proof For 2-Query Conjunction}.
 Consider a complex decision support query $Q$ composed of two atomic aggregate threshold queries $Q=Q_1 \cap Q_2$. $Q_1$ and $Q_2$ are answered by TSLM $M_1$, $M_2$ respectively, and mechanism $M = M_1 \cap M_2$ is used to answer query $Q$. If $M_1$ and $M_2$ guarantee an FPR bound of $\alpha_1$ and $\alpha_2$ respectively, then $M$ has an FPR bound of $\alpha_1 + \alpha_2$. If we set $\alpha_i = \alpha/n$, we obtain a $\alpha$ bound on FPR.

\noindent
\begin{proof}
$\forall \lambda_i \in \Lambda,$ \\
Let $A$ be the event that $\lambda_i \in M_1(D)$, $B$ be the event that $\lambda_i \in M_2(D)$, $C$ be the event that $\lambda_i \in Q_1(D)$ and $D$ be the event that $\lambda_i \in Q_2(D)$. The probability of false positives can be derived as:
\begin{eqnarray}
FPR &=& P[\lambda_i  \in M_1(D) \cap M_2(D)| \lambda_i  \not \in Q_1(D)\cap Q_2(D)]\nonumber\\
&=&P[AB| \overline{CD}]\nonumber=P[AB|\overline{C}\vee \overline{D}]\nonumber=P[AB|\overline{C}D \vee C\overline{D} \vee \overline{CD}] \nonumber \\
&=&\frac{P[AB \wedge \overline{C}D \vee C\overline{D} \vee \overline{CD}]}{P[CD]}\nonumber\\
&=& \frac{P[(AB \wedge \overline{C}D) \vee (AB \wedge C\overline{D}) \vee(AB\wedge \overline{CD})]}{P[CD]}
\end{eqnarray}
Given that the three clauses are mutually exclusive events
\begin{eqnarray*}
&=&\frac{P[AB \wedge \overline{C}D] + P[AB \wedge C\overline{D}] + P[AB\wedge \overline{CD}]}{P[CD]}\\
 &=& \frac{P[AB | \bar{C}D] P[\bar{C}D] }{ P[CD]}
 + \frac{P[AB | C\bar{D}] P[C\bar{D}]}{ P[CD]} \\
 &+& \frac{P[AB| \overline{CD}] P[\overline{CD}]}{ P[CD]} \\
 &\leq& P[AB | \bar{C}D] 
 + P[AB | C\bar{D}] 
 + P[AB| \overline{CD}] \\
\end{eqnarray*}
We know that $M_1$ and $M_2$ are mechanisms of independent randomness, i.e. add random noise. This means that $A$ is independent of $B$ given $CD$ and vice versa. Therefore we can rewrite $P[AB|\overline{C}D] = P[A|\overline{C}D] P[B|\overline{C}D]$. We obtain:
\begin{eqnarray*}
    &=& P[A|\overline{C}D] P[B|\overline{C}D] + P[A|C\overline{D}]P[B|C\overline{D}] + P[A|\overline{CD}]P[B|\overline{CD}]
\end{eqnarray*}
There is a conditional independence between $A$ and $D$ given $C$ due to the nature of the randomized mechanisms. Similarly there is a conditional independence between $B$ and $C$ given $D$. Thus, we can rewrite $P[A|\overline{C}D] P[B|\overline{C}D] = P[A|\overline{C}] P[B|D] $. Therefore,
\begin{eqnarray*}
&=& P[\overline{A}|C]P[B|D] + P[A|C]P[\overline{B}|\overline{D}] + P[A|\overline{C}]P[B|\overline{D}]
\end{eqnarray*}
 Knowing that  $FPR_1 = P[A|\overline{C}]$, $FPR_2 = P[B|\overline{D}]$, $TPR_1 = P[A|C]$ and $TPR_2 = P[B|D]$, we substitute:
 \begin{eqnarray}
     &=& FPR_1 \cdot TPR_2 + TPR_1 \cdot FPR_2 + FPR_1 \cdot FPR_2 \nonumber \\
     &=&  FPR_1 \cdot TPR_2  +FPR_2(FPR_1 + TPR_1) \nonumber\\
     &\leq& FPR_1 + FPR_2 \leq \alpha_1 + \alpha_2
 \end{eqnarray}
 So if we set $\alpha_1=\alpha_2= \alpha/2$, then $FPR \leq \alpha$ .
\end{proof}

\noindent
\textbf{2. Proof For 2-Query Disjunction}.
 Consider a complex decision support query $Q$ composed of two atomic aggregate threshold queries $Q=Q_1 \cup Q_2$. $Q_1$ and $Q_2$ are answered by TSLM $M_1$, $M_2$ respectively, and mechanism $M = M_1 \cup M_2$ is used to answer query $Q$. If $M_1$ and $M_2$ guarantee an FPR bound of $\alpha_1$ and $\alpha_2$ respectively, then $M$ has an FPR bound of $\alpha_1 + \alpha_2$. If we set $\alpha_i = \alpha/n$, we obtain a $\alpha$ bound on FPR.

\noindent
\begin{proof}
$\forall \lambda_i \in \Lambda,$ \\
Let $A$ be the event that $\lambda_i \in M_1(D)$, $B$ be the event that $\lambda_i \in M_2(D)$, $C$ be the event that $\lambda_i \in Q_1(D)$ and $D$ be the event that $\lambda_i \in Q_2(D)$. The probability of false positives can be derived as:
\begin{eqnarray}
    FPR &=& P[\lambda_i \in M_1(D) \cap M_2(D)| \lambda_i   \not \in Q_1(D)\cap Q_2(D)]\nonumber\\  
    &=& P[A\vee B| \overline{C\vee D}]\nonumber\\
    &=& \frac{P[(\overline{A}B \vee A\overline{B} \vee AB) \wedge \overline{CD}]}{ P[\overline{C\vee D}]} \nonumber
\end{eqnarray}
Given that the three clauses are mutually exclusive events
\begin{eqnarray*}
    &=& \frac{P[\overline{A}B \wedge \overline{CD}]
 + P[A\overline{B} \wedge \overline{CD}]
 + P[AB \wedge \overline{CD}]}{ P[\overline{C\vee D}]} \\
 % &=& \frac{P[(\overline{AB}) | \overline{CD}] P[\bar{C}D] }{ P[\overline{C\vee D}]}
 % + \frac{P[(\overline{AB}) | C\bar{D}] P[C\bar{D}]}{ P[\overline{C\vee D}]} \\
 % &+& \frac{P[(\overline{AB})| CD] P[CD]}{ P[\overline{C\vee D}]} \\
 &=& P[\overline{A}B | \overline{CD}] 
 + P[A\overline{B}  | \overline{CD}] 
 + P[AB| \overline{CD}] \\
 \end{eqnarray*}
 We know that $M_1$ and $M_2$ are mechanisms of independent randomness, i.e. add random noise. This means that $A$ is independent of $B$ given $CD$ and vice versa. Therefore we can rewrite $P[\overline{A}B | \overline{CD}]  = P[\overline{A}|\overline{CD}] P[B|\overline{CD}]$. We obtain:
 \begin{eqnarray*}
     &=& P[\overline{A}|\overline{CD}] P[B|\overline{CD}] + P[A|\overline{CD}]P[\overline{B}|\overline{CD}] + P[A|CD]P[B|CD]
 \end{eqnarray*}
 There is a conditional independence between $A$ and $D$ given $C$ due to the nature of the randomized mechanisms. Similarly there is a conditional independence between $B$ and $C$ given $D$. Thus, we can rewrite $ P[\overline{A}|\overline{CD}] P[B|\overline{CD}]= P[\overline{A}|\overline{C}] P[B|\overline{D}] $. Therefore,
 \begin{eqnarray*}
 &=& P[\overline{A}|\overline{C}] P[B|\overline{D}] + P[A|\overline{C}]P[\overline{B}|\overline{D}] + P[A|\overline{C}]P[B|\overline{D}]
\end{eqnarray*}
 Knowing that  $FPR_1 = P[A|\overline{C}]$, $FPR_2 = P[B|\overline{D}]$, $TNR_1 = P[\overline{A}|\overline{C}]$ and $TNR_2 = P[\overline{B}|\overline{D}]$, we substitute:
 \begin{eqnarray}
    &=& TNR_1 \cdot FPR_2 + FPR_1 \cdot TNR_2 + FPR_1 \cdot FPR_2 \nonumber \\
    &=& (1-FPR_1) FPR_2 + FPR_1 ( 1-FPR_2) + FPR_1 FPR_2 \nonumber \\
    &\leq& FPR_1 + FPR_2 - FPR_1 FPR_2 \leq FPR_1 + FPR_2 
 \leq \alpha_1 + \alpha_2
\end{eqnarray}
So if we set $\alpha_1=\alpha_2= \alpha/2$, then $FPR \leq \alpha$.
\end{proof}

\noindent
\textbf{Generalization of FPR bound}. By using a similar proof by induction as in Appendix~\ref{app:genbound} (Proof for Theorem 4.5), we obtain the bound on FPR given any query T, where each sub-query $Q_i$ of $n$ sub-queries, with an associated $o_i$ number of occurrences, as follows:
\begin{equation*}
    FPR \leq \sum_{i=1}^n o_i\alpha_i
\end{equation*}
If we set $\alpha_i=\alpha/2o_i$ for each sub-query, then $FPR \leq \alpha$.

\subsection{Proof of ProBE Algorithm}
\label{app:algo}
% \textbf{Proof for Theorem \ref{def:tslm}}. Algorithm 1 satisfies $\epsilon_{\max}$-DP and a $\beta$-bound on the False Negative Rate.
% \begin{proof}
    
%     First, we show the proof that TSLM offers the guarantee that for an atomic aggregate threshold query $Q_{a_i}$, setting the privacy budget to $\epsilon_i=\frac{\Delta g_i\ln(1/(2\beta_i)}{u_i}$ guarantees the FNR is bounded by $\beta_i$ as defined in \cite{mide}. For all predicates $\lambda_i \in \Lambda$:
%     \begin{eqnarray*}
%         && P[\lambda_i \not \in M(D) | \lambda_i \in Q(D)] \\
%         &=& P[g(D^f_{\lambda_i}) + \eta_i \leq c_i - u_i | g(D^f_{\lambda_i}) > c_i ] \\
%         &\leq& P[\eta_i \leq -u_i]  \leq \frac{e^{-ln(1/2\beta)}}{2} \leq \beta 
%     \end{eqnarray*}

%     Second, from the generalization proof in Appendix A.1, we know that $\sum_{i=1}^n \beta_i \leq \beta$. Therefore the FNR for query $Q^{\Lambda,F}$ is bound by $\beta$.

%     Third, the algorithm satisfies $\epsilon_{max}$-DP as it checks that the current privacy budget $\epsilon_i$ does not exceed the limit of $\epsilon_{max}$, and adds noise from the Laplace Distribution with a mean of 0 and a standard deviation of $1/\epsilon_i$. 
% \end{proof}
\noindent
\textbf{Proof for Theorem \ref{def:mspwlm}}. Algorithm 2 satisfies $\epsilon_{\max}$-DP, a $\beta$-bound on the False Negative Rate and a $\alpha$ bound on the False Positive Rate.
\begin{proof}
      First, we show the proof that TSLM offers the guarantee that for an atomic aggregate threshold query $Q_{a_i}$, setting the privacy budget to $\epsilon_i=\frac{\Delta g_i\ln(1/(2\beta_i)}{u_i}$ guarantees the FNR is bounded by $\beta_i$ as defined in \cite{mide}. For all predicates $\lambda_i \in \Lambda$:
    \begin{eqnarray*}
        && P[\lambda_i \not \in M(D) | \lambda_i \in Q(D)] \\
        &=& P[g(D^f_{\lambda_i}) + \eta_i \leq c_i - u_i | g(D^f_{\lambda_i}) > c_i ] \\
        &\leq& P[\eta_i \leq -u_i]  \leq \frac{e^{-ln(1/2\beta)}}{2} \leq \beta 
    \end{eqnarray*}
     Second, we prove that setting $\beta$ per iteration to $\beta/2$ upholds a $\beta$ bound on the overall mechanism's FNR. 
     
     We know that the FNR is bounded as $FNR \leq \sum_{i=1}^n \beta_i$. Therefore, if we set $\beta_{step} = \beta_2$ where $step = i \in \{1,2\}$, then we obtain $FNR \leq \beta/2 + \beta/2 = \beta$. Therefore the FNR is bounded by $\beta$.
     
     Third, the algorithm satisfies $\epsilon_{max}$-DP.
     
     (i) Phase One of ProBE adds noise from the Laplace Distribution according to the Laplace Mechanism\cite{Dwork:2014:AFD:2693052.2693053}, i.e. with a mean of 0 and a standard deviation of $1/\epsilon_i$. This means that Phase one is $\epsilon_i$-differentially private.
     (ii) Phase Two of ProBE similarly adds noise from the Laplace Distribution according to the Laplace Mechanism with a newly computed $\epsilon_j$ privacy budget, making phase two  $\epsilon_j$-differentially private. 
     (iii) Because the two phases are executed sequentially, their composition (i.e. the overall ProBE mechanism) is $\epsilon_i + \epsilon_j$-differentially private. (iv) It follows that ProBE satisfies $\epsilon_{max}$-DP as it checks that the current privacy budget (i.e. $\epsilon_{i} + \epsilon_{j}$) does not exceed the limit of $\epsilon_{max}$.
     % it checks that the current privacy budget $\epsilon_i$ does not exceed the limit of $\epsilon_{m}$ per iteration which is the ex-post DP cost that is smaller than the maximum privacy budget allowed $\epsilon_m < \epsilon_{max}$, and adds noise from the Laplace Distribution with a mean of 0 and a standard deviation of $1/\epsilon_i$.
    
     Fourth, the algorithm satisfies the $\alpha$-bound on FPR because we estimate an upper bound on the FPR by deriving an upper bound on $FP$ and a lower bound on $N$. Thus, since $FPR \leq FPR_{est} $, then enforcing a bound on the upper bound $FPR_{est} \leq \alpha$ (by exiting the algorithm if it is exceeded) will enforce a bound on the actual $FPR$. i.e. $FPR  \leq FPR_{est} \leq \alpha \Rightarrow FPR \leq \alpha$. 
     
\end{proof}
\section{ProBE Computational Complexity Analysis}
% \subsection{}
We analyze the computational complexity of ProBE as it relates to regular SQL queries by using the example query below:
\begin{verbatim}
SELECT disease, count(*) FROM PATIENT_DATA
WHERE disease_type ='viral'
GROUP BY disease
HAVING (count(*) > c1 AND avg(age) > 65) 
\end{verbatim}

If executed over an SQL system such as MySQL, the query evaluation plan would be:  
\begin{enumerate}
    \item Full Table Scan on $PATIENT\_DATA$
    \item GROUP BY disease attribute
    \item Compute aggregations on $count(*)$ and on avg(age) based on the tuples in a group per group and filter matching groups based on the count(*) > c1 and avg(age) > 65 conditions
    \item Return matching groups
\end{enumerate}

In ProBE, if we break the query into sub-queries and execute them separately then apply the conjunction or disjunction as discussed in the paper, the approach will correspond to the following plan:

For Q1:
\begin{enumerate}
    \item Full Table Scan on $PATIENT\_DATA$ 
    \item GROUP BY disease attribute
    \item Compute aggregations with noise on count(*)  based on the tuples in a group per group and filter matching groups based on count(*) > c1 condition 
\end{enumerate}

For Q2:
\begin{enumerate}
    \item Full Table Scan on $PATIENT\_DATA$ 
    \item GROUP BY disease attribute
    \item Compute aggregations with noise on avg(age)  based on the tuples in a group per group and filter matching groups based on avg(age) > 65 condition

\end{enumerate}

And lastly:
\begin{enumerate}
    \item Compute conjunction of Q1 and Q2
    \item Return matching groups
\end{enumerate}

In the execution plan above, if we have $k$ operators (i.e. conjunctions and/or disjunctions), the technique will result in a computational complexity of almost $k$ times. 
Note that when we scan for each of the sub-queries, we do not need to do this scan independently. Indeed, a better plan would be:

\begin{enumerate}
    \item Full Table Scan on $PATIENT\_DATA$
    \item GROUP BY disease attribute
    \item Compute all aggregations with noise i.e. count(*) and avg(age) and filter matching groups based on the count(*) > c1 and avg(age) >65 conditions 
    \item Compute conjunction of Q1 and Q2
    \item Return matching groups
\end{enumerate}

The above execution plan would meet our requirement, and has complexity similar to the original SQL plan, but will require an in-database implementation to be slightly modified to achieve the above. Our goal in this paper is primarily to develop the foundation for answering complex queries in a differentially private manner and not on optimizing the query performance in terms of execution time. Nonetheless, this aspect of optimizing the execution of our algorithm is a very interesting direction of future work.

Note that implementation of the GROUP BY operation has to be special in that all possible groups to be returned (i.e. all the diseases) must be pre-determined, as opposed to the SQL GROUP BY implementation which does not have knowledge of the existing groups. This issue applies to both second and third plan.

\section{Detailed Multi-Step Entropy-based Algorithm}
\label{app:probeent}
The two-step algorithm ProBE assigns different privacy levels to different predicates due to the early elimination of data points at the first step, meaning predicates that go on to the second step have a higher privacy loss $\epsilon$. This concept is captured through the definition of Predicate-wise DP (PWDP)\cite{mide}, a fine-grained extension of differential privacy which quantifies the different levels of privacy loss data points may have in multi-step algorithms. Formally,
 \begin{definition}[Predicate-wise Differential Privacy (PWDP)]
    Given a set of mutually exclusive predicates and their corresponding privacy budgets $\Theta = \{(\lambda_1,\epsilon_1),...,(\lambda_k,\epsilon_k)\}$, we say a randomized mechanism $M$ satisfies $\theta$-PWDP if for all i, for any neighboring databases $D$ and $D'$ differing in at most one record that satisfies $\lambda_i$, the following holds:
    \begin{equation}
        P[M(D) \in O] \leq e^{\epsilon_i} P[M(D') \in O]
    \end{equation}
\end{definition}
Naturally, a measure to quantify this new definition of privacy is needed Previous work tackles this problem by proposing a new privacy metric for PWDP entitled \textit{Min-Entropy} \cite{mide} $\mathcal{H}_{min}$ which measures a lower bound on the level of uncertainty given the set of predicates and their respective privacy levels $\Theta = \{(\lambda_1,\epsilon_1),...,(\lambda_k,\epsilon_k)\}$. Formally,
\begin{definition}
[PWDP Min-Entropy]
The Min-Entropy of a $\Theta$-predicate-wise differentially private mechanism with $\Theta = \{(\lambda_1,\epsilon_1),$ $... ,(\lambda_k,\epsilon_k)\}$ is defined as:

\begin{equation*}
  \mathcal{H}_{min}(\Theta) = min \ \Sigma^k_{i=1} - \hat{p_i}\log\hat{p_i}
\end{equation*}
% \vspace{-10}
\begin{equation}
     s.t. \frac{e^{-\epsilon_i}}{\Sigma_ie^{\epsilon_i}} \leq \hat{p_i} \leq \frac{e^{\epsilon_i}}{\Sigma_ie^{-\epsilon_i}} \  \forall i \in [1,k] \  and \  \Sigma_i \hat{p_i} =1 
\end{equation}
 where $\hat{p_i}$ is the probability that a random tuple x will take a value $t$ that satisfies $\lambda_i$.
 \label{def:entropy}
\end{definition}

the Data Dependent Predicate-Wise Laplace Mechanism (DDPWLM) also introduced in \cite{mide} maximizes min-entropy (i.e. maximizing the lower bound on uncertainty) in a multi-step algorithm by also using the noisy aggregate values obtained from previous iterations to compute the best privacy level $\epsilon$ at which Min-Entropy is maximized for the set of elements in the uncertain region. This algorithm similarly guarantees a $\beta$-bound on FNR for a single aggregate threshold query by setting the privacy loss to $\epsilon = \frac{\Delta g\ln(1/(2\beta))}{u}$, but it does so by setting a starting privacy level $\epsilon_s$ as well as a maximum level $\epsilon_m$ prior to execution which is spent across a fixed number of iterations $m$ in a way that maximizes min-entropy. In each iteration, the privacy budget is further distributed across fine-grained steps $m_f$ and the privacy level with the highest min-entropy is chosen as the next iteration's budget. It follows that the uncertain region parameter $u$ is, again, chosen statically at the beginning, then chosen to minimize privacy loss rather than in a way that can provide a bound on false positives.

We modify this algorithm by integrating our two-step algorithm ProBE into its framework in order to not only answer complex DS queries, but also to provide a post-facto bound on false positives. Our modified algorithm is entitled ProBE-Ent as described in Algorithm \ref{algo:ddpwlm}. Instead of having a fixed starting $\epsilon_s$, we internally choose a starting $u_0$ and compute the initial privacy level (line 7). Within the first step of DDPWLM, we run the FP estimation algorithm (Function in Algorithm \ref{algo:secondstep}) in order to compute the optimal uncertain region parameter $u_{opt}$. We use this new uncertain region as an upper bound for the algorithm, i.e. we compute the $\epsilon_m$ upper bound that represents the exit condition for the algorithm. Additionally, we provide a $\beta$ budget optimization which exploits the multi-step nature of the algorithm in a way that provides a potentially higher budget if previous sub-queries exit early.

% Instead of predetermining the privacy budget of each iteration, DDPWLM chooses the next iteration's $\epsilon$ by calculating the probabilities of each predicate being higher/lower than the shifted threshold using previous noisy aggregated values. These probabilities 
% are used to calculate the min-entropy, and the highest value is used to choose the optimal privacy budget $\epsilon$. This algorithm similarly bounds the $\beta$ per iteration equally to $\beta/m$, therefore setting the worse-case privacy loss per iteration to be $\epsilon_m = \frac{\Delta g\ln(1/(2\beta/m))}{u}$.

\noindent
\textbf{$\beta$-Budget Redistribution}. In the case of multi-step algorithms, we exploit their iterative nature to further optimize $\beta$ budget allocation during execution. Given that certain queries may terminate early (if an iteration $j<m$ yields no predicates within the uncertain region), the assigned $\beta_i$ for such a query may not be fully used and may thus be wasted. We exploit the progressive nature of such algorithms to further optimize the privacy budget: if the previous sub-query terminates before the $m$-th step of its run, we can redistribute the leftover false negative rate budget to subsequent sub-queries, thus fully utilizing the false negative rate bound given and consequently minimizing the overall privacy loss. For each sub-query, we calculate the remaining $\beta$ budget after running the MSPWLM algorithm by subtracting the used $\beta_i$ budget from the overall $\beta$  (line 9). For subsequent sub-queries, we check if the remaining $\beta$ budget $\beta_{rem}$ from the previous iteration is higher than 0. If this is not the case, then we use the ProBE optimization with the original $\beta$ and all the sub-queries. Otherwise, we update the overall $\beta$ to the remaining $\beta_{rem}$ and re-run the ProBE optimization using only the subsequent sub-queries $Q_i, \ldots ,Q_n$ (lines 4-6). To illustrate this algorithm, we use the example below.
\begin{exmp}Consider a 3-step PPWLM run with a conjunction query composed of 4 aggregate threshold queries; if the first query finishes executing after its 1st step, its leftover overall false negative rate will be $\beta - \Delta\beta_1$, where  $\Delta\beta_1=\frac{1}{3}\beta_1$. Therefore we re-compute the false negative rates for subsequent queries $\beta_2,\beta_3,\beta_4$ using Eq. (\ref{eq:gen_beta_occ}) with $\beta' = \beta - \Delta\beta_1$ and with the $Q_2,Q_3,Q_4$ leftover queries.
\end{exmp}
\begin{algorithm}[t] 
\begin{algorithmic}[1]
\Procedure{EntProbe}{$Q, D, u, \beta,o,\Delta g, \epsilon_{max},m,m_f,t$}
\State Let $O_f \gets \{\},\epsilon_f \gets 0,\beta_{rem} \gets 0$
\For{$i=1,...,n$}
\If{$\beta_{rem} > 0 $}
\State update overall FNR budget $\beta \gets \beta_{rem}$
\State $Q \gets \{Q_{i},...,Q_n\}$
\EndIf
\State $\beta_i \gets \frac{\Delta g_i \beta  \prod_{x=1}^{n,x\neq i } (u_0 )}
{o_i\sum_{y=1}^{n}\prod_{x=1}^{n,x\neq y } (u_0 \Delta g_y )}$,$\epsilon_s = \frac{\Delta g_i\ln(1/(2\beta_i/m))}{u_0}$
\State $O_i, \epsilon_{i}, \beta_{used} \gets$ \textsc{Ddpwlm}($Q_i,u_i,\beta_i,\epsilon_{max},\epsilon_{s},m,m_f$)
\State $\beta_{rem} \gets 0$ if $\beta_i=\beta_{used}$ \textbf{else} $\beta - \beta_{used}$ 
\State $\epsilon_f \gets \epsilon_f + \epsilon_i$
\If{$\epsilon_f \leq \epsilon_{max}$}
\If{$type = 0$}
\State $O_f \gets O_{i}$ if $O_f = \emptyset$ \textbf{else} $O_f \gets O_f \cap O_{i}$
\State \Return $O_f,\epsilon_f$ if $O_{i} = \emptyset$

\ElsIf{$type =1$}
\State $ O_f \gets O_f \cup O_{i}$
\EndIf
\Else
\State \Return `Query Denied'
\EndIf
\State \Return $O_f,\epsilon_f$

\EndFor
\EndProcedure
\end{algorithmic}
\caption{ProBE Entropy-based Mechanism. $Q=\{Q_1,...,Q_n\}$, $u=\{u_1,...,u_n\}$, $o= \{o_1,...,o_n\}$, $\Delta g=\{\Delta g_1,...,\Delta g_n\}$, $t=\{t_1,...,t_{n-1}\}$  where $t_i=0$ if conjunction, $1$ if disjunction}\label{algo:ddpwlm}
\end{algorithm}

\section{DETAILED RELATED WORK}
Differential Privacy \cite{Dwork:2014:AFD:2693052.2693053,dwork2} has become a well-studied standard for privacy-preserving data exploration and analysis \cite{relatedDP,relatedDP2,relatedDP3,relatedDP4}. Various work has proposed frameworks for answering specific query types, such as range queries \cite{rangequery1,rangequery2,rangequery3}, and linear counting queries \cite{countquery1,countquery2,matrix}. This body of work, however, is not applicable to the aggregate threshold queries that our solution considers, nor does it take the approach of accuracy-first, which aims to minimize privacy loss given an accuracy constraint. Join queries may be more applicable to such queries, as conjunction-only queries may be written as join queries. However, such an approach complicates the queries at hand: the join operator is known for having high sensitivity resulting in low accuracy when answering such queries. Additionally, recent algorithms tackling join queries \cite{join1,join2,join3,join4} do not provide accuracy guarantees in terms of FNR or FPR, and their data-dependent nature make it difficult to successfully enforce error bounds.

 Multiple accuracy-constrained systems for differentially private data analysis have been proposed in recent years \cite{Apex:2019:RQP:1007568.1007642,dpella,cache,ex-postdp}, which allow data analysts to interactively specify accuracy requirements over their queries while providing a formal privacy guarantee. However, these solutions do not specifically focus on decision support queries and do not take into account their specific accuracy requirements. CacheDP \cite{cache} proposes a query engine that uses previous answers stored in a differentially private cache to lower the overall privacy budget, but this framework does not support aggregate threshold queries, and hence is not applicable to our problem of complex DS queries. The system APEx \cite{Apex:2019:RQP:1007568.1007642}, and programming framework DPella \cite{dpella} both allow data analysts to provide accuracy bounds for differentially private query answering, and both support aggregate threshold queries (also referred to as iceberg counting queries), but neither support complex decision support queries which are compositions of atomic queries, nor do they incorporate the asymmetric utility characteristic (i.e. the importance of the false negative rate) that decision support queries have. Ligett et al. \cite{ex-postdp}, which we extend in our work through the use of the ex-post DP notion, uses empirical error to determine the best privacy budget $\epsilon$. Such an approach is not suitable for aggregate threshold queries, as the sensitivity of the error would result in a high privacy loss, and the testing for empirical error incurs an additional cost to privacy. 

The problem of incorporating differential privacy in decision support has been tackled in MIDE \cite{mide}, which our work extends. MIDE \cite{mide} makes use of the asymmetric utility feature in DS applications to formally define strict accuracy guarantees in terms of the false negative rate $\beta$, and proposes novel mechanisms to answer DS queries at a minimal privacy loss while upholding such guarantees. However, this work only addressed simple atomic queries, rather than complex DS queries composed of multiple aggregate statistics evaluated against their respective thresholds. Other work such as Fioretto et al. \cite{fairness_ds} also studies decision support on differentially private data, but does so from a fairness lens. It tackles aggregate threshold queries, but proposes solutions to mitigate bias resulting from making decisions on such data, rather than optimizing privacy loss or providing formal accuracy guarantees.

\section{LAGRANGE MULTIPLIERS METHOD}
The Lagrange Multipliers method \cite{lagrange} is a optimization technique used to maximize or minimize multivariate functions subject to equality or inequality constraints. Given an objective function $f$, the goal is to find the extremum of such a function given a constraint function $g$. The intuition behind this method is that if $f$ and $g$ attain an extremum at $x^\ast$, then one of the level curves of $f$ and $g$ are tangent at $x^\ast$, meaning that their gradients $\nabla f$ and $\nabla g$ are parallel. This implies that the gradient of the constraint function $\nabla g$ is a multiple of the objective function gradient $\nabla f$, and this multiple is referred to as a Lagrange Multiplier $\lambda$. The relationship between the gradients and the multiplier is depicted in the Lagrangian Function, which is the equation:
\begin{equation*}
    \nabla f(x)= \lambda \nabla g(x)
\end{equation*}
Solving this equation for the multiplier $\lambda$ thus yields the optimal variable expression which can be substituted into the constraint function to obtain the corresponding extrema. Formally,
\begin{theorem}
[Lagrange Multipliers Method] Let $f: \mathbb{R}^d \rightarrow \mathbb{R}$ be the objective function, and $g: \mathbb{R}^d \rightarrow \mathbb{R}^n$ be the constraint function where both functions are $C^1$. If $f$ attains a local extremum at $x^{\ast}$ such that $rank(D(g(x^\ast)) = n$ given the constraint function, then there exists a unique multiplier $\lambda$ such that:
\begin{equation}
    \nabla f(x^\ast)= \lambda \nabla g(x^\ast)
\end{equation}
\end{theorem}

\section{ADDITIONAL EXPERIMENTS}
% \textcolor{blue}{As we have included the results of ProBE-Ent in the main paper in terms of ex-post DP loss $\epsilon$, we include the Min-Entropy results in the experiment below.}

\begin{table*}[!h]
\small
% \resizebox{\textwidth}{!}{
\begin{tabular} { |c|c|c|c|c|c| }
\hline
{\bf Overall $\beta$ }& {\bf Phase One $\beta$ $ \% $} & {\bf Phase Two $\beta $ $\%$} & {\bf Privacy loss $\epsilon$} & {\bf FNR} & {\bf FPR} \\
\hline
\multirow{5}{*}{$5 \cdot 10^{-2}$}& 10 & 90 & $4.55 \cdot 10^{-2}$ & 0 & $1.81 \cdot 10^{-3}$
 \\ 
\cline{2-6}
& 40 & 60 & $4.56 \cdot 10^{-2}$ & 0 & $3.03 \cdot 10^{-3}$
\\ 
\cline{2-6}
& 50 & 50 & $4.44 \cdot 10^{-2}$ & 0 & $2.40 \cdot 10^{-3}$ 
\\
\cline{2-6}
& 60 & 40 &  $4.80 \cdot 10^{-2}$ & 0 & $3.01 \cdot 10^{-3}$ 
\\
\cline{2-6}
& 90 & 10 &  $4.16 \cdot 10^{-2}$& 0 & $5.45 \cdot 10^{-3}$ 
\\
\hline
\multirow{5}{*}{$5 \cdot 10^{-4}$}& 10 & 90 & $5.46 \cdot 10^{-2}$ & 0 & $1.21 \cdot 10^{-3}$
 \\ 
\cline{2-6}
& 40 & 60 & $5.66 \cdot 10^{-2}$ & 0 & $1.81 \cdot 10^{-3}$
\\ 
\cline{2-6}
& 50 & 50 & $5.12 \cdot 10^{-2}$ & 0 & $4.84 \cdot 10^{-3}$ 
\\
\cline{2-6}
& 60 & 40 &  $5.32 \cdot 10^{-2}$ & 0 & $3.63 \cdot 10^{-3}$ 
\\
\cline{2-6}
& 90 & 10 &  $5.15 \cdot 10^{-2}$& 0 & $1.21 \cdot 10^{-3}$ 
\\
\hline
\multirow{5}{*}{$5 \cdot 10^{-6}$}& 10 & 90 & $8.15 \cdot 10^{-2}$ & 0 & $1.2 \cdot 10^{-3}$
 \\ 
\cline{2-6}
& 40 & 60 & $8.36 \cdot 10^{-2}$ & 0 & $3.07 \cdot 10^{-4}$
\\ 
\cline{2-6}
& 50 & 50 & $8.12 \cdot 10^{-2}$ & 0 & $1.21 \cdot 10^{-3}$ 
\\
\cline{2-6}
& 60 & 40 &  $8.18 \cdot 10^{-2}$ & 0 & $6.06 \cdot 10^{-4}$ 
\\
\cline{2-6}
& 90 & 10 &  $8.09 \cdot 10^{-2}$& 0 & $1.12 \cdot 10^{-3}$ 
\\
\hline
\end{tabular}
\caption{Privacy loss and Accuracy when varying the $\beta$ budget distribution across the two ProBE Phases}
\label{tab:varbeta}
\end{table*}

\textbf{Min-Entropy Privacy Results}. We extend our experiments to include Min-Entropy denoted $\mathcal{H}_{min}(\Theta)$ as part of our privacy metrics used to evaluate the performance of all algorithms. We use the same parameters as the main paper: we set a FNR bound of $\beta = 0.05$, a FPR bound of $\alpha= 0.1$, a maximum privacy loss of $\epsilon_{max} = 5$, a starting uncertain region of  $u_0=30\%|D|$ for ProBE-based algorithms (the original ProBE algorithm, the ProBE-Naive variation with the Naive first step, and the ProBE-Ent entropy-based variation) and a default $u = 12\%$ for the Naive algorithm. We run each algorithm over a 100 iterations and vary the number of sub-queries from 1 to 6. 
\begin{figure*}
\captionsetup[subfigure]{justification=centering}
    \centering
\begin{subfigure}[b]{.80\textwidth}
    \centering
    \includegraphics[width=0.25\textwidth]{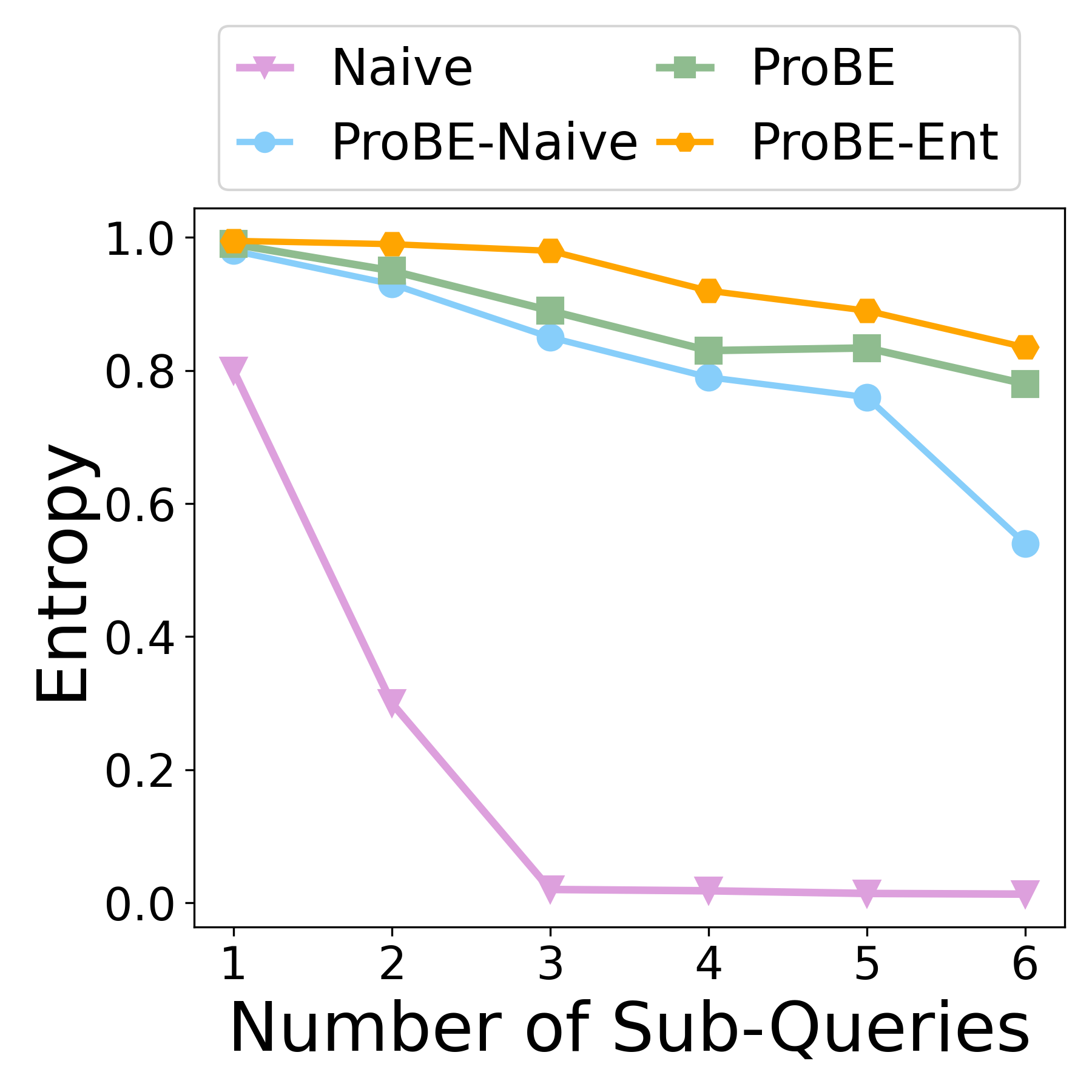}
    % \end{subfigure}
    %  \begin{subfigure}[b]{.25\textwidth}
    % \centering
     \includegraphics[width=0.25\textwidth]{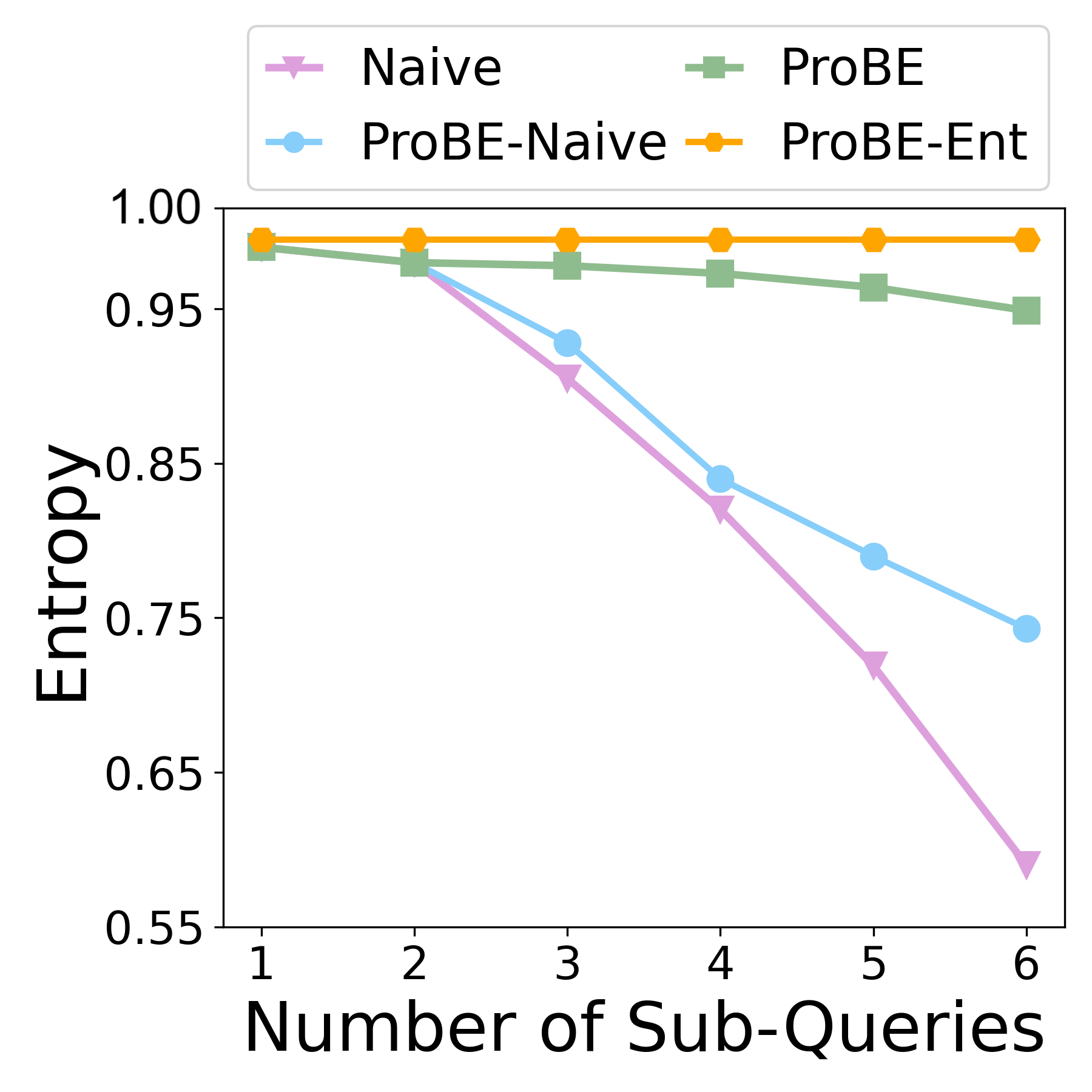}
     % \end{subfigure}
     % \begin{subfigure}[b]{.25\textwidth}
     % \centering
     \includegraphics[width=0.25\textwidth]{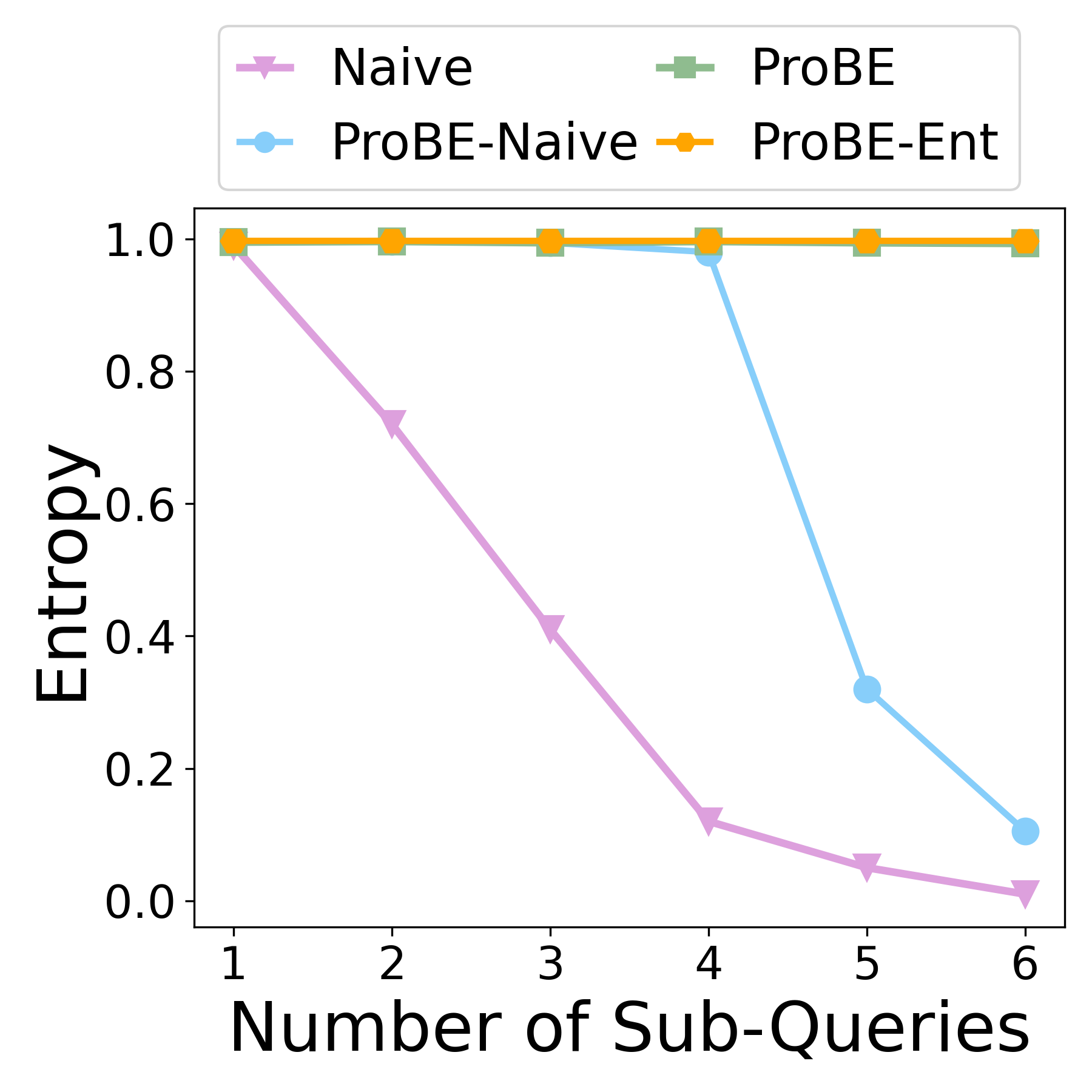}
    % \caption{NYTaxi data Epsilon (top) and \\ Min-Entropy (bottom)}
    \end{subfigure}
    
        \begin{subfigure}[b]{.80\textwidth}
    \centering
    \includegraphics[width=0.25\textwidth]{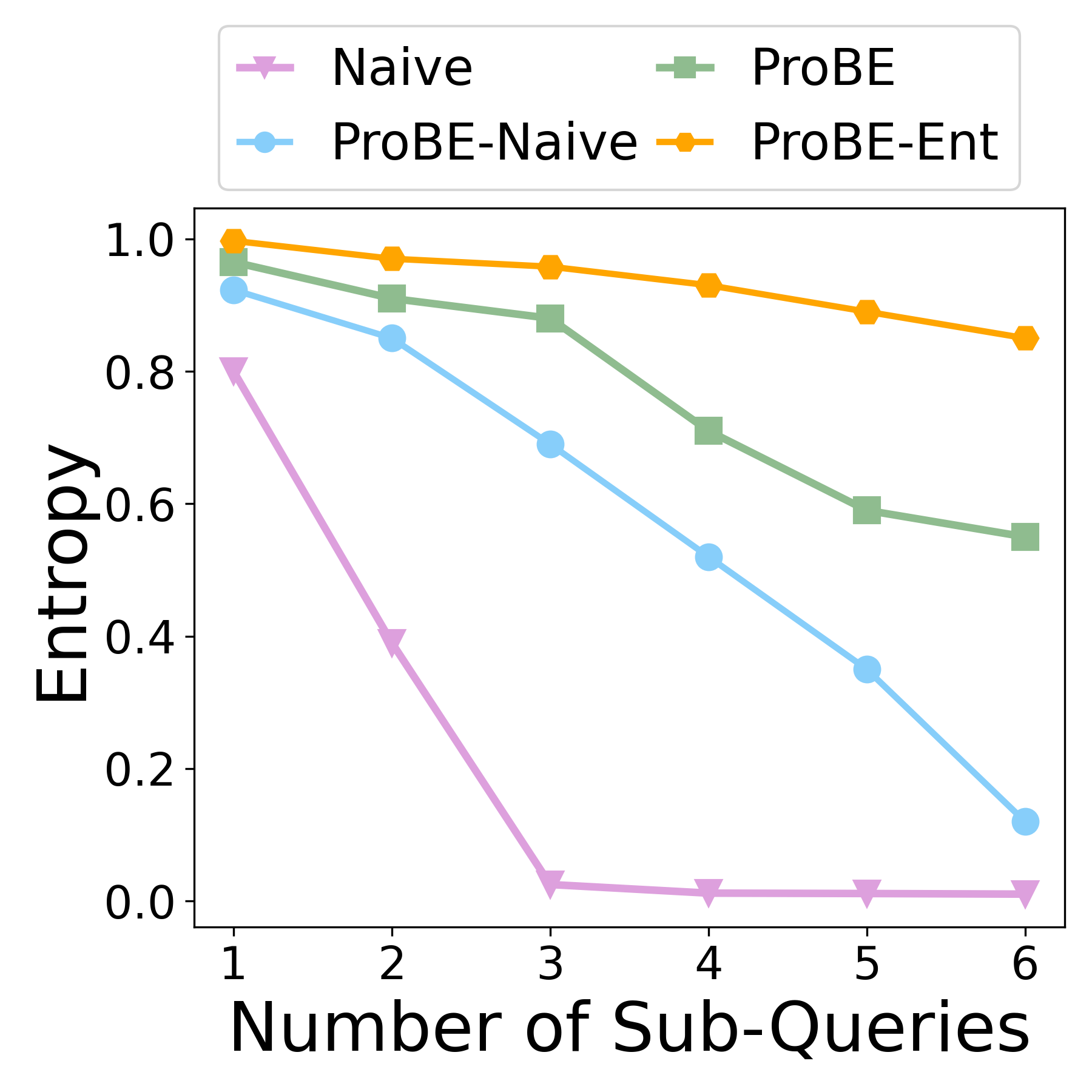}
     % \end{subfigure}
     % \begin{subfigure}[b]{.25\textwidth}
     % \centering
     \includegraphics[width=0.25\textwidth]{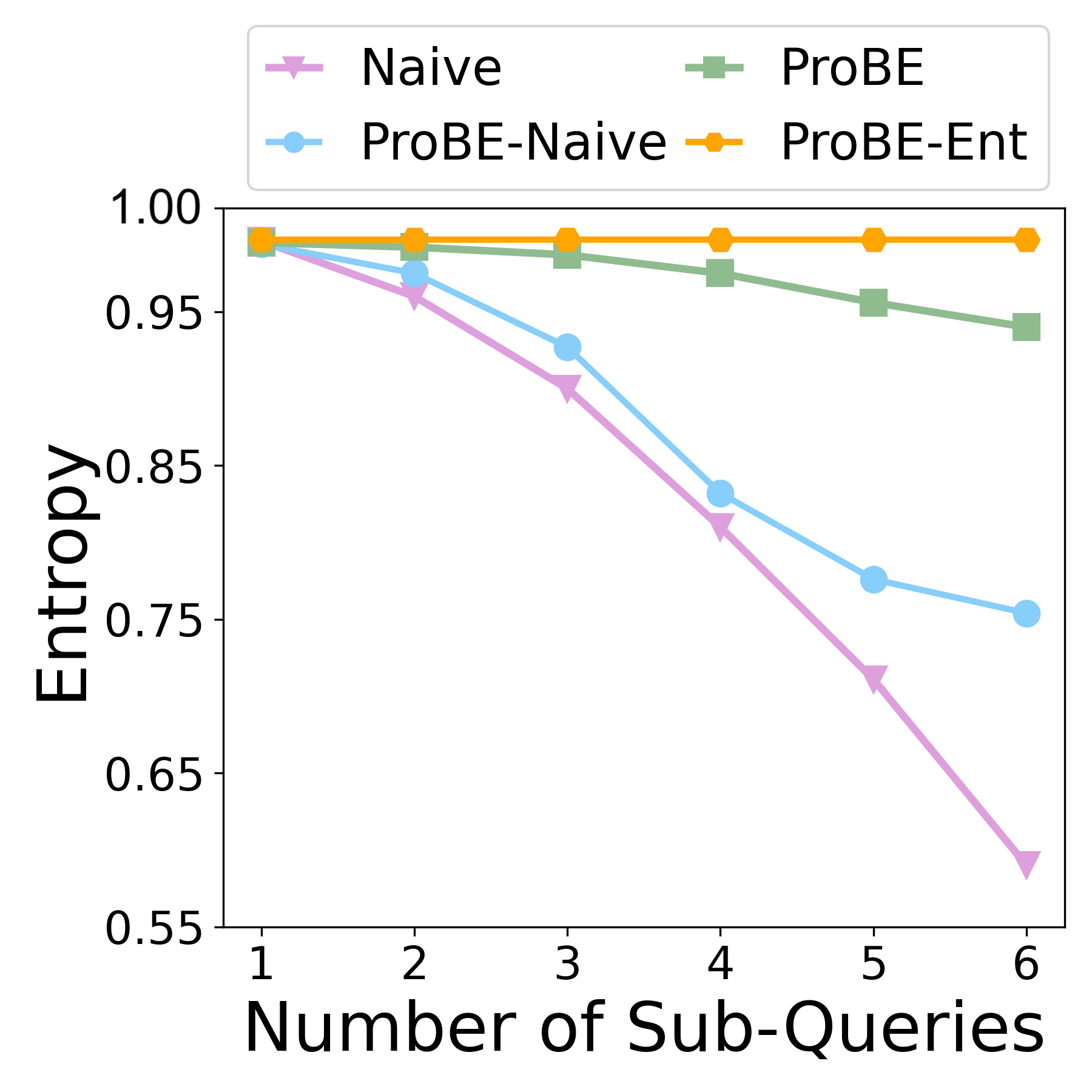}
     %  \end{subfigure}
     % \begin{subfigure}[b]{.25\textwidth}
     % \centering
     \includegraphics[width=0.25\textwidth]{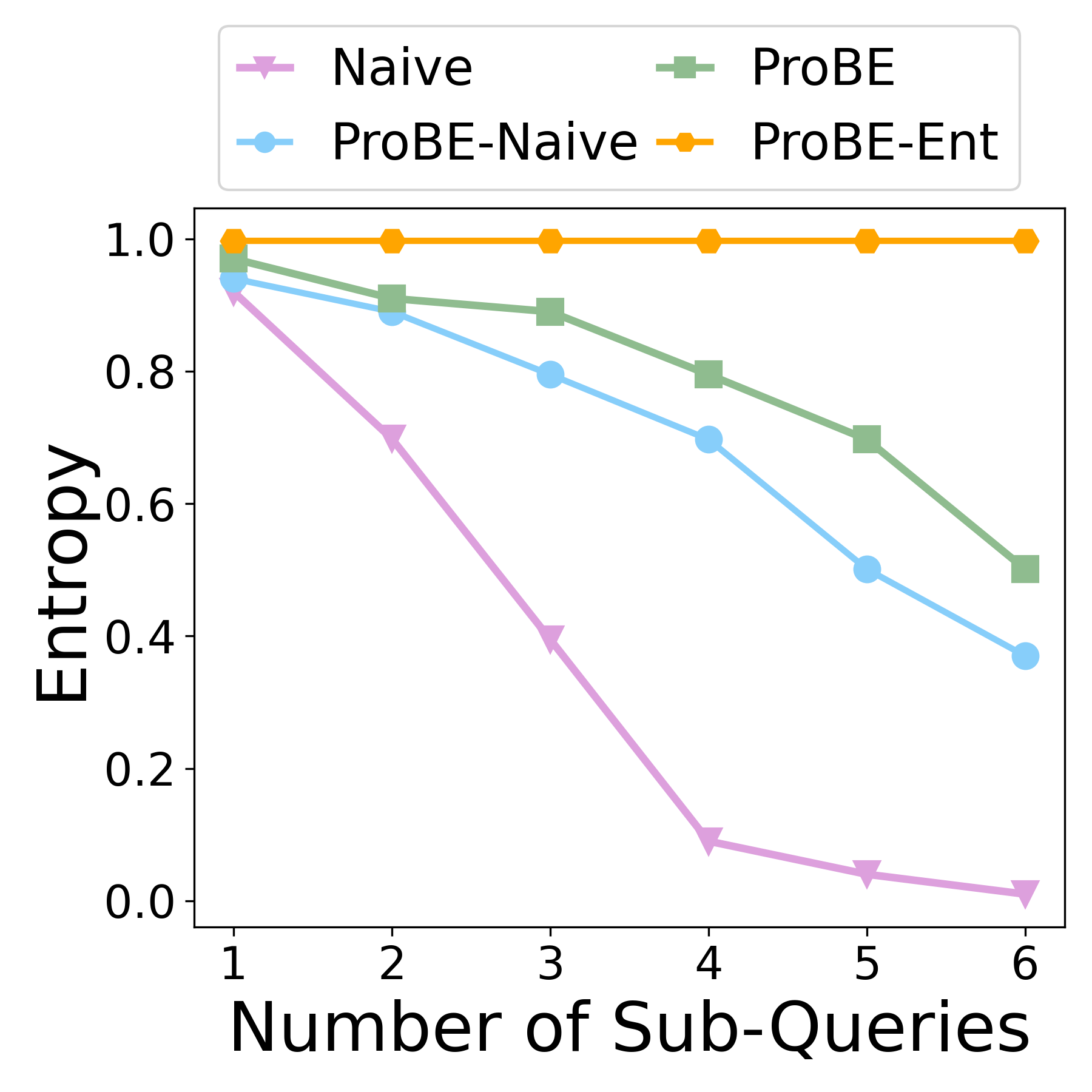}
    % \caption{NYTaxi data Epsilon (top) and \\ Min-Entropy (bottom)}
    \end{subfigure}
    \caption{Privacy Loss in terms of Min-Entropy $\mathcal{H}_{min}(\Theta)$ for 1-6 sub-queries at $\beta=0.05$ and $\alpha=0.1$ for Conjunction (row 1) and Combined Conjunction/Disjunction (row 2) using NYTaxi (column 1), Sales (column 2) and UCI (column 3) data. }
    \label{fig:privacyresapp}
\end{figure*}

In terms of Min-entropy $\mathcal{H}_{min}(\Theta)$, the closer to $1$ the results are (meaning that the lower bound on uncertainty is extremely high) the better. The results in Figure \ref{fig:privacyresapp} thus show that ProBE-Ent outperforms other algorithms and stays constant or suffers only from a small decrease as the number of sub-queries increases. This directly follows from the fact that ProBE-Ent is designed to maximize min-entropy at each iteration, regardless of the number of sub-queries. Other algorithms suffer from a decrease in min-entropy as the number of sub-queries increase, as it is inversely correlated to the ex-post DP loss $\epsilon$.

\textbf{Varying $\beta$ budget split across phases}. We evaluate the privacy loss and accuracy measures of our proposed ProBE algorithm when varying the distribution of the overall $\beta$ bound budget across the two phases of the algorithm. We use the following combinations of percentages:  $\{10\%-90\%$, $90\%-10\%$, $50\%-50\%$, $40\%-60\%$, $60\%-40\%$, $\}$ where the first percentage is allocated to Phase One of ProBE and the second is allocated to Phase Two. We use a simple disjunction query $Q_1 \cup Q_2$ on the Taxi dataset with the fixed parameter of $\alpha=0.1$ and vary the $\beta$ value from the default $\beta=5 \cdot 10^{-2}$ to a much smaller value of $5 \cdot 10^{-6}$. Our results are shown in Table \ref{tab:varbeta}. The different distributions used for the $\beta$ budget do not significantly impact the privacy loss $\epsilon$. There is no direct correlation observed between privacy loss and the $\beta$ distribution across the two phases, as we observe that although the lowest privacy loss achieved is when $90\%$ of $\beta$ is allocated to the Phase One, this is not the case for the $60\%$ allocation. We additionally note that the FPR is the highest when Phase One is allocated $90\%$ of $\beta$ and lowest when $90\%$ is allocated to Phase Two, but the FPR bound of $\alpha = 0.1$ is upheld regardless of the split. As predicted from the inverse relationship between privacy loss and the FNR bound of $\beta$, the smaller the $\beta$ value, the higher privacy loss $\epsilon$ is as seen in Table \ref{tab:varbeta}. When using smaller values of $\beta$, we note that the FNR as well as FPR bounds are still upheld, but the optimal distribution of the $\beta$ budget across phases cannot be discerned despite varying $\beta$ values and remains a non-trivial optimization problem.

% \textbf{Disjunction-only privacy and accuracy results for a varying number of sub-queries}. We use the same privacy metrics (ex-post privacy loss $\epsilon$ and min-entropy $\itbeta(\Theta)$) and accuracy metrics (FNR and FPR) to assess the performance of the Naive, Probe-based TSLM and Probe-based MSPWLM using both PPWLM and DDPWLM when varying the number of sub-queries from 1 to 6. We use the same default values for experimental parameters, $u=12$ and $\beta=0.005$, as well as $m=4$ steps for the progressive algorithm MSPWLM and $m_f=3$ fine-grained steps for the data dependent version across the Sales, UCI and NYTaxi datasets. Results for disjunction-only queries are consistent with other query types, with the data-dependent MSPWLM achieving the lowest or near lowest privacy loss and highest entropy out of all algorithms as the number of sub-queries increases. The false negative rate (FNR) achieves the bound of $\beta=0.005$ for all algorithms and the trade-off between the FNR and FPR is at most 0.20, with the FPR being higher for MSPWLM due to the trade-off between FPR and privacy loss $\epsilon$.

\begin{figure*}
\captionsetup[subfigure]{justification=centering}
    \centering
    \begin{subfigure}[b]{.22\textwidth}
    \centering
    \includegraphics[width=\textwidth]{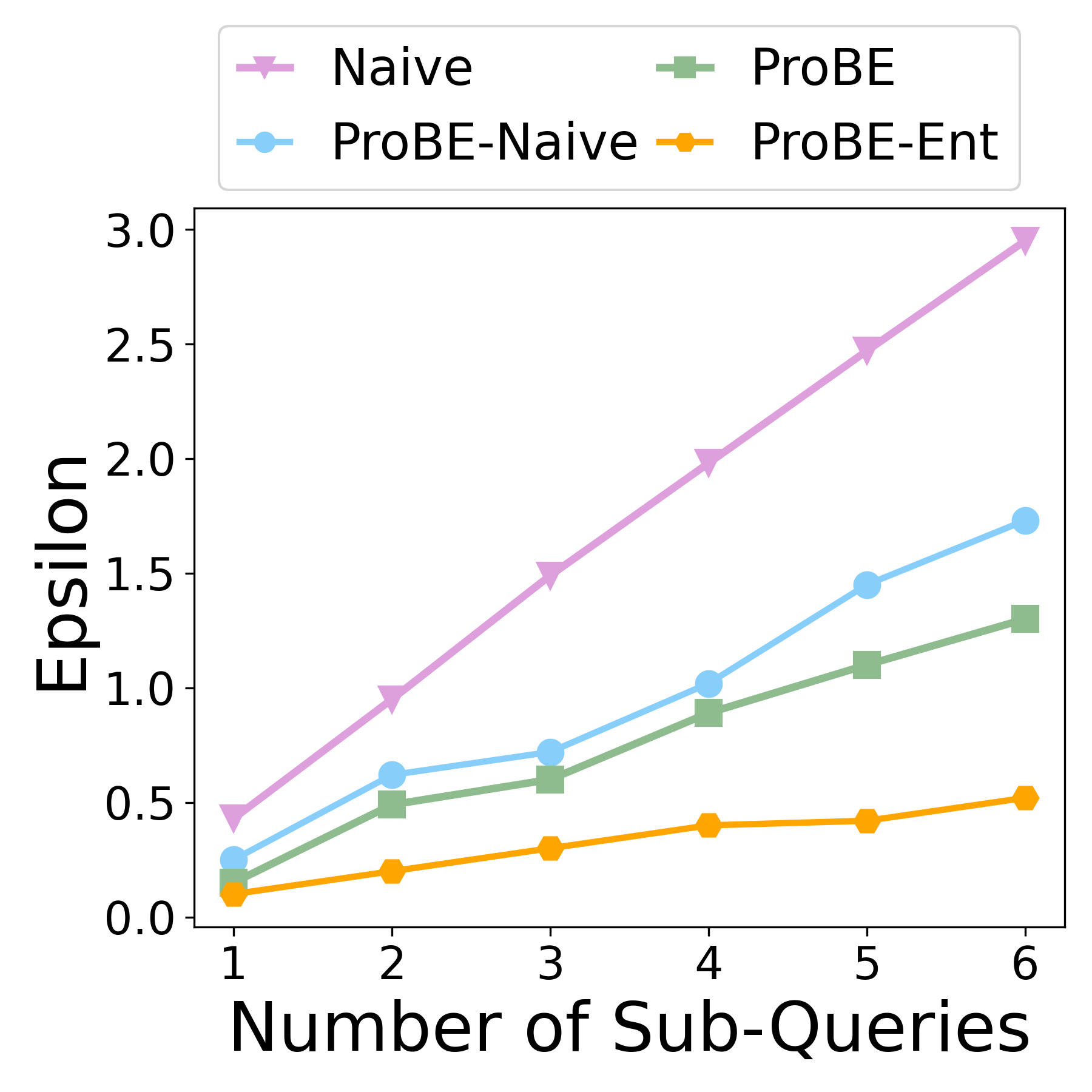}
    \caption{Ex-Post DP $\epsilon$}
    \label{fig:avgEp}
    \end{subfigure}
    \begin{subfigure}[b]{.220\textwidth}
    \centering
    \includegraphics[width=\textwidth]{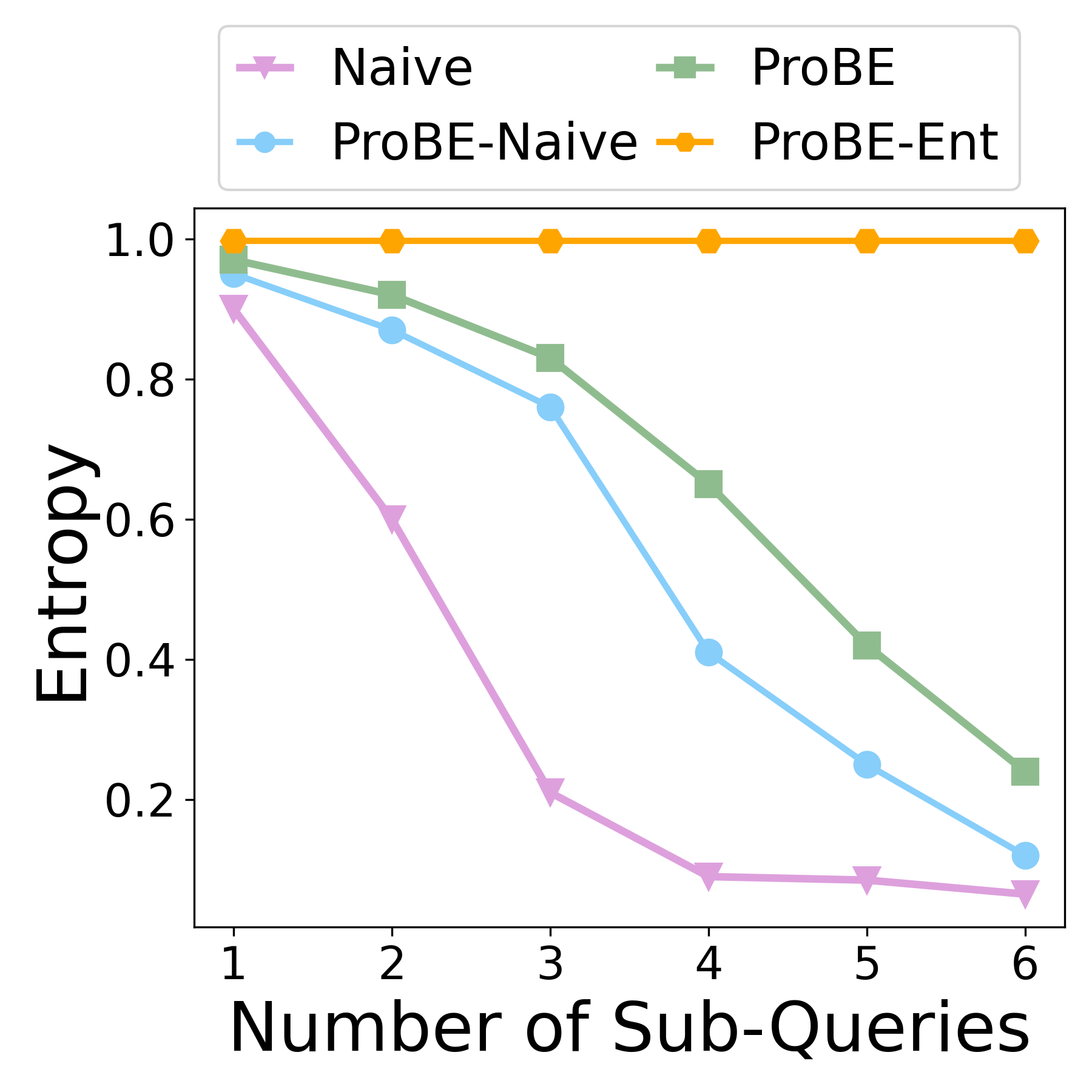}
    \caption{Min-Entropy $\itbeta(\Theta)$}
    \label{fig:avgEnt}
    \end{subfigure}
    \begin{subfigure}[b]{.25\textwidth}
    
    \centering
    \includegraphics[width=\textwidth]{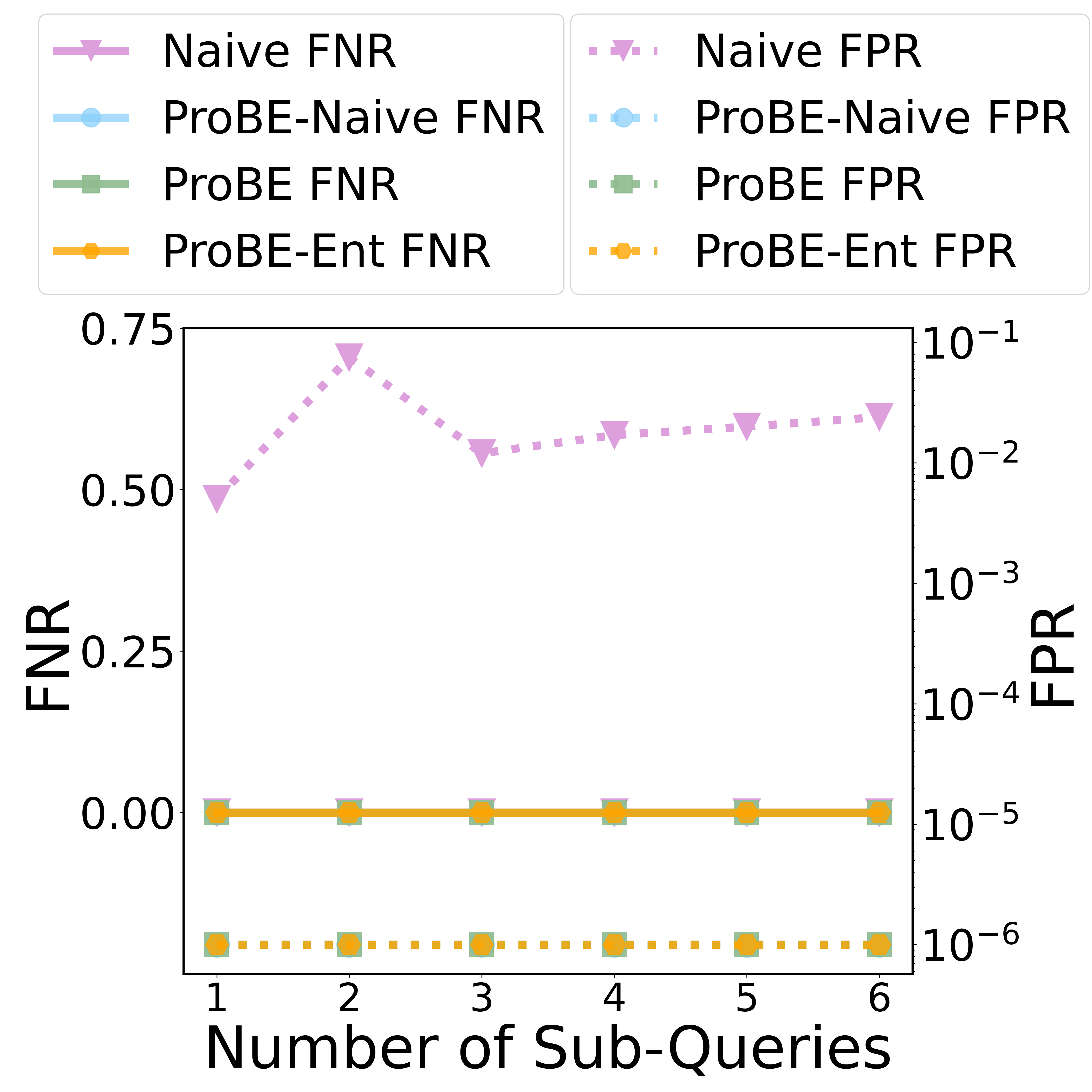}\\[0.5mm]
   \caption{FNR and FPR}
   \label{fig:avgFNRFPR}
   
    \end{subfigure}
    \caption{Privacy ($\epsilon$,$\itbeta(\Theta)$) and Accuracy (FNR, FPR) Results for AVG functions at $\beta=0.05$ and $\alpha=0.1$ on the Sales dataset.}
    \label{fig:avg}
    
\end{figure*}

% \noindent
\textbf{AVERAGE aggregate function conjunction-only privacy and accuracy results}. We model queries for the Sales dataset which use the AVERAGE aggregate function based on meaningful KPIs typically used in decision support applications (e.g. average transaction value). We include the results for conjunction-only complex queries in Figure \ref{fig:avg} with a varying number of sub-queries from 1 to 6. We use the same default parameters to evaluate the performance of the Naive, ProBE, ProBE-Naive and ProBE-Ent algorithms. We set $u=12\%$ for single-step algorithms, $u_0=30\%$ for multi-step algorithms and $\beta=0.05$, $\alpha=0.1$, as well as $m=4$ steps and $m_f=3$ fine-grained steps for ProBE-Ent. Our privacy results in terms of ex-post privacy loss $\epsilon$ and min-entropy $\mathcal{H}_{min}$ show that AVERAGE queries have similar results as COUNT queries but with a higher range of values for $\epsilon$, with ProBE-Ent having the lowest privacy loss and highest min-entropy. The higher average for privacy loss is due to the sensitivity of the query being higher (AVG often has a higher sensitivity than COUNT). The accuracy results in terms of false negative rate (FNR) and false positive rate (FPR) show that all algorithms again achieve the bounded guarantees of $\beta=0.05$, and the multi-step algorithms achieve the FPR bound of $\alpha$.

\end{document}